\documentclass[aps,pre,reprint,superscriptaddress,nofootinbib]{revtex4-2}

\usepackage{graphicx}
\graphicspath{{figures/}}
\usepackage{float} 
\usepackage[T1]{fontenc} 
\usepackage{lmodern}  
\usepackage{hyperref}
\usepackage{csquotes}
\usepackage{amsmath}
\usepackage{amsfonts} 
\usepackage{amssymb}
\usepackage{textcmds}
\usepackage{dsfont}
\usepackage{caption}
\usepackage{subcaption}
\usepackage{ragged2e}
\usepackage[percent]{overpic}
\usepackage{xcolor}
\usepackage{multirow}
\makeatletter
\renewcommand{\@makecaption}[2]{%
  \vskip\abovecaptionskip
  \parbox{\linewidth}{\justifying #1: #2\par}
  \vskip\belowcaptionskip
}
\makeatother
\usepackage{booktabs}

\usepackage[usenames,dvipsnames]{xcolor}
\hypersetup{colorlinks=true, linkcolor=BrickRed, urlcolor=blue!50!black, citecolor=blue!50!black}

\usepackage{amsmath,amsthm,amsfonts,amssymb,times,bbm,graphicx,color,epsfig,bm}
\usepackage{url} 
\usepackage{hyperref}
\renewcommand{\vec}[1]{\textbf{\textrm{#1}}}

\usepackage[normalem]{ulem}

\DeclareMathAlphabet\mathbfcal{OMS}{cmsy}{b}{n}

\renewcommand{\H}{\mathcal{H}}
\newcommand{\Hr}{\H_r}

\usepackage{xcolor}

\newif\ifshownotes
\shownotestrue 

\begin{document}

\title{Energy landscape and dissipation of sliding magnetic rotor arrays}
\author{Johannes Krotz}
\email{jkrotz@nd.edu}
\affiliation{Department of Aerospace and Mechanical Engineering, University of Notre Dame, Notre Dame, USA}
\author{Anton L\"uders}
\email{anton.lueders@uibk.ac.at}
\affiliation{Institut f\"ur Theoretische Physik, Universit\"at Innsbruck, Technikerstra{\ss}e, 21A, A-6020 Innsbruck, Austria}

\begin{abstract}
Magnetically coupled damped rotational degrees of freedom can give rise to tribological loss when excited by a relative sliding motion, linking friction to the dynamics of magnetic moments. We analytically investigate such magnetic friction by studying a recently introduced simplified model for a rigid magnetic rotor array sliding over a commensurate magnetic substrate. Here, the array rotors can rotate about an axis perpendicular to the sliding direction and are damped by microscopic shaft friction, while the substrate magnets have a fixed in-plane orientation. The simplified model reduces the collective rotor dynamics of this setup to a set of coupled nonlinear differential equations, which can be studied by the corresponding energy landscape in the quasi-static regime. We find that the parameter plane spanned by layer separation and time can be divided into distinct chambers, in which the structure of the energy landscape remains invariant. Using these chambers, we identify the layer separation intervals corresponding to different dynamical states and recover the emergence of a regime where the moment alignment alternates globally while exhibiting peak dissipation. Additionally, we derive the leading nontrivial orders of the collective magnetic friction in the regimes of small and large gaps between the array and the substrate. For the alternating regime, we find that friction can be computed by tracking energy jumps from unstable to stable critical points at the chamber boundaries. These results provide an analytical foundation for the observed dynamics and dissipation in sliding rotor arrays. Because the simplified model is scale-free, our conclusions transfer across a broad range of microscopic and macroscopic length scales.
\end{abstract}

\maketitle

\section{Introduction}

Sliding friction occurs when the surfaces of two layers are in direct contact or interact via fields while moving relative to each other. The corresponding dissipated energy is fundamentally encoded in the morphology of the layer materials and configurational patterns on the surfaces. As friction is therefore intrinsically linked to structure, surprising phenomena and dissipation anomalies can emerge when particle configurations and internal degrees of freedom suddenly or continuously change during sliding. For instance, an increase in frictional forces has been predicted~\cite{Benassi2011, Igloi2011} and experimentally measured via scanning probe microscopy for substrate layers that are close to continuous~\cite{Kisiel2015} and first-order~\cite{Panizon2018, Wang2023} phase transitions.

Systems where such structural changes can be linked to reorientations of rotational degrees of freedom are magnetic materials. Here, the phenomenon of magnetic friction~\cite{Heinrich2012, Wolter2012, Li2018, Xiaolin2016} can emerge when magnetic interactions between tip-substrate setups or sliding layers result in spin excitations, domain wall motion, or collective dynamics~\cite{Gruetter1997, Fusco2008, Magiera2009, Demery2010, Rissanen2019, Gu2026}. The associated energy dissipation and nonequilibrium phase behavior have been extensively explored in theoretical studies, often using model systems such as Ising layers~\cite{Kadau2008, Komatsu2020, Komatsu2021, Komatsu2023} or Heisenberg spin assemblies~\cite{Magiera2011, Magiera2011-2, Magiera2013, Magiera2014}. In these approaches, the materials are typically driven out of equilibrium either by relative motion, such as sliding~\cite{Li2016, Rissanen2018, Komatsu2019, Sugimoto2019} and rotation~\cite{Hilhorst2011}, or by imposed shear~\cite{Hucht2009, Angst2012, Hucht2012}.

A recently introduced macroscopic model system enables the study of magnetic friction on a length scale accessible with the naked eye~\cite{Gu2026, Nigues2026}. The setup consists of an array of damped magnetic rotors (the slider) that moves relative to a substrate containing a commensurate lattice of millimeter-sized magnets~\cite{Gu2026}. Each rotor on the slider is itself a millimeter-scale magnet that can freely rotate under microscopic shaft friction about an axis perpendicular to the sliding direction. In contrast, the magnets embedded in the substrate have fixed orientations with in-plane magnetization. A detailed schematic of the experimental configuration can be found in Ref.~\cite{Gu2026} and is shown in Fig.~\ref{fig:sket}.

Sliding the rotor array relative to the substrate reveals a pronounced peak in magnetic friction with a distinctly non-monotonic dependence on the layer separation~\cite{Gu2026, Nigues2026}. Interestingly, maximum friction occurs when the slider magnets undergo frequent changes in the collective order, which aligns with the idea of dissipation anomalies at configurational transitions~\cite{Benassi2011, Igloi2011}. At small separations, the magnetic moments of the slider align parallel, forming a ferromagnetic state that rotates synchronously with the sliding motion; in this regime, the friction remains weak. At large separations, the slider settles into a static antiferromagnetic configuration in which neighboring moments are parallel along the sliding direction but antiparallel across it, resulting in negligible rotor motion and correspondingly low friction. At intermediate separations, however, the system repeatedly switches between the ferromagnetic and antiferromagnetic states, leading to hysteresis, which dissipates energy.

To support the experimental observations and numerical simulations, a simplified theoretical model was introduced that reduces the system to two effective degrees of freedom~\cite{Gu2026}. The model retains only nearest-neighbor interactions and groups the magnetic moments into two sublattices. All moments within each sublattice are assumed to be identical, which substantially reduces the complexity of the problem. Despite its minimal construction, the resulting equations of motion successfully capture the three characteristic regimes observed in the full system.

So far, the simplified model was mostly studied numerically, and a comprehensive analytical treatment of the model is still missing. In this work, we provide such an analysis by examining the energy landscape of the simplified rotor-array model. In particular, we show that the parameter plane spanned by the layer separation and time can be rigorously divided into distinct chambers. Within each chamber, the energy landscape remains structurally invariant: the number and the Morse type~\cite{Milnor1963MorseTheory} of critical points, as well as the relative orientation of the magnetic moments corresponding to a particular critical point, are constant. Using this chamber decomposition, we can infer the dynamical behavior of the slider, mapping out the exact layer separation intervals that correspond to each dynamical regime. In addition, we analytically derive the nontrivial leading order terms of the magnetic friction for the regimes of particularly small and large gaps between the slider and the substrate. In the regime of an alternating magnetic arrangement, we derive that the magnetic friction can be calculated by tracking the perturbation-triggered energy jumps from unstable to stable critical points at the boundaries of the different chambers.

This work is structured as follows: First, Sec.~\ref{SEC:found} introduces a generalized, dimensionless version of the simplified model from Ref.~\cite{Gu2026}, which uses $N$ degrees of freedom for an $N\times N$ rotor array. For readers unfamiliar with the findings of Ref.~\cite{Gu2026}, Sec.~\ref{SEC:regimes} provides a brief summary of the expected dynamical behavior and magnetic friction for the rotor array-substrate setup. In Sec.~\ref{SEC:LANDSCAPE}, we derive the chamber decomposition of the energy landscape for a two-degree-of-freedom model by calculating both the bifurcation curves of the critical points and the boundaries where the relative alignment of the magnetic moments changes. We also validate these analytical findings numerically. Then, we derive estimates for the magnetic friction across the different dynamical regimes in Sec.~\ref{SEC:FRICTION}. Finally, Sec.~\ref{SEC:DISCUSSION} concludes the paper with a discussion of our results and their broader implications.

\section{Foundations of the simplified model} \label{SEC:found}

\begin{figure}[ht!]
\includegraphics[width=0.5\textwidth]{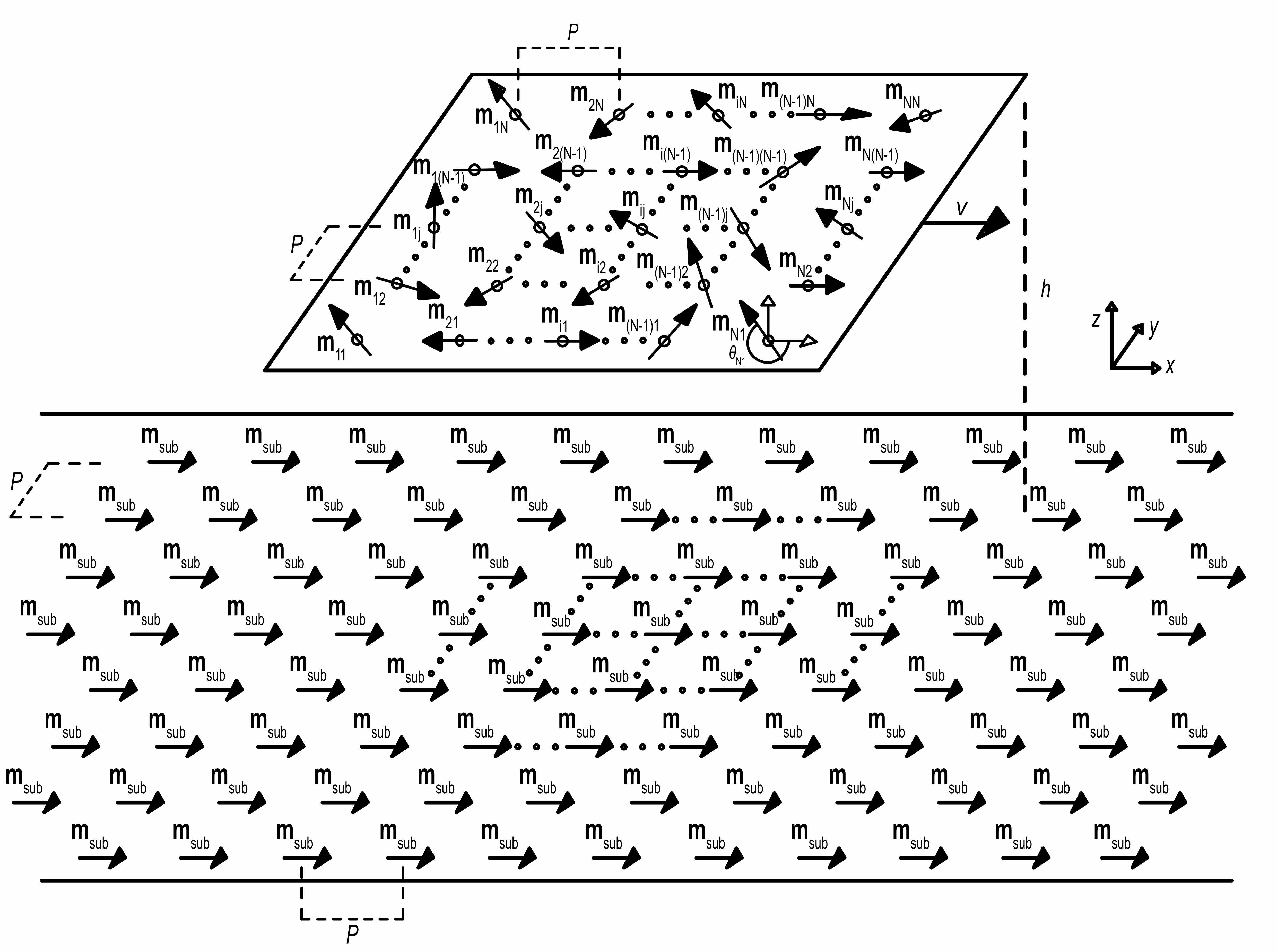}
\caption{Sketch of the slider-substrate setup considered in this work. Two magnetic layers, each containing a square lattice of magnetic moments with lattice constant $P$, move relative to one another with velocity $v$. The substrate moments $\vec{m}_{\textrm{sub}}$ have a fixed orientation along the positive $x$-direction, while the slider moments $\vec{m}_{ij}$ can rotate about the $y$-axis. The index $i$ labels the position along the $x$-axis, and the index $j$ labels the position along the $y$-axis. The layer separation is denoted by $h$, and the rotational degrees of freedom of the slider moments are described by the angles $\theta_{ij}$.}
\label{fig:sket}
\end{figure}

We consider the slider-substrate setup as described in Ref.~\cite{Gu2026}, which consists of a square array of magnets with lattice constant $P$ and a commensurate magnetic \textit{substrate} (See Fig.~\ref{fig:sket} for a sketch of the setup). The \textit{slider} (i.e., the rotor array) contains $N \times N$ magnetic dipoles with magnetic moments $\vec{m}_{ij}$, each of which can rotate about the $y$-axis. Here, the first index $i$ labels the lattice position along the $x$-direction, which is also the direction of the slider's translation, while the second index $j$ labels the position along the $y$-direction. The dipole orientations are described by angles $\theta_{ij}$, measured relative to the positive $x$-axis and increasing under clockwise rotation [i.e., $ \vec{m}_{ij} \propto (\cos(\theta_{ij}),0,-\sin(\theta_{ij}))^{\textrm{T}} $ looking at the $ xz $ plane with the $ x $-axis pointing from left to right]. The internal magnetic energy of the slider is given by
\begin{equation}
    \mathcal{H}_I = \frac{1}{2}\sum\limits_{\substack{ijkm\\(i,j) \neq (k,m)}} E(\vec{m}_{ij}, \vec{m}_{km})
\end{equation}
with the interaction energy corresponding to two magnetic dipoles with center separation vector $ \vec{r} $
\begin{align}
    &E(\vec{m}_{ij}, \vec{m}_{km}) \nonumber\\
    &= \frac{\mu_0}{4\pi | \vec{r}|^3}
    \left[\vec{m}_{ij} \cdot \vec{m}_{km} -
        \frac{3 (\vec{m}_{ij}\cdot \vec{r})(\vec{m}_{km} \cdot \vec{r})}{|\vec{r}|^2}
    \right].
\end{align}
Although dipolar interactions are long-ranged, the results of Ref.~\cite{Gu2026} indicate that restricting the interaction to nearest neighbors is sufficient to qualitatively reproduce both the characteristic dynamical behavior and the non-monotonic peak in sliding friction. Additionally, the corresponding experiments already demonstrated that, for fixed $j$, the angles $\theta_{ij}$ are approximately equal when the system occupies the relevant minima of the energy landscape during sliding. In the following, we make use of these two observations to derive a simplified form of $\mathcal{H}_I$: we truncate the interaction energy at the nearest neighbors and reduce the number of degrees of freedom in direct analogy with Ref.~\cite{Gu2026}.

First, we simplify the interaction energy of the magnetic moments within the same row $j$. Assuming nearest-neighbor interactions and parallel magnetic moments, one readily finds
\begin{equation}
    \sum\limits_{i} E(\vec{m}_{ij}, \vec{m}_{(i+1)j}) 
    = \frac{\mu_0 m_0^2 N}{4\pi P^3} \left[ 3\sin^2(\theta_j) - 2 \right],
\end{equation}
where $\theta_{ij} = \theta_j$ for all $i$, $m_0 = |\vec{m}_{ij}|$ for all $i,j$, and we assume an even $N$ together with periodic boundary conditions $\vec{m}_{(N+1)j} = \vec{m}_{1j}$ for simplicity. The total interaction energy between two neighboring rows $j$ and $j+1$ can be simplified in the same manner, which yields
\begin{equation}
    \sum_{i} E(\vec{m}_{ij}, \vec{m}_{i(j+1)})
    = \frac{\mu_0 m_0^2 N}{4\pi P^3} \cos(\theta_j - \theta_{j+1}),
\end{equation}
again with periodic boundaries $\vec{m}_{i(N+1)} = \vec{m}_{i1}$. Consequently, the full internal interaction energy of the slider reduces to
\begin{align}
    \frac{\mathcal{H}_I}{\varepsilon}
    &= \sum_j \left[ 3\sin^2(\theta_j) + \cos(\theta_j - \theta_{j+1}) \right] + \text{const.} \nonumber \\
    &= \sum_j \left[ -\frac{3}{2}\cos(2\theta_j) + \cos(\theta_j - \theta_{j+1}) \right] + \text{const.}, \label{eq:HINN}
\end{align}
where $\varepsilon = \mu_0 m_0^2 N/4\pi P^3$ is the interaction energy scale.

The slider moves with constant velocity $v$ in the $x$-direction while hovering above a periodic substrate at a fixed separation $h$. The substrate consists of a square array of magnets with lattice constant $P$, whose magnetic moments are aligned along the $x$-direction. As argued in Ref.~\cite{Gu2026}, the interaction between the slider moments and the substrate moments can be effectively captured by a periodic driving term of the form~\footnote{Under the assumption that the substrate can be modeled as a rotating external field $ \vec{B}{\textrm{sub}} = -B_0(\cos(k v t),0,-\sin(kvt))^{\textrm{T}} $, the corresponding Zeeman-like energy term is given by $ - \vec{m}_{ij} \cdot \vec{B}{\textrm{sub}} = B_0\cos(\theta_{ij}-kvt) $. If we picture the substrate as a lattice of magnetic moments parallel to the $x$-direction, the external field is chosen such that the rotor magnets are anti-aligned at $ t = 0$.}
\begin{equation}
    \mathcal{H}_{\textrm{sub}} 
    = \sum\limits_j 
    \frac{\mu_0 m_0 m_{\textrm{sub}} N}{4\pi h^3}
    \cos(\theta_j - k v t).
\end{equation}
Here, $k = 2\pi/P$ and $m_{\textrm{sub}}$ denotes the magnitude of the substrate magnetic moments using a dipole approximation. The resulting simplified total energy is
\begin{align}
    \frac{\mathcal{H}}{\varepsilon}
    &= \sum\limits_j 
    \left[
        -\frac{3}{2}\cos(2\theta_j)
        + \cos(\theta_j - \theta_{j+1})
        + A\cos(\theta_j - \tilde{v} t)
    \right] \nonumber \\
    &+ \text{const.}, \label{eq:FULLH}
\end{align}
with $A = (P/h)^3 (m_{\textrm{sub}}/m_0)$ and $\tilde{v} = k v$~\footnote{Note that $ \tilde{v} $ has the physical dimension of an angular velocity and it is equal to the definition of $ \omega $ in Ref.~\cite{Gu2026}}.

The three terms in Eq.~\eqref{eq:FULLH} have distinct physical roles: The term proportional to $ -\cos(2\theta_j)$ favors alignment of each row along the $x$-direction, the coupling term $\cos(\theta_j-\theta_{j+1})$ favors alternating orientations between neighboring rows, and the final term couples each row to the phase of the drive, which mimics the substrate moving relative to the slider. The dimensionless \textit{separation parameter} $A$ therefore measures the relative strength of the substrate contribution to the total energy compared to the internal magnetic interaction of the slider. Large $A$ corresponds to small layer separation $ h $ and strong substrate coupling, while small $A$ corresponds to large separation $ h $ and weak substrate coupling.

Assuming overdamped dynamics and modeling dissipation due to the microscopic shaft friction via a Stokes drag, the equations of motion are
\begin{equation}
    \dot{\theta}_j
    = -\frac{1}{\Gamma}
    \frac{\partial \mathcal{H}}{\partial \theta_j}, \label{eq:gradient_flow}
\end{equation}
where $\Gamma = \varepsilon \gamma$~\footnote{Do not confuse $ \gamma = \Gamma/\varepsilon $ used in this work with the microscopic shaft friction coefficient of a single magnetic moment used in the simulations of Ref.~\cite{Gu2026}, which use the same symbol.} is the effective friction coefficient describing the total dissipation associated with the rotation of all moments in row $j$. In detail, assuming the microscopic shaft-friction coefficient for a single rotor is $ \xi $, the effective coefficient for a full line of moments is $ \Gamma = N \xi $~\footnote{In Ref.~\cite{Gu2026}, the microscopic shaft-friction coefficient $ \xi $ is denoted by the symbol $ \gamma $ instead.}.

Inserting the total energy function, the simplified equations of motion become
\begin{align}
    \gamma \dot{\theta}_j 
    & = -3\sin(2\theta_j)
    + \sin(\theta_j - \theta_{j+1})
    + \sin(\theta_j - \theta_{j-1}) \nonumber \\
    & \quad + A \sin(\theta_j - \tilde{v} t), \label{eq:EOMFULL}
\end{align}
which describe a set of overdamped, trigonometrically coupled mathematical pendulums driven by a traveling wave. Under the additional assumption of only two degrees of freedom --- namely, $\theta_j = \theta_{j+2} \equiv \theta_e$ for even $j$ and $\theta_j = \theta_{j+2} \equiv \theta_o$ for odd $j$ --- these equations reduce to a dimensionless form of the simplified model proposed in Ref.~\cite{Gu2026} (where $ \gamma $ can be absorbed in the time variable to set a dimensionless time-scale if needed)~\footnote{A full dimensionless formulation of the equation of motion is given by $ \partial \theta_j /\partial \tilde{t}  = -3\sin(2\theta_j) + \sin(\theta_j - \theta_{j+1}) + \sin(\theta_j - \theta_{j-1}) + A \sin(\theta_j - \tilde{v}^*\, \tilde{t}) $ with dimensionless time $ \tilde{t} = t/\gamma $ and phase $ \tilde{v}^* = \gamma \tilde{v} $.}

Notice that the derived equations of motion~\eqref{eq:EOMFULL} share the structure of a \textit{Frenkel-Kontorova model}~\cite{Braun1999,Floria1996}, featuring trigonometric coupling potentials together with a trigonometric driving term. Strikingly, the energy contribution associated with the internal energy of each row $j$ plays the role typically attributed to the substrate potential in a conventional Frenkel-Kontorova model, whereas the magnetic substrate in our setup effectively acts as the external driving. Additionally, the simplified model and its energy function~\eqref{eq:FULLH} resemble the Hamiltonian of the recently introduced \textit{extended Stoner--Wohlfarth model} for generalized van der Waals magnetic materials, such as A-type antiferromagnets~\cite{Wang2026}, in the limit of zero interlayer stiffness and for a rotating external field. See also other commonly used models for (anti)ferromagnetic materials, which exhibit similarly structured energy landscapes~\cite{Negulescu2011, Yang2021}. These similarities to established model systems highlight the versatility and generality of our simplified model despite the strong assumption. Under this broader view, our subsequent derivations apply to any system with a Hamiltonian of this form.

\begin{figure*}[ht!]
    \centering

    \begin{minipage}[t]{0.25\textwidth}
        \centering
        \begin{overpic}[height=1.8in]{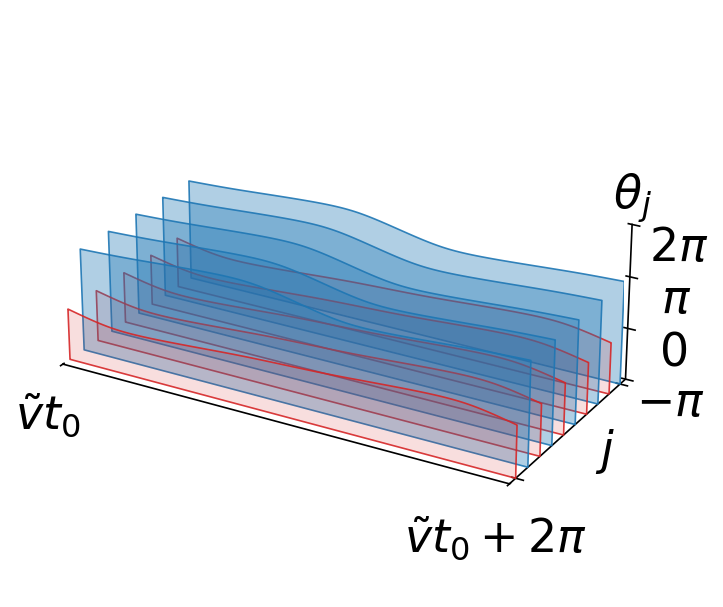}
            \put(4,60){\colorbox{white}{\textbf{(a)}}}
        \end{overpic}
    \end{minipage}%
    \hspace{1.5cm}
    \begin{minipage}[t]{0.25\textwidth}
        \centering
        \begin{overpic}[height=1.8in]{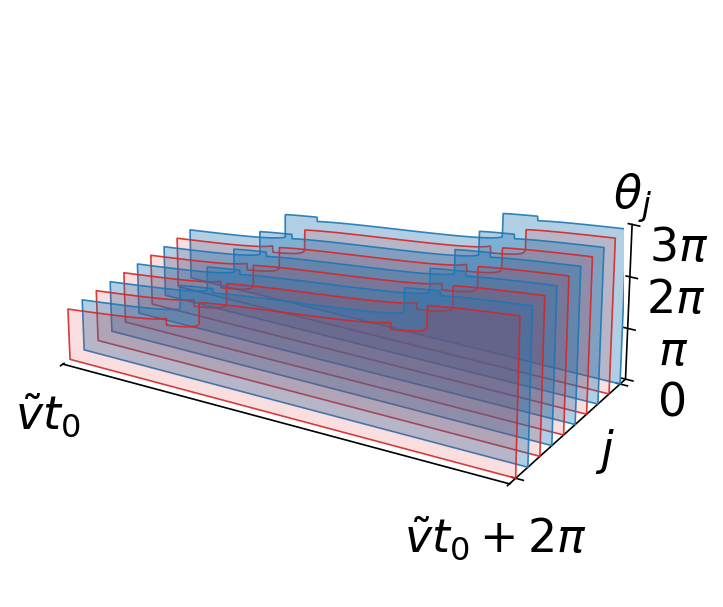}
            \put(4,60){\colorbox{white}{\textbf{(b)}}}
        \end{overpic}
    \end{minipage}%
    \hspace{1.5cm}
    \begin{minipage}[t]{0.25\textwidth}
        \centering
        \begin{overpic}[height=1.8in]{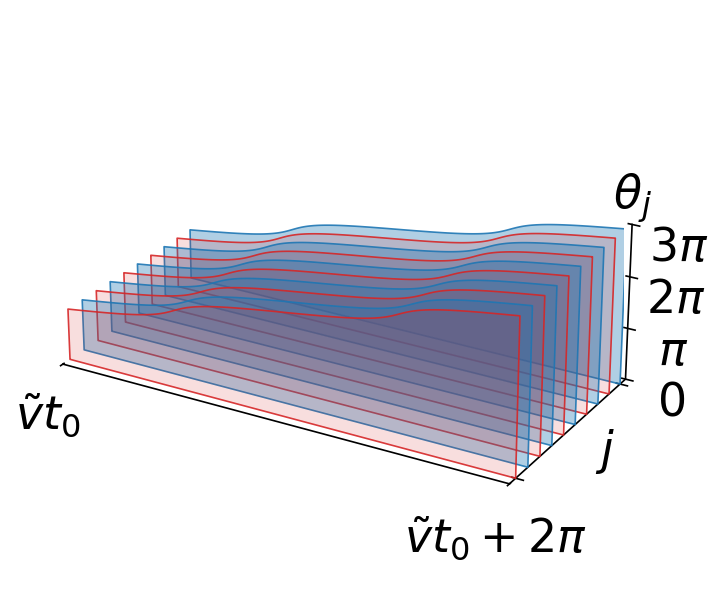}
            \put(4,60){\colorbox{white}{\textbf{(c)}}}
        \end{overpic}
    \end{minipage}


    \begin{minipage}[t]{0.25\textwidth}
        \centering
        \begin{overpic}[height=1.2in]{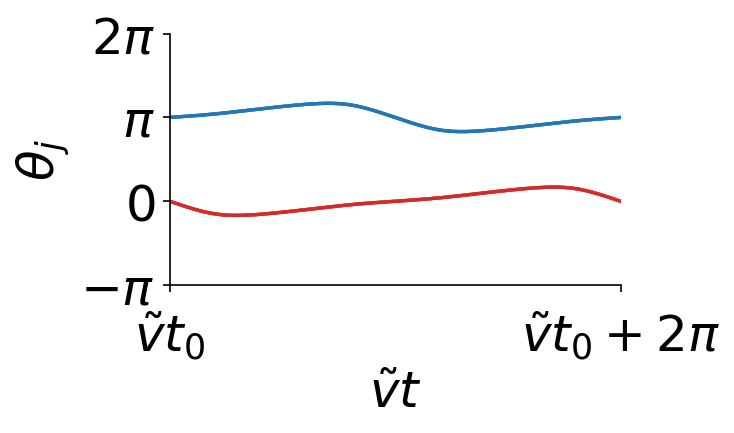}
            \put(4,60){\colorbox{white}{\textbf{(d)}}}
        \end{overpic}
    \end{minipage}%
    \hspace{1.5cm}
    \begin{minipage}[t]{0.25\textwidth}
        \centering
        \begin{overpic}[height=1.2in]{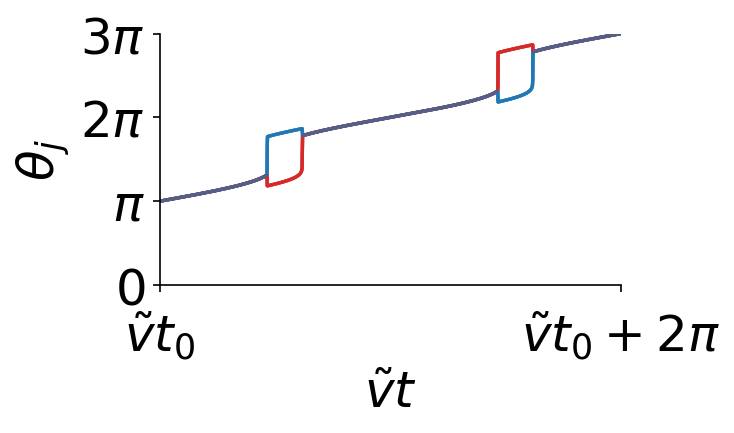}
            \put(4,60){\colorbox{white}{\textbf{(e)}}}
        \end{overpic}
    \end{minipage}%
    \hspace{1.5cm}
    \begin{minipage}[t]{0.25\textwidth}
        \centering
        \begin{overpic}[height=1.2in]{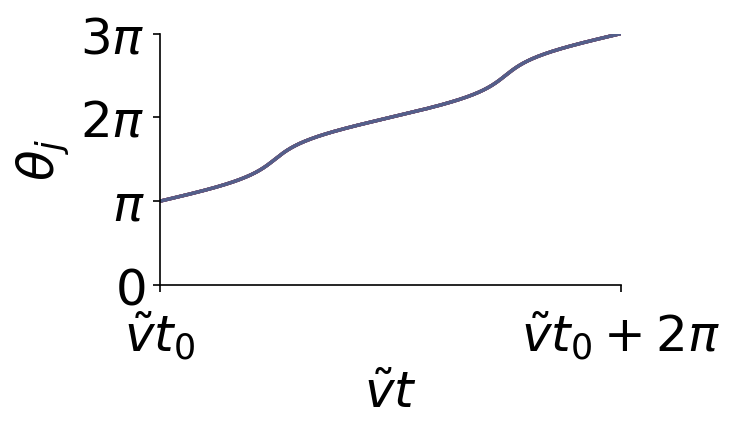}
            \put(4,60){\colorbox{white}{\textbf{(f)}}}
        \end{overpic}
    \end{minipage}
    
    \caption{ Example trajectories for the three dynamical regimes. (a)--(c) Individual trajectories for the antiferromagnetic (AFM, $A=4$), competitive (CP, $A=6$), and ferromagnetic (FM, $A=11$) regimes, respectively. (d)--(f) Overlay of the corresponding trajectories projected onto the same plane. All trajectories are obtained by numerical integration of the equations of motion and use $N=10$, and $\gamma \tilde{v} = 0.001 $. The initial conditions were $\theta_{j}=0+\eta_j$ for odd $j$ and $\theta_{j}=\pi+\eta_j$ for even $j$ where $\eta_j\in [-0.01,0.01]$ is a uniform random number ($ \tilde{v} $ is set to unity for the numerical integration). Trajectories depicted start after a relaxation cycle.  }
    \label{fig:three_regimes}
\end{figure*}

\begin{figure}
    \centering
    \includegraphics[width=0.95\linewidth]{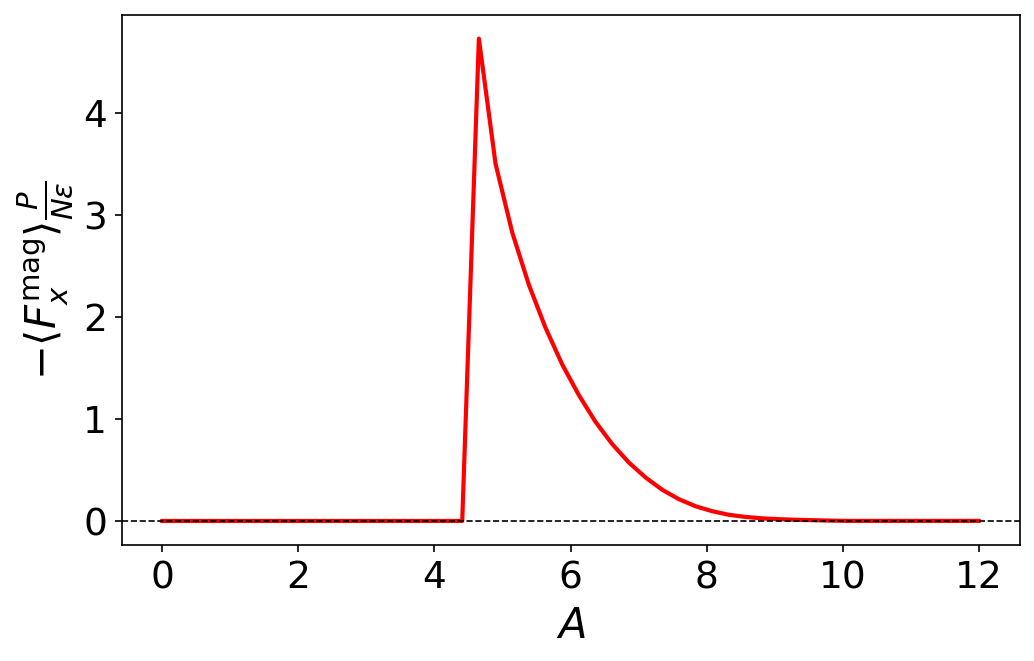}
    \caption{Magnetic friction as a function of the layer-separation parameter $A$. The data was obtained using numerical integration of the equations of motion for $N=20$ and $\gamma \tilde{v} = 0.001 $ initialized with $0=\theta_j=\theta_{j+1}+\pi$ perturbed by uniform random numbers $\eta_j \in [-0.01,0.01]$ ($ \tilde{v} $ is set to unity for the numerical integration).}
    \label{fig:friction}
\end{figure}

Throughout this work we focus on the \textit{quasi-static regime}. By this we mean that the energy landscape changes slowly compared to the relaxation of the rotor angles, so that the system remains close to a stable minimum of the instantaneous energy landscape. In the simplified model, $\gamma$ sets the relaxation time of the rotor angles, while $(2\pi/\tilde v) $ sets the time scale on which the substrate phase changes. The quasi-static limit is therefore characterized by $\gamma \tilde v \ll 2\pi$. This limit allows us to infer the driven dynamics from the structure of the energy landscape: the system follows a stable minimum until it disappears or loses stability, at which point it relaxes to another available minimum. This is the standard behavior of slowly driven dissipative systems near dynamic bifurcations~\cite{Berglund2000}.

We study the behavior of the simplified model for both ferromagnetic [$\theta_{j+1}(0) = \theta_{j}(0)$] and antiferromagnetic [$\theta_{j+1}(0) = \theta_j(0) +\pi$] initial conditions with $ \theta_1(0) = 0 $ or $ \theta_1(0) = \pi $, which are the meaningful choices with respect to experimental observations~\cite{Gu2026} and generic time-periodic steady states (a brief description of the behavior for other initial conditions is given in Appendix~\ref{app:InitialCheck}). However, we consistently assume that the initial angles are slightly perturbed unless stated otherwise. Specifically, at instances where we numerically integrate the equations of motion for comparison, we add small random numbers to $\theta_j(0)$ to break symmetry-protected solutions that are not expected to be excited in physical realizations of the slider-substrate setup. For example, when starting from perfectly ferromagnetic initial conditions, Eq.~\eqref{eq:EOMFULL} decouples for the individual $\theta_j(t)$, causing the magnetic moments to remain identical at all times, even as the system passes through highly unstable magnetic configurations (see below for a detailed discussion of such states). We do not consider such symmetric states in detail within the main text. Note that uniqueness of the solutions of Eq.~\eqref{eq:EOMFULL} ensures that such perfectly symmetric case can only be reached asymptotically, unless we start in a perfect initial configuration. This avoids the decoupling of the different degrees of freedom during sliding.

As confirmed in the next section, the simplified model qualitatively captures the essential dynamical regimes found in Ref.~\cite{Gu2026}, but relies on several idealizations. We outline the main limitations in Appendix~\ref{APP:Limitations}. In particular, we explain why they are reasonable within our scope and discuss how they affect the interpretation of our results compared to experimental realizations of the slider-substrate setup.

\section{The three dynamical regimes} \label{SEC:regimes}

To provide an overview of the different dynamical states that emerge in our simplified $N$-degree-of-freedom model, we briefly reproduce the findings of Ref.~\cite{Gu2026} by numerically integrating the equations of motion~\eqref{eq:EOMFULL}. To this end, we solve the system within \textit{Python} using $\texttt{scipy.integrate.solve\_ivp}$ with $\texttt{method='Radau'} $~\cite{Virtanen2020}. This option implements an implicit fifth-order Runge--Kutta method of the Radau IIA family and is suited for stiff systems~\cite[Sec.~IV.8]{HairerWanner1996}. Relative and
absolute error tolerances were set to $\texttt{rtol}=10^{-12}$ and $\texttt{atol}=10^{-14}$, respectively. Note that a more detailed numerical analysis follows below. This section focuses on providing sufficient background for readers unfamiliar with the findings of Ref.~\cite{Gu2026}.

Depending on the separation between the layers, encoded in the parameter $A$ in our dimensionless formulation, the system exhibits three characteristic dynamical regimes (provided the sliding velocity is sufficiently small for the system to be considered quasi-statically). For large $A$ (small separations), the magnetic moments of the slider align parallel to one another ($\theta_j(t) \approx \theta_{j+1}(t)$ for all $j$) and rotate synchronously with the periodic driving imposed by the substrate. Owing to the parallel alignment, this regime is referred to as the \textit{ferromagnetic (FM) regime}. For small $A$ (large separations), the slider moments are effectively decoupled from the substrate and remain in an antiferromagnetic configuration with $\theta_j(t) \approx \theta_{j+1}(t) + \pi$, defining the \textit{antiferromagnetic (AFM) regime}. Finally, for intermediate values of $A$, the system alternates between a parallel and a roughly antiparallel configuration, giving rise to the \textit{competing (CP) regime}. Such dynamical behavior that emerges due to intermittent stability of different arrangements of the magnetic moments has also been identified in magnetic tip-substrate setups, where magnetic vortices drive the dissipation~\cite{Magiera2014}. Note also that jumps between different stable states are quite natural in complex nonlinear systems~\cite{Gilmore1993, Strogatz2018}. Sample trajectories for each of the three observed regimes using the parameters $N = 10$ and $\gamma \tilde{v}=0.001$ are shown in Fig.~\ref{fig:three_regimes} (if not stated otherwise, $ \tilde{v} $ is always set to unity in the numerical integrations).

During sliding, the rotation of the magnetic moments on the slider dissipates energy, leading to frictional forces. The resulting magnetic friction can be derived by~\cite{Gu2026}
\begin{align}
\langle F_x^{\mathrm{mag}} \rangle = - \varepsilon \frac{A}{P} \sum_j  \int_0^{2\pi/\tilde{v}} \sin(\theta_j - \tilde{v} t) \, \dot{\theta}_j(t) \, \textrm{d} t.  \label{eq:MAGFRIC}
\end{align}
Here, $ \langle \cdot \rangle $ indicates the time average over a full period $ 2\pi/\tilde{v} $ in the time-periodic steady state. See Appendix~\ref{APP:MF} for a derivation of this relation and alternative formulations. Hence, the magnetic friction is fully determined by the torques exerted on the rotor array by the substrate field and by the dynamics of the slider moments.

Depending on the dynamical regime, the system exhibits distinct frictional responses, giving rise to the reported non-monotonic behavior. In the FM regime, dissipation is dominated by the Stokes-like drag term acting on the smooth, nearly uniform rotations of the moments. In the AFM regime, the friction $ \langle F_x^{\textrm{mag}} \rangle $ is essentially zero, with only small residual contributions arising from weak oscillatory perturbations around the perfect AFM configuration. In the CP regime, a pronounced peak in the dissipation appears, which is associated with hysteresis loops in the torque. This friction, calculated for a sample system with $N=20$ and $\gamma \tilde{v} = 0.001 $ for various values of $A$, is depicted in Fig.~\ref{fig:friction}. 

The friction peak emerges between $A \approx 4.5$ and $A \approx 9$. Through tests, we find that the dependence of the friction on $A$, as well as the boundaries of the layer separation interval of the friction peak, are insensitive to changes in $N$ for the chosen initial conditions. Additionally, we find that the lower boundary is invariant under changes in $\gamma \tilde{v}$, while the upper boundary varies slightly with $\gamma \tilde{v}$ and saturates at roughly $A \approx 8.8$ as $\gamma \tilde{v} \rightarrow 0$. The dependence of the upper boundary on $\gamma \tilde{v}$ is discussed below. Note that the range of layer separations with high magnetic friction corresponds to a regime of sudden changes in the relative orientation of the slider magnets, which we use to formally define the CP regime in the following.

\section{Rigorous prediction of the dynamical states via classification of the energy landscape} \label{SEC:LANDSCAPE}

In the following, we aim to analytically derive the expected dynamical behavior of the simplified model for a given $A$, focusing on the three characteristic regimes identified above. In the quasi-static regime, where the energy landscape evolves more slowly than the relaxation of the magnetic moments, the dynamical behavior is fully determined by the instantaneous energy landscape. The system is therefore expected to remain close to a stable minimum of this landscape as long as that minimum persists. A sudden qualitative change in the relative orientation of the magnetic moments during sliding occurs when the followed minimum disappears or loses stability, at which point the system must relax toward another available stable state~\footnote{More precisely, the relevant comparison is between the time scale on which a critical point is unstable and the time scale defined by the slope of the energy function in its vicinity. If the instability is weak or short-lived, trajectories may remain close to the unstable critical point for a finite time, even though it is not attracting.}.

This interpretation is consistent with the gradient-flow structure~\cite{Smale1961} of Eq.~\eqref{eq:gradient_flow}. If the explicit time dependence were removed, for instance, by freezing the drive at some fixed time $t^\ast$ and considering the energy function $\mathcal{H}((\theta_j)_j, t^\ast;A)$, then generic solutions of Eq.~\eqref{eq:gradient_flow} would relax toward local minima of $\mathcal{H}((\theta_j)_j, t^\ast; A)$. Solutions initialized exactly on unstable critical points may instead remain there or converge to a saddle point, but such special cases are not expected to play a role in real physical realizations.

Once the explicit time dependence is restored, we can introduce the concept of a \textit{branch}. We define it as the connected subset of the $(A,\tilde{v}t)$-parameter plane associated with a particular critical point of the energy landscape~\footnote{If two critical points combine during the time evolution, the merged critical point is treated as a new, distinct one, and the branches of the two originals terminate.}. In other words, within such a branch, an idealized system can follow the same critical point if $A$ and $\tilde{v}t$ are varied arbitrarily slowly.

Although we refer to a quasi-static regime, even a vanishingly small sliding velocity subtly modifies the system behavior. In fact, during the time evolution, the system should not be expected to lie exactly at an instantaneous minimum. Because the energy landscape evolves continuously, its extrema move in time, and a finite relaxation rate produces a small dynamical lag. This lag is essential, for example, to sustain the driving torque required for a smoothly rotating state, as well as to facilitate collective dissipation in the FM regime (see Appendix~\ref{APP:Lag}), and has been discussed previously in the context of magnetic friction~\cite{Magiera2009}. It also matters near unstable critical points: the finite dynamical lag offsets the instantaneous state from the followed stationary point as it evolves, acting as an effective perturbation that can facilitate relaxation toward nearby stable critical points when the followed stationary point becomes unstable.

\begin{figure}[ht!]
\centering
\includegraphics[width=0.95\linewidth]{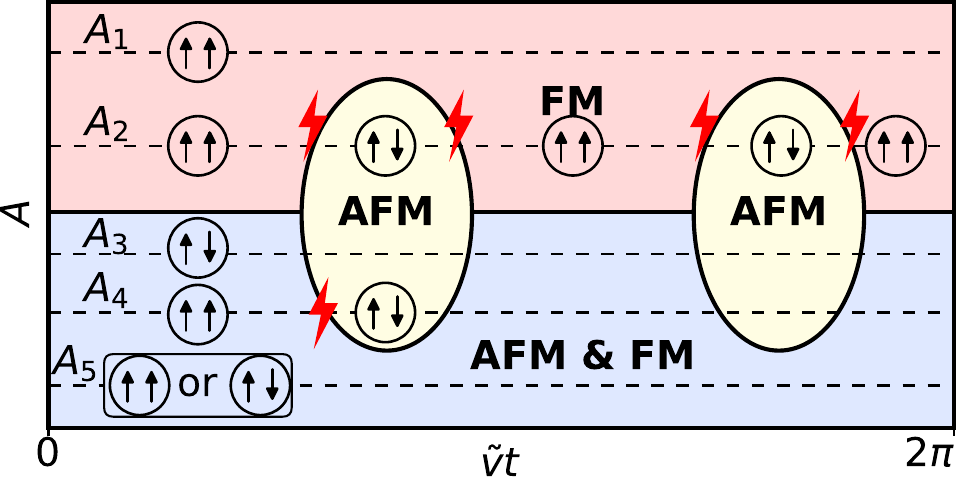}
\caption{Schematic decomposition of the $(A,\tilde{v}t)$-parameter plane into chambers. Within each chamber, the available stable minima have fixed relative alignment of the slider moments. Since $A$ is fixed during the dynamics, time evolution corresponds to a horizontal line-out.}
\label{fig:Minsmag_only}
\end{figure}

Tests in which we numerically solved the equations of motion showed that unstable critical points can be persistently followed only in symmetry-protected or effectively decoupled situations, where perturbations along the unstable direction are suppressed (see Appendix~\ref{APP:Protected} for an example). For the symmetry-breaking initial conditions chosen in this work and under arbitrarily small perturbations (even at the scale of numerical accuracy), the dynamically relevant states are the local minima. This implies that we can fully predict the behavior of the system once we identify the subsets of the branches where the corresponding critical points are minima and determine the relative orientation of the slider moments associated with each of them.

The sketch in Fig.~\ref{fig:Minsmag_only} illustrates this idea. Suppose we know the branches of all minima and the subsets of these branches that correspond to the same relative orientation of the magnetic moments on the slider. This allows us to partition the $(A,\tilde{v}t)$-parameter plane into areas with a fixed number of minima and a fixed relative configuration. For instance, in the blue region of the example system depicted in Fig.~\ref{fig:Minsmag_only}, the energy landscape contains two distinct minima: one with parallel and one with antiparallel alignment of the rows of the slider. For simplicity, we refer to these configurations as the \textit{FM state} and the \textit{AFM state}, respectively. In the red region, the landscape has a single stable minimum corresponding to the FM state, while the yellow regions contain a single stable minimum corresponding to the AFM state.

Since the layer-separation parameter $A$ remains fixed during sliding, the time evolution corresponds to a horizontal line-out through Fig.~\ref{fig:Minsmag_only}. We can use this to predict the dynamics: For example, along the line-out $A_1$, the system remains close to the same stable minimum corresponding to the FM state, and the slider moments stay parallel throughout. In this case, the system is clearly in the FM regime. Line-out $A_2$ alternately crosses regions with a single FM-state minimum and a single AFM-state minimum. Here, the system must switch the relative orientation of the moments, placing it in the CP regime. On the other hand, the line-out $A_3$ begins in an AFM-state minimum. Although it crosses regions with different numbers of minima (the blue and yellow areas), the relative orientation can remain in the AFM state throughout. In contrast, the line-out $A_4$ differs from $A_3$ only by the initial minimum it occupies: it starts in an FM-state minimum. Upon entering the first yellow region, the system must flip into the AFM state, where it remains for the rest of the time evolution. Therefore, both $A_3$ and $A_4$ correspond to the AFM regime. Finally, $A_5$ is a line-out that lies entirely within the blue region. In this case, the system stays in whichever minimum it occupies at the start of the evolution. Typically, this is the AFM state, although the FM state is also a stable local minimum.

The central goal of this section is to analytically construct such a decomposition of the $(A,\tilde{v}t)$-plane, which serves as a state diagram for the instantaneous relative orientation of the slider moments. To derive this decomposition, we identify all critical points of the energy function for each parameter set $(A,\tilde{v}t)$ and classify them by their Morse type~\cite{Milnor1963MorseTheory} (i.e., we identify whether they are minima, saddles, or maxima). This yields what we call the \textit{Morse signature} of a parameter set, namely the number of minima, saddles, and maxima at a given point $(A,\tilde{v}t)$. We additionally classify each critical point by the relative orientation of the slider moments, distinguishing FM from AFM states. In the following, we refer to connected regions in the $(A,\tilde{v}t)$-plane with a fixed number of critical points, constant Morse signature, and a fixed relative orientation of the magnetic moments as \textit{chambers}.

By definition, the chambers form a disjoint decomposition of the $(A,\tilde{v}t)$-plane. Their boundaries consist of \textit{bifurcation curves}, which determine where the number or the Morse type of critical points changes, and \textit{alignment curves}, which mark where the relative orientation of the magnetic moments associated with a given critical point switches. From this general treatment, we can then extract information about the possible dynamical states of the system.

Naturally, our strategy conceptually aligns with classic approaches that analyze the underlying energy landscape and map the corresponding basins of attraction~\cite{Mauro2021, Wales1969}. Frameworks of this type are widely applied in condensed matter physics~\cite{Suzuki2025}, particularly for describing glassy systems~\cite{Angelani2001}. In this context, a jump from one stationary point to another represents a structural transition of the system, which, in our setup, corresponds to a switch between FM and AFM states. More generally, the emergence of bifurcation points as control parameters vary is frequently invoked to explain phase transitions and regime changes, as demonstrated recently in active~\cite{Dauchot2019, Rusch2024} and soft matter~\cite{Rusch2026} systems. In our model, the dimensionless time $\tilde{v}t$ acts efficiently as a control parameter that continuously reshapes the energy landscape and its basins of attraction. Consequently, we treat it as such throughout the following analysis. In the context of magnetic friction, expecting maximum dissipation near bifurcation points conceptually parallels findings that demonstrate peak friction near the critical temperature in sliding Ising spin layers~\cite{Kadau2008}.

To make the analytical decomposition tractable, we work within the \textit{two-angle reduction} (i.e., the two-degree-of-freedom simplification used in Ref.~\cite{Gu2026}) defined by $\theta_j=\theta_e$ for even $j$ and $\theta_j=\theta_o$ for odd $j$ (with periodic boundary conditions). Accordingly, we reduce $\mathcal{H}(\theta_1,\ldots,\theta_N;A, \tilde{v}t)$ to $\mathcal{H}(\theta_e,\theta_o; A, \tilde{v}t)$. Inserting this simplification, the energy function is proportional to the reduced energy, which we define as
\begin{align}
    &\Hr(\theta_e, \theta_o; A,\tilde{v}t) \nonumber \\
   &= -\frac{3}{2}\left[\cos(2\theta_e)+\cos(2\theta_o)\right]
     + 2\cos(\theta_e-\theta_o) \nonumber \\
    &\quad + A\left[\cos(\theta_e-\tilde{v}t)+\cos(\theta_o-\tilde{v}t)\right].
\end{align}
Note that we restrict ourselves to $A>0$ and $ \tilde{v} > 0 $ in order to remain within the physically meaningful parameter range. Moreover, since
\begin{align}
\Hr(\theta_e,\theta_o; A,\tilde{v}t)
&=
\Hr(\theta_e+\pi,\theta_o+\pi;A,\tilde{v}t+\pi),\\
\Hr(\theta_e,\theta_o;A,\tilde{v}t)
&=
\Hr(\pi-\theta_e,\pi-\theta_o;A,\pi-\tilde{v}t),
\end{align}
it is sufficient to further restrict the analysis to $\tilde{v}t\in[0,\pi/2)$. The energy landscape for any other value of $\tilde{v}t$ can then be recovered from a reference value $\tilde{v}t^\ast\in[0,\pi/2)$ by repeated use of these symmetries. Even though this appears to be a strong restriction, the reduced energy landscape is a sufficient tool to describe generic time-periodic steady states of the full system (see Appendix~\ref{app:InitialCheck}).

Two further symmetries simplify the classification: First, $\theta_e$ and $\theta_o$ are $2\pi$-periodic. Hence, the relevant angle space is a compact set; equivalently, $(\theta_e,\theta_o)\in\mathbb{T}^2 = \mathbb{R}^2/(2\pi\mathbb{Z})^2$ lies on the torus $\mathbb{T}^2$. Second, the reduced energy is invariant under interchanging the two sublattice angles, $\Hr(\theta_e,\theta_o;A,\tilde{v}t)=\Hr(\theta_o,\theta_e;A,\tilde{v}t)$. Therefore, if $(\theta_e,\theta_o)$ is a critical point, then $(\theta_o,\theta_e)$ is also a critical point with the same energy and the same Morse type. When $\theta_e\neq\theta_o$ modulo $2\pi$, these two critical points are distinct as points in angle space, but they differ only by relabeling the even and odd sublattices. They therefore represent the same physical configuration for the purposes of our classification. In the following, we use this symmetry only as a bookkeeping device: symmetry-related critical points are counted once rather than twice.

\subsection{Bifurcation curves}~\label{sec:bifu}

We begin by identifying the bifurcation curves (or \textit{bifurcation set}) in the $(A,\tilde{v}t)$-plane at which the number or the Morse type of the critical points of $\Hr$ may change. We define the abbreviation $\vec{p}=(\theta_e,\theta_o)$. A critical point is a solution of the first-order stationarity condition
\begin{equation}
F(\vec{p};A,\tilde{v}t):=\nabla_p\Hr(\vec{p};A,\tilde{v}t)=0,
\label{eq:critical_map}
\end{equation}
where $\nabla_p$ denotes the gradient with respect to the components of $\vec{p}$. It is called nondegenerate if the quadratic expansion of $\Hr$ around this point has no flat direction, i.e., if the Hessian matrix $ D_p^2\Hr(\vec{p};A,\tilde{v}t)=D_pF(\vec{p};A,\tilde{v}t)
$ is invertible. Here, $D_p$ is the differential operator with respect to the components of $\vec{p}$. Conversely, a degenerate critical point is a stationary point at which the Hessian has at least one zero eigenvalue.

By the \textit{implicit function theorem}, the boundaries of the chambers formed by the bifurcation curves are fully determined by the presence of at least one degenerate critical point (see Appendix~\ref{APP:Implicite} for a detailed discussion). We can therefore compute the bifurcation curves by solving the stationarity equations together with the Hessian-degeneracy condition
\begin{align}
\nabla_p\Hr(\vec{p};A,\tilde{v}t)&=0,
\label{eq:bif_stationarity}
\\
\det\!\bigl(D_p^2\Hr(\vec{p};A,\tilde{v}t)\bigr)&=0.
\label{eq:bif_degeneracy}
\end{align}
Here, $D_p^2\,\Hr$ is the corresponding $2\times 2$ Hessian matrix of second derivatives with respect to the two angular degrees of freedom. Since the determinant is the product of a matrix's eigenvalues, Eq.~\eqref{eq:bif_degeneracy} is equivalent to requiring that at least one Hessian eigenvalue vanishes.

Although tedious and quite technical for our simplified model, calculating the bifurcation curves is straightforward: by simultaneously solving the stationarity equations and the Hessian-degeneracy condition, one identifies the state vectors $\vec{p}$ in angle space that correspond to bifurcation points by eliminating the parameters $A$ and $\tilde{v}t$. Reapplying the stationarity equations then yields the corresponding parameter sets $(A,\tilde{v}t)$, defining the bifurcation curves in the $(A,\tilde{v}t)$ parameter plane. For completeness, the explicit derivation of these bifurcation curves is detailed below. Readers primarily interested in the physical properties of the system may skip directly to the passage following Eq.~\eqref{eq:FM_param_curves}.

The analytical derivation of the bifurcation curves can be simplified by introducing the auxiliary variables~\footnote{Note that $\chi_+$ and $\chi_-$ were defined without the factor $1/2$ in Ref.~\cite{Gu2026}.}
\begin{equation}
\chi_+ = \frac{\theta_e+\theta_o}{2},
\qquad
\chi_- = \frac{\theta_e-\theta_o}{2},
\qquad
\zeta = \cos(\chi_-) .
\label{eq:chi_zeta_def}
\end{equation}
Using the angle-exchange symmetry $\theta_e \leftrightarrow \theta_o$, we choose representatives with $\chi_- \in [0,\pi/2]$, so that $\zeta \in [0,1]$.

In the $(\theta_e,\theta_o)$-plane, the line $\theta_e=\theta_o$ modulo $2\pi$ is the diagonal. Points on this diagonal have identical even- and odd-sublattice angles and therefore correspond to a perfect FM state. Equivalently, they satisfy $\chi_-=0$ and $\zeta=1$. We refer to these points as \textit{diagonal points}, or as the \textit{strict FM set}. Points with $\theta_e\neq\theta_o$ modulo $2\pi$ lie away from this diagonal and are called \textit{off-diagonal points}. In terms of the auxiliary variables, off-diagonal points are characterized by $0<\chi_-\leq\pi/2$, or, equivalently, $0\leq\zeta<1$. The endpoint $\zeta=0$ (i.e., $\chi_-=\pi/2$) corresponds to a perfect, or strict, AFM state.

We do not treat the endpoint $\zeta=0$ as a separate set analogous to the strict FM set because perfect AFM stationarity would require both $\partial_\zeta\Hr = 2A\cos(\chi_+ - \tilde{v}t) = 0$ and $\partial_{\chi_+}\Hr = -6\sin(2\chi_+) = 0$. These conditions can only be satisfied simultaneously at isolated values of $\tilde{v}t$, not along a finite time interval. Such isolated parameter values do not define a chamber boundary relevant for the time evolution of the system. The dynamical AFM regime found for the simplified model is therefore characterized by \textit{AFM-like states}, represented by the interior off-diagonal solutions with $0<\zeta<1$, which approach the AFM limit where appropriate.

After straightforward trigonometric simplifications, the reduced energy can be expressed in terms of the auxiliary variables as
\begin{align}
\Hr(\chi_+,\zeta;A,\tilde{v}t)
&=
\bigl(4-6\cos(2\chi_+)\bigr)\zeta^2
\nonumber\\
&
+2A\cos(\chi_+-\tilde{v}t)\,\zeta
+3\cos(2\chi_+)-2.
\label{eq:Hr_chi_zeta}
\end{align}
The useful feature of this formulation is that the energy is quadratic in $\zeta$. Because $\zeta$ ranges over $[0,1]$, we must consider interior critical points ($0<\zeta<1$, where easy coordinate transformations for the stationarity equations are possible) and boundary critical points ($\zeta=0$ or $1$), which require separate treatment.

We first consider the off-diagonal critical points defined on the interior range $0<\zeta<1$. In this regime, both the stationarity equations and the Hessian-degeneracy condition can be translated directly into the auxiliary variables. The linear part of the coordinate transformation between the angle coordinates $(\theta_e,\theta_o)$ and $(\chi_+,\chi_-)$ is
\begin{equation}
\begin{pmatrix}
\chi_+\\
\chi_-
\end{pmatrix}
=
\frac12
\begin{pmatrix}
1 & 1\\
1 & -1
\end{pmatrix}
\begin{pmatrix}
\theta_e\\
\theta_o
\end{pmatrix},
\qquad
\begin{pmatrix}
\theta_e\\
\theta_o
\end{pmatrix}
=
\begin{pmatrix}
1 & 1\\
1 & -1
\end{pmatrix}
\begin{pmatrix}
\chi_+\\
\chi_-
\end{pmatrix}.
\label{eq:linear_angle_transform}
\end{equation}
Both matrices have nonzero determinant, so the map between $(\theta_e,\theta_o)$ and $(\chi_+,\chi_-)$ is locally invertible; either set of variables provides a complete description of the system. In the off-diagonal interior $0<\chi_-<\pi/2$, the additional map $\zeta=\cos(\chi_-)$ is also locally invertible because $\partial\zeta/\partial\chi_-=-\sin(\chi_-)\neq 0$. In detail, the local inverse is $\vec{p}(\vec{q}) = (\chi_+ + \arccos(\zeta), \chi_+ - \arccos(\zeta))$, whose Jacobian is
\begin{equation}
J(\vec{q})=\frac{\partial \vec{p}}{\partial \vec{q}}
=
\begin{pmatrix}
1 & -1/\sqrt{1-\zeta^2}\\
1 & 1/\sqrt{1-\zeta^2}
\end{pmatrix},
\label{eq:jacobian_chi_zeta}
\end{equation}
which is invertible for $0<\zeta<1$. Here, we define the abbreviation $\vec{q}=(\chi_+,\zeta)$. As the coordinate map is locally invertible, no critical points are lost or introduced through singular or ill-defined behavior of the coordinate maps when describing the system via $(\chi_+,\zeta)$ instead of $ (\theta_e,\theta_o)$ in the off-diagonal interior. Notice also that
\begin{equation}
\partial_{\chi_-}\Hr = -\sin(\chi_-)\,\partial_\zeta\Hr,
\label{eq:chi_minus_zeta_derivative}
\end{equation}
so $\partial_{\zeta}\Hr=0$ is equivalent to the condition $\partial_{\chi_-}\Hr=0$ whenever $0<\chi_-<\pi/2$.

Adapting the Hessian-degeneracy condition of the interior range to the auxiliary variables $\vec{q}=(\chi_+,\zeta)$ can also be done using the Jacobian of the coordinate map. At a critical point, the Hessians in the two coordinate systems are related by
\begin{equation}
D_q^2\Hr(\vec{q};A,\tilde{v}t)
=
J(\vec{q})^\textrm{T}\, D_p^2\Hr(\vec{p}(\vec{q});A,\tilde{v}t)\, J(\vec{q}),
\label{eq:hessian_coordinate_transform}
\end{equation}
where $J(\vec{q})$ is the Jacobian from Eq.~\eqref{eq:jacobian_chi_zeta}. The usual additional terms in the second-derivative transformation vanish at a critical point because $\nabla_p\Hr=0$. Since $J(\vec{q})$ is invertible in the interior region, $D_q^2\Hr$ has a zero eigenvalue if and only if $D_p^2\Hr$ has a zero eigenvalue. We therefore impose the transformed Hessian-degeneracy condition $\det D_q^2\Hr=0$, where $D_q^2$ denotes the second derivative with respect to the components of $\vec{q}$.

We can now apply the modified stationarity and Hessian-degeneracy conditions to our system. Differentiating with respect to $\vec{q}$ gives
\begin{align}
\partial_{\zeta}\Hr
&=
2A\cos(\chi_+-\tilde{v}t)
+4\zeta\bigl(2-3\cos(2\chi_+)\bigr),
\label{eq:dH_dzeta}
\\
\partial_{\chi_+}\Hr
&=
-2A\zeta\sin(\chi_+-\tilde{v}t)
+6(2\zeta^2-1)\sin(2\chi_+).
\label{eq:dH_dchip}
\end{align}
Thus, every off-diagonal critical point satisfies
\begin{align}
A\cos(\chi_+-\tilde{v}t)
&=
2\zeta\bigl(3\cos(2\chi_+)-2\bigr),
\label{eq:interior_cp_1}
\\
A\zeta\sin(\chi_+-\tilde{v}t)
&=
3(2\zeta^2-1)\sin(2\chi_+).
\label{eq:interior_cp_2}
\end{align}
These are the stationarity equations for the off-diagonal critical points.

Using Eqs.~\eqref{eq:interior_cp_1} and \eqref{eq:interior_cp_2} to simplify the second derivatives at a critical point, one finds
\begin{align}
\partial_{\zeta\zeta}\Hr
&=
8 - 12\cos(2\chi_+),
\label{eq:hess_zz}
\\
\partial_{\chi_+\zeta}\Hr
&=
6\sin(2\chi_+)\left(2\zeta + \frac{1}{\zeta}\right),
\label{eq:hess_chip_z}
\\
\partial_{\chi_+\chi_+}\Hr
&=
12\cos(2\chi_+)(\zeta^2 - 1) + 8\zeta^2.
\label{eq:hess_chip_chip}
\end{align}
Hence, the transformed Hessian-degeneracy condition reduces, after simplification, to a quadratic equation in $\cos(2\chi_+)$, namely
\begin{align}
&\bigl(72\zeta^2 + 9\bigr)\cos^2(2\chi_+)
- 24\zeta^2\cos(2\chi_+)
\nonumber\\
&\qquad
- 20\zeta^4 - 36\zeta^2 - 9
= 0.
\label{eq:offdiag_degeneracy_quadratic}
\end{align}
Solving this quadratic equation yields two algebraic candidate curves, 
\begin{equation}
x_{\pm}(\zeta) := \cos(2\chi_+)
=
\frac{
4\zeta^2 \pm \sqrt{160\zeta^6 + 324\zeta^4 + 108\zeta^2 + 9}
}{
3(8\zeta^2 + 1)
}.
\label{eq:offdiag_xpm}
\end{equation}
Both signs must be considered: For a given $0<\zeta<1$, each sign yields a possible bifurcation curve whenever $x_{\pm}(\zeta)\in[-1,1]$. The minus branch is admissible throughout $0<\zeta<1$, while the plus branch is admissible only for those values of $\zeta$ for which $x_+(\zeta)\leq 1$. For each allowed value $x_{\pm}(\zeta)$, the corresponding values of $\chi_+$ follow from $\cos(2\chi_+)=x_{\pm}(\zeta)$, keeping only the nonequivalent solutions modulo the angular symmetries.

We have now identified the degenerate critical points in the angle space. To obtain the corresponding bifurcation curves in the $(A,\tilde{v}t)$-plane, we extract the corresponding parameters $A$ and $\tilde{v}t$ directly from Eqs.~\eqref{eq:interior_cp_1} and \eqref{eq:interior_cp_2}. To make this explicit, we define
\begin{align}
C &= 2\zeta\bigl(3\cos(2\chi_+)-2\bigr),
\label{eq:C_def}
\\
S &= \frac{3(2\zeta^2-1)\sin(2\chi_+)}{\zeta}.
\label{eq:S_def}
\end{align}
Then the stationarity equations are equivalent to $A\cos(\chi_+-\tilde{v}t)=C$ and $A\sin(\chi_+-\tilde{v}t)=S$. Therefore,
\begin{align}
A &= \sqrt{C^2+S^2},
\label{eq:offdiag_param_A}
\\
\tilde{v}t &= \chi_+ - \operatorname{atan2}(S,C)
\label{eq:offdiag_param_t}
\end{align}
modulo $2\pi$, after which $\tilde{v}t$ is mapped back to the fundamental interval $[0,\pi/2)$ using the time symmetries stated above. Thus, the off-diagonal part of the bifurcation set is given by two analytic parametric curves, corresponding to the two signs in $x_{\pm}(\zeta)$, with only the physically admissible parameter values retained.

The derivation of the bifurcation curve corresponding to the diagonal points ($\zeta=1$) is simpler. However, because the map $\zeta=\cos(\chi_-)$ is not locally invertible at $\chi_-=0$, we must work with the coordinates $(\chi_+,\chi_-)$. In these variables, the reduced energy is
\begin{align}
\Hr(\chi_+,\chi_-;A,\tilde{v}t)
&=
-3\cos(2\chi_+)\cos(2\chi_-)
+2\cos(2\chi_-)
\nonumber\\
&\quad
+2A\cos(\chi_+-\tilde{v}t)\cos\chi_-.
\label{eq:Hr_chi_plus_minus}
\end{align}
By definition, the diagonal points correspond to $\chi_-=0$. Restricting the energy to this condition gives
\begin{equation}
\Hr^{\rm FM}(\chi_+;A,\tilde{v}t)
= 2 - 3\cos(2\chi_+) + 2A\cos(\chi_+-\tilde{v}t),
\label{eq:Hr_FM_branch}
\end{equation}
and the corresponding stationarity equation is
\begin{equation}
A\sin(\chi_+-\tilde{v}t)=3\sin(2\chi_+).
\label{eq:FM_cp_condition}
\end{equation}

At a diagonal critical point, the Hessian in the $(\chi_+,\chi_-)$ variables is diagonal,
\begin{equation}
D^2_{\chi_+,\chi_-}\Hr\big|_{\chi_-=0}
=
\begin{pmatrix}
\lambda_\parallel & 0\\
0 & \lambda_\perp
\end{pmatrix},
\label{eq:FM_hessian_diag}
\end{equation}
with
\begin{align}
\lambda_\parallel
&=
12\cos(2\chi_+)-2A\cos(\chi_+-\tilde{v}t),
\label{eq:lambda_parallel}
\\
\lambda_\perp
&=
12\cos(2\chi_+)-8-2A\cos(\chi_+-\tilde{v}t).
\label{eq:lambda_perp}
\end{align}
The $\chi_+$ direction changes both sublattice angles in the same way and is therefore tangent to the diagonal $\theta_e=\theta_o$ in the $(\theta_e,\theta_o)$-plane. The $\chi_-$ direction changes the difference between the sublattice angles and thus points away from the diagonal into the off-diagonal region. Consequently, the two possible diagonal bifurcation curves (defined by the Hessian-degeneracy condition) are distinguished by whether the curvature in the $\chi_+$ direction or in the $\chi_-$ direction vanishes.

\begin{figure*}[ht!] 
\centering 
\begin{overpic}[height=0.43\linewidth] {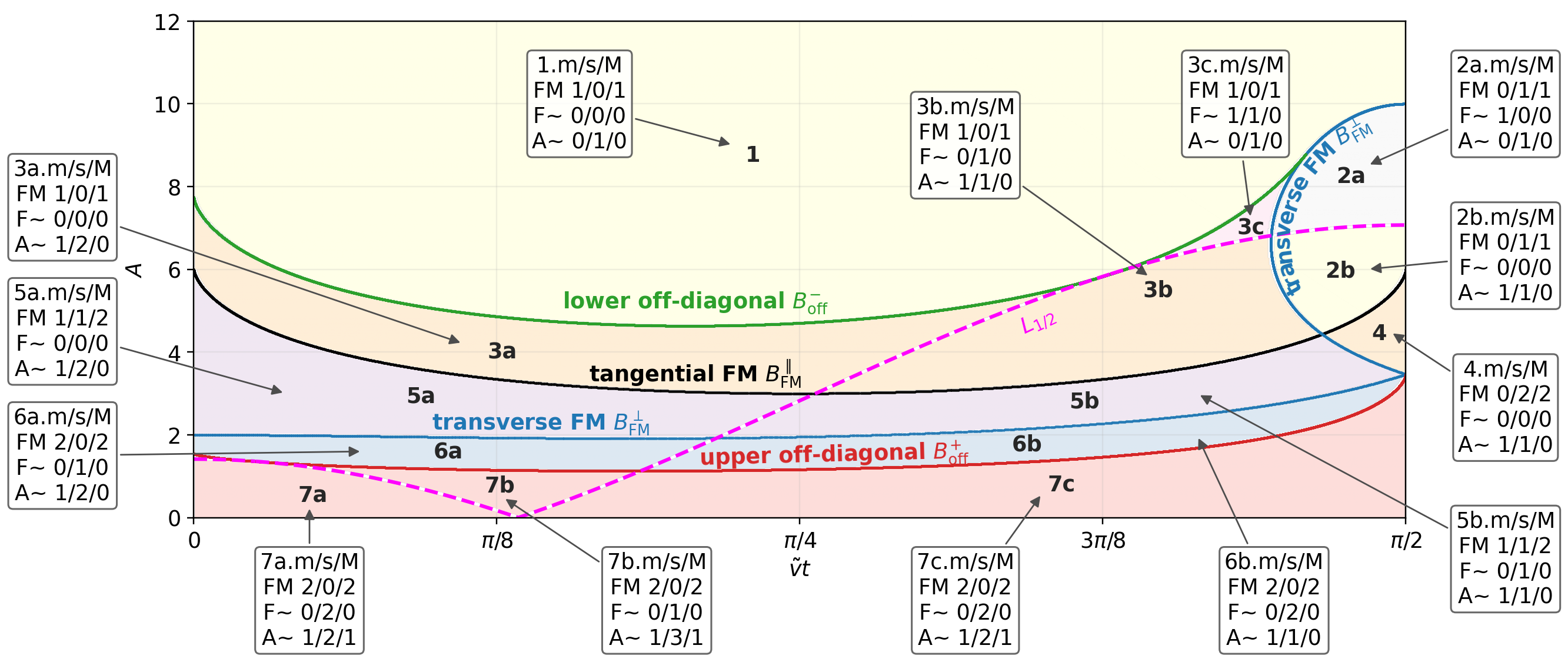} \put(3,40){\sffamily\bfseries (a)} 
\end{overpic} 
\phantom{.}\hspace{1cm}\begin{overpic}[height=0.43\linewidth] {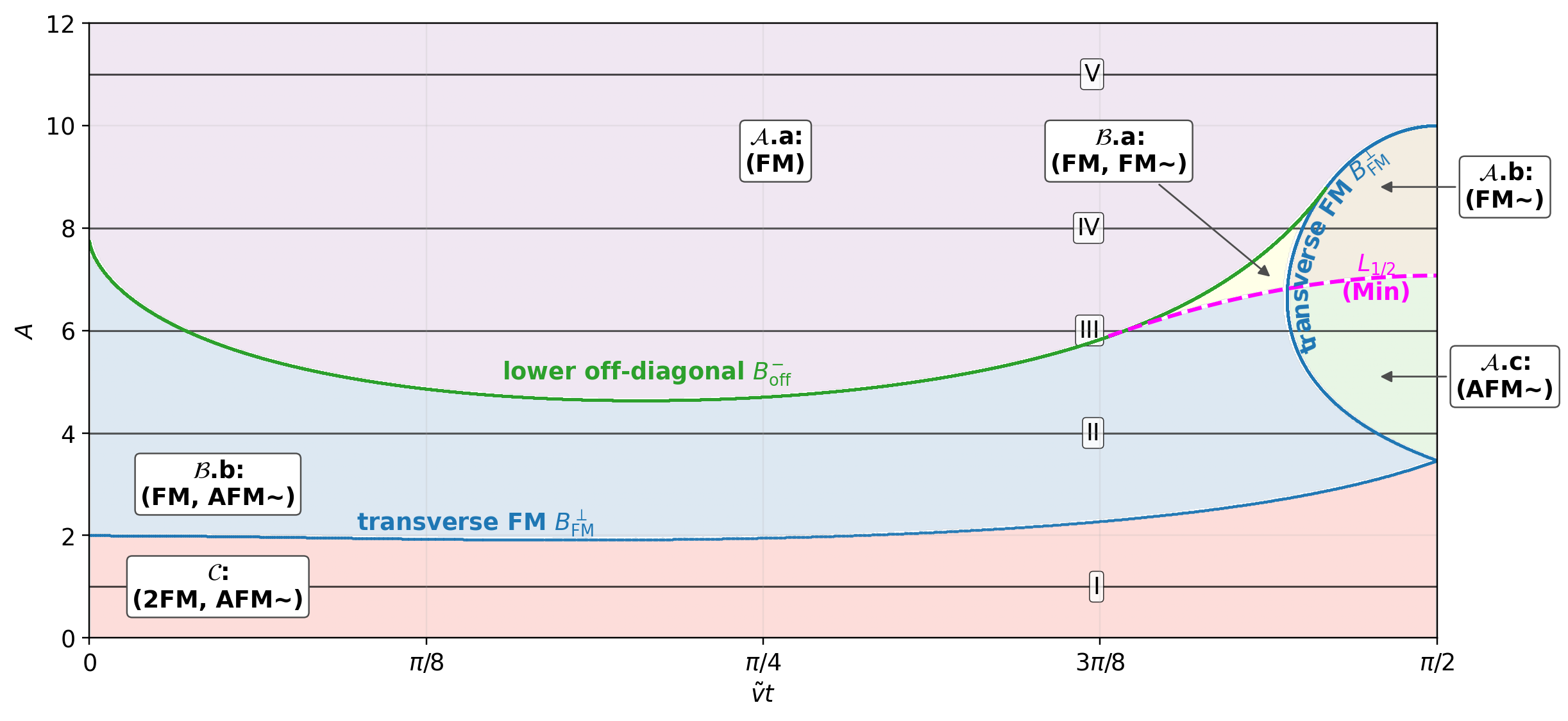} \put(-2,47){\sffamily\bfseries (b)} 
\end{overpic} 
\caption{
Chamber decomposition of the $(A,\tilde vt)$-parameter plane. (a) Full decomposition by Morse type and alignment state. Within each chamber, the number of minima, saddles, and maxima is fixed, and every critical point has a fixed alignment classification: strict FM, FM-like, or AFM-like. The chamber boundaries consist of the analytically derived bifurcation curves and the alignment curve $L_{1/2}$.
(b) Reduced decomposition retaining only the stable minima and their magnetic alignment. This representation identifies which stable FM, FM-like, and AFM-like states are available at each point in the parameter plane.} \label{fig:BuL} \end{figure*}

The \textit{tangential strict FM bifurcation curve} is obtained by setting $\lambda_\parallel=0$. Together with the diagonal stationarity condition Eq.~\eqref{eq:FM_cp_condition}, this yields
\begin{align}
A\sin(\chi_+-\tilde{v}t) &= 3\sin(2\chi_+),
\label{eq:fm_tangential_curve_sin}
\\
A\cos(\chi_+-\tilde{v}t) &= 6\cos(2\chi_+).
\label{eq:fm_tangential_curve_cos}
\end{align}
Again, this bifurcation is tangential because the vanishing curvature occurs along the strict FM set itself.

The \textit{transverse strict FM bifurcation curve} is obtained by setting $\lambda_\perp=0$. Together with the diagonal stationarity condition Eq.~\eqref{eq:FM_cp_condition}, this gives
\begin{align}
A\sin(\chi_+-\tilde{v}t) &= 3\sin(2\chi_+),
\label{eq:fm_transverse_curve_sin}
\\
A\cos(\chi_+-\tilde{v}t) &= 2\left[3\cos(2\chi_+)-2\right].
\label{eq:fm_transverse_curve_cos}
\end{align}
This bifurcation is called transverse because the vanishing curvature occurs in the direction that makes the two sublattice angles unequal. Physically, it marks the point where a strict FM critical point gains or loses stability against off-diagonal perturbations.

Equations~\eqref{eq:fm_tangential_curve_sin} and
\eqref{eq:fm_tangential_curve_cos}, as well as
Eqs.~\eqref{eq:fm_transverse_curve_sin} and
\eqref{eq:fm_transverse_curve_cos}, are implicit equations for the two corresponding diagonal bifurcation curves in the $(A,\tilde{v}t)$-plane. They are nevertheless exact and can be evaluated directly using $\chi_+$ as a parameter. For each value of $\chi_+$, we define
\begin{align}
S_{\rm FM} &= 3\sin(2\chi_+),\\
C_\parallel &= 6\cos(2\chi_+),\\
C_\perp &= 2\left[3\cos(2\chi_+)-2\right].
\label{eq:FM_param_defs}
\end{align}
Then,
\begin{align}
A_\alpha &= \sqrt{S_{\rm FM}^2 + C_\alpha^2},\\
\tilde{v}t_\alpha &= \chi_+ - \operatorname{atan2}(S_{\rm FM},C_\alpha),
\label{eq:FM_param_curves}
\end{align}
with $\alpha\in{\parallel,\perp}$, after which $\tilde{v}t_\alpha$ is mapped back to the fundamental interval $[0,\pi/2)$ using the symmetries stated above.

We have now calculated four bifurcation curves: the tangential FM bifurcation curve $B_{\rm FM}^{\parallel}$, defined by Eqs.~\eqref{eq:fm_tangential_curve_sin} and \eqref{eq:fm_tangential_curve_cos}; the transverse FM bifurcation curve $B_{\rm FM}^{\perp}$, defined by Eqs.~\eqref{eq:fm_transverse_curve_sin} and \eqref{eq:fm_transverse_curve_cos}; and the two off-diagonal bifurcation curves $B_{\rm off}^{+}$ and $B_{\rm off}^{-}$, defined by Eq.~\eqref{eq:offdiag_xpm} with Eqs.~\eqref{eq:offdiag_param_A} and \eqref{eq:offdiag_param_t}.
\footnote{Generically, the off-diagonal bifurcation curves $B_{\mathrm{off}}^\pm$ and the tangential FM curve $B_{\mathrm{FM}}^\parallel$ correspond to saddle-node (fold) bifurcations, at which pairs of critical points appear or disappear. The transverse FM curve $B_{\mathrm{FM}}^\perp$ corresponds to a symmetry-breaking pitchfork bifurcation: the strict-FM critical point persists across the curve but changes Morse type, while symmetry-related off-diagonal critical points emerge from or merge into it. }

 Together, they form the bifurcation set $B$. These curves are depicted in Fig.\ref{fig:BuL}~a). They decompose the $(A,\tilde{v}t)$-plane of our simplified model into seven subsets of constant Morse signature. These chambers are numbered in Fig.~\ref{fig:BuL}~a) as zones $1$ through $7$.

The stationarity equations and the Hessian-degeneracy condition successfully identify the bifurcation set. However, they do not determine the actual Morse types of the critical points within each chamber. Fortunately, the number of critical points in each chamber is finite, which makes the characterization of the Morse signature feasible (see Appendix~\ref{APP:Number} for an explanation). In fact, the classification reduces to a finite number of steps. It suffices to determine the critical points and their Morse types at one representative parameter value in each chamber. This is exactly how we determine the Morse signature of the decomposition: for each chamber, we select a single representative point $(A,\tilde{v}t)$, compute the finite set of critical points at that parameter value, and classify their Morse types by evaluating the corresponding curvatures. The derived Morse signature of the simplified model is also provided for each of the $7$ zones in Fig.~\ref{fig:BuL}~a).

\subsection{Alignment curve and relative orientation of the slider moments}

The previous subsection showed how the parameter domain $(A,\tilde{v}t)\in[0,\infty)\times[0,\pi/2)$ is divided into regions of constant Morse signature. This already distinguishes the strict FM set ($\zeta=1$) from the off-diagonal behavior ($0<\zeta<1$), but it does not yet determine which relative orientations of the moments correspond to which off-diagonal critical point. For a given parameter set $(A,\tilde{v}t)$, an off-diagonal critical point may be close to FM, close to AFM, or somewhere in between. To clarify this behavior and finalize the decomposition of the $(A,\tilde{v}t)$-plane into distinct chambers, we now quantitatively define the different states of relative orientation and determine the alignment curve, which separates chambers with the same Morse signature but different rotor configurations.

To quantitatively define the different possible configurations of the magnetic moments on the slider (i.e., \textit{the alignment state}), we introduce the alignment parameter
\begin{equation}
\rho := \zeta^2 = \cos^2(\chi_-)\in[0,1].
\end{equation}
We argued earlier that, within the symmetry constraints of our problem, $\chi_-\in[0,\pi/2]$, and hence $\rho$ provides a convenient scalar measure: Throughout this work, $\rho=1$ corresponds to \textit{strict FM alignment}, while smaller values of $\rho$ correspond to larger angular separation. In particular, we call $\rho>1/2$ (i.e., $\chi_-<\pi/4$) \textit{FM-like}, and $\rho<1/2$ (i.e., $\chi_->\pi/4$) \textit{AFM-like}~\footnote{This definition of the alignment states is an alternative to the distinction into FM and AFM states based on the order parameter of Ref.~\cite{Gu2026}. For our analysis based on the decomposition of the $(A,\tilde{v}t)$-plane, the distinction based on $\rho$ is more convenient.}. The threshold between these two behaviors is therefore $\rho=1/2$.

We define the alignment curve as the set of off-diagonal critical points that satisfy $\rho=1/2$ (i.e., $\zeta=1/\sqrt{2}$). Substituting this into Eqs.\eqref{eq:interior_cp_1} and \eqref{eq:interior_cp_2} yields
\begin{align}
A\cos(\chi_+-\tilde{v}t)
&=
\sqrt{2}\,\bigl(3\cos(2\chi_+)-2\bigr),
\label{eq:align_1}\\
A\sin(\chi_+-\tilde{v}t)
&=
0.
\label{eq:align_2}
\end{align}
Since $A>0$, the second equation implies $\chi_+-\tilde{v}t\in\pi\mathbb{Z}$, so that $\chi_+=\tilde{v}t$ or $\chi_+=\tilde{v}t+\pi$ modulo $2\pi$. In either case, $\cos(2\chi_+)=\cos(2\tilde{v}t)$, and therefore Eq.\eqref{eq:align_1} reduces to the explicit function
\begin{align}
A(t)=\sqrt{2}\,\bigl|3\cos(2\tilde{v}t)-2\bigr|,
\qquad
\tilde{v}t\in[0,\pi/2).
\label{eq:alignment_curve}
\end{align}
We denote this curve by $L_{1/2}$. It defines the boundary at which an off-diagonal critical point changes between AFM-like and FM-like behavior. Note that alternative alignment curves $L_\lambda$, separating regions with $\rho<\lambda$ and $\rho>\lambda$, can be derived in the same way for any $\lambda\in(0,1)$, although for $\lambda\neq 1/2$ the resulting expressions are no longer available in such a simple explicit form.

The alignment curve $L_{1/2}$ allows us to finally determine the full chamber decomposition of the $(A,\tilde{v}t)$-plane. In theory, the alignment states of the different critical points within each chamber could be found by checking a single representative parameter set $(A,\tilde{v}t)$, similar to the approach used before to identify the Morse signature. However, while the decomposition via the bifurcation set offers convenient test points (such as $A=0$, $\tilde{v}t=0$, or $\tilde{v}t=\pi/2$) for each relevant subset, the full decomposition, taking the alignment curve into account, lacks such readily available representatives. Instead, it is often more practical to characterize the chambers by tracking which critical points change their alignment state when crossing $L_{1/2}$. Consequently, we determine the full alignment information by testing representative points where convenient and tracking transitions across $L_{1/2}$ elsewhere. In Appendix~\ref{APP:Alignment}, we outline the framework needed to track these transitions.

We have now constructed the full decomposition of the parameter plane of the studied simplified model. The four bifurcation curves determine where the number of critical points or their Morse types can change. The alignment curve $L_{1/2}$ does not create or destroy critical points; instead, it identifies where an already existing off-diagonal critical point crosses the threshold between FM-like and AFM-like behavior. It follows that each connected component of $ \bigl([0,\infty)\times[0,\pi/2)\bigr)\setminus\bigl(B\cup L_{1/2}\bigr) $ is a chamber of fixed Morse signature and fixed relative orientation of the slider moments.

We now label every critical point in each chamber as either strict FM (denoted as FM), FM-like ($\mathrm{FM}\sim$), or AFM-like ($\mathrm{AFM}\sim$), and simultaneously record whether it is a minimum, saddle, or maximum. The full decomposition obtained in this way is shown in Fig.~\ref{fig:BuL}~a). Here, the seven zones determined by the bifurcation set are further subdivided into subzones (e.g., 7a, 7b, and 7c) according to their alignment states.

Figure.~\ref{fig:BuL}~b) takes the full decomposition shown in the left panel, and reduces it to a classification based solely on the number of minima and on the alignment state. This is the decomposition that we ultimately aimed for: it tells us, for every parameter pair $(A,\tilde{v}t)$, which stable states are possible, directly paralleling the structure of the introductory toy example.

\subsection{Prediction of the dynamical behavior using the chamber decomposition}

With the rigorously derived chamber decomposition of the $(A,\tilde{v}t)$-plane for the simplified model under study, we now have access to all the information needed to determine the dynamical behavior at given $ A $. In particular, using the reduced decomposition that retains only the minima of the energy landscape [Fig.~\ref{fig:BuL}~b)], we can directly infer the relative orientation of the magnetic moments during time evolution and thus determine whether the system is in the FM, AFM, or CP regime. By incorporating the additional information provided by the full chamber decomposition [Fig.~\ref{fig:BuL}~a)], we also gain physical insight into the nature of the transitions that occur while the system exhibits alternating configurations (which we can later use to derive the magnetic friction in the CP regime).

To understand the dynamics over a full period ($\tilde{v}t=2\pi$), we can exploit the symmetries of the simplified model. For the interval $\pi/2 \le \tilde{v}t < \pi$, the time evolution effectively corresponds to a reflection at the right boundary of the decompositions in Fig.~\ref{fig:BuL}, with the system moving from right to left through Fig.~\ref{fig:BuL} thereafter. Similarly, for $\pi \le \tilde{v}t < 3\pi/2$, the evolution reflects at the left boundary, moving from left to right after reaching the reflection. This symmetric pattern can be extended indefinitely to determine the dynamics over any time interval. 

\begin{figure}[ht!]
\centering
\includegraphics[width=\linewidth]{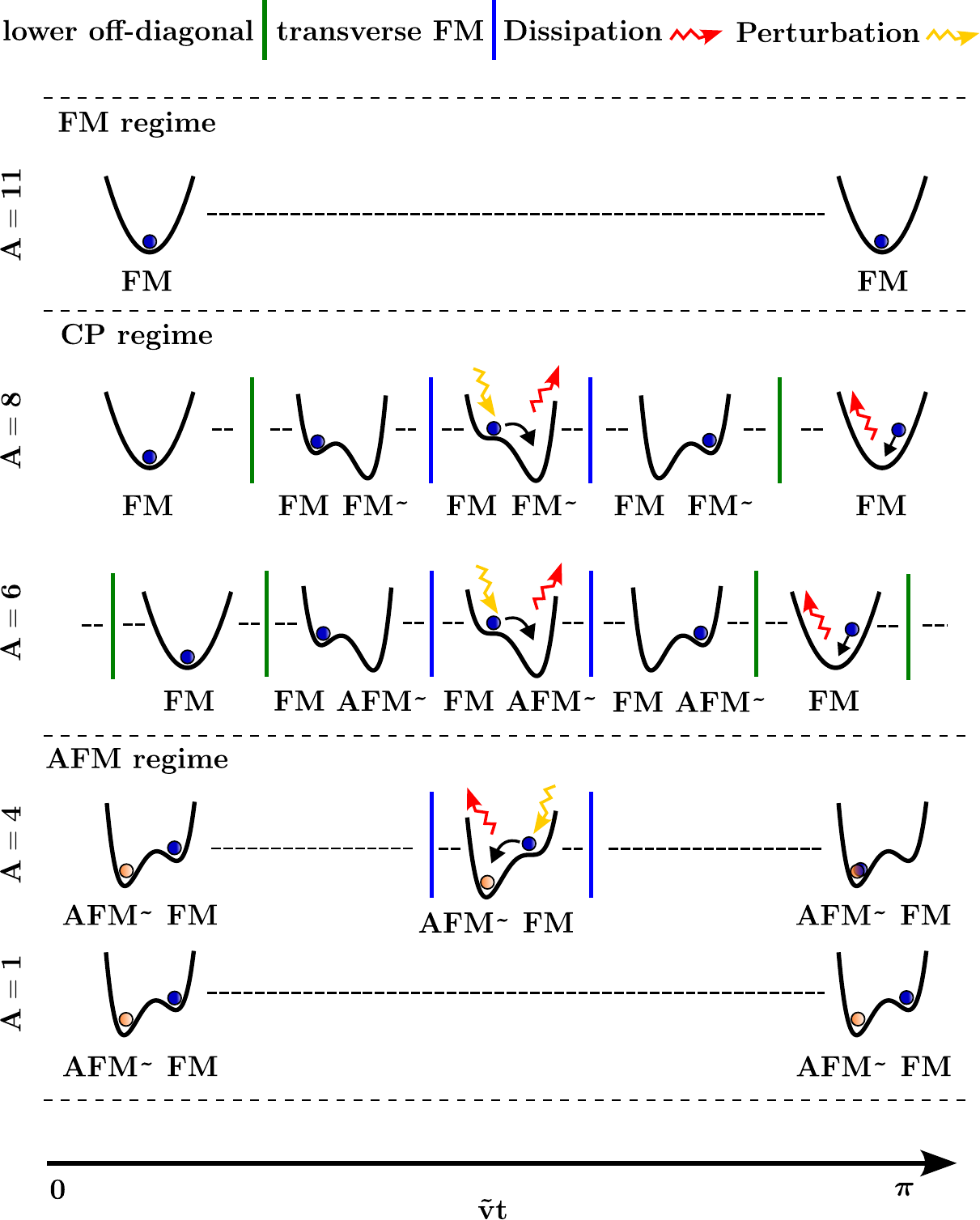}
\caption{Schematic sketch of the relevant critical points along different line-outs through the $ (A,\tilde{v}t) $-parameter plane. The sketch shows the time-evolution in the interval $ \tilde{v}t \in[0,\pi] $. The green lines indicate the crossing of the lower-off diagonal bifurcation curve. The blue lines indicate crossings of the transverse FM bifurcation curve. The yellow lightning symbols indicate perturbations which trigger the instability transitions to a new local minimum of the energy landscape with a different relative alignment of the slider moments. The red lightning symbols highlight where a distinct amount of energy is dissipated due to changes in the energy landscapes which result in jumps between critical points and relaxation processes. The blue and orange circles indicate systems with different initial conditions. For $ A = 6$, the initial and final state of the depicted time interval are not shown to avoid overcrowding.}
\label{fig:Landscape}
\end{figure}

Using the same concept discussed for the schematic decomposition in Fig.~\ref{fig:Minsmag_only} at the beginning of this section, we can illustrate the predictive power of the chamber decomposition with several example line-outs at fixed $A$. To corroborate the discussion, Fig.~\ref{fig:Landscape} provides sketches of the energy landscape during the time evolution over the interval $0 \le \tilde{v}t \le \pi$. To ensure clarity, these sketches show only the critical points that are relevant for the understanding the underlying mechanisms.

The most straightforward cases occur for large $A>10$ and small $A<2$. In these regimes, the line-outs do not cross any chamber boundaries, and the system simply follows the minimum in which it was initialized. For instance, consider the line-out at $A=11$ (top row of Fig.~\ref{fig:Landscape}). Here, only one stable minimum exists, namely a strict FM minimum, and the system is clearly in the FM regime~\footnote{Note that the studied systems are never perfectly in a strict FM minimum during time evolution, as such a state is only reached asymptotically when starting from perturbed initial conditions. Hence, while $\rho \to 1$, $\rho \neq 1$ strictly holds within our analysis.}. At $A=1$ (bottom row of Fig.~\ref{fig:Landscape}), the system may occupy three minima: two strict FM minima and one AFM-like minimum. All three persist throughout the entire time evolution (in Fig.~\ref{fig:Landscape}, the two strict FM minima are depicted as a single state). In a physical realization, as in the setup of Ref.~\cite{Gu2026}, the system is typically initialized in the AFM-like minimum, so the line-out at $A=1$ corresponds to the AFM regime.

More intriguing is the line-out at $A=4$ (fourth row of Fig.~\ref{fig:Landscape}). Here, the system can be initialized close to either a strict FM or an AFM-like minimum. However, this line-out crosses the transverse FM bifurcation curve, where the strict FM minimum turns into a saddle point (see the full decomposition in the left panel of Fig.~\ref{fig:BuL}). Hence, a small symmetry-breaking perturbation or a sufficiently large dynamical lag can trigger a transition from the strict FM state to the AFM-like minimum, which is the only stable configuration after the first crossing of this curve. Upon reflection at the right boundary, the system crosses the transverse FM bifurcation curve once more. However, since the AFM-like minimum continues to exist even after passing the bifurcation set again, the system can still follow this minimum indefinitely. Consequently, the line-out at $A=4$ is expected to correspond to an AFM-like state once the time-periodic steady state is reached and $ A = 4 $ lies in the AFM regime.

The example line-outs at $A=6$ (third row of Fig.~\ref{fig:Landscape}) and $A=8$ (second row of Fig.~\ref{fig:Landscape}) correspond to what we have so far intuitively referred to as the CP regime. However, the details underlying the corresponding dynamical behavior differ slightly. This can be understood from the distinct chambers that the system traverses. Let us begin with the line-out at $A=6$, which represents the more generic CP behavior discussed in Ref.~\cite{Gu2026}. Although the system can be initialized close to either a strict FM minimum or an AFM-like minimum, only the former is relevant, since the system quickly crosses the lower off-diagonal bifurcation curve where the AFM-like minimum ceases to exist. After crossing this curve, the system has only one stable minimum, namely, a strict FM minimum.

As the time evolution continues, the system crosses the lower off-diagonal bifurcation curve again. At that point, a new AFM-like minimum emerges; however, the system remains close to the strict FM minimum, since it is still stable. This changes shortly afterwards when the line-out crosses the transverse FM bifurcation curve. Here, the strict FM critical point becomes a saddle, and a perturbation or sufficiently large dynamical lag causes a rapid relaxation into the AFM-like minimum, which is the only stable minimum upon the first crossing of the transverse FM curve.

The system remains in this AFM-like state until the trajectory crosses the lower off-diagonal bifurcation curve again, after reflection at the right boundary of the decomposition in Fig.~\ref{fig:BuL}. At that point, the AFM-like minimum disappears, and the system relaxes into the strict FM minimum that has reappeared after the second crossing of the transverse FM curve. This periodic switching between the strict FM minimum and the AFM-like minimum continues indefinitely, thus the line-out at $A=6$ corresponds to the CP regime.

The mechanism corresponding to the line-out at $A=8$ is slightly different. Here, the system starts above the lower off-diagonal bifurcation curve, which means that the only stable minimum at the beginning is a strict FM minimum. The system crosses the lower off-diagonal bifurcation curve for the first time between $\tilde{v}t=3\pi/8$ and $\tilde{v}t=\pi/2$. At this crossing, an FM-like minimum emerges (in contrast to the AFM-like minimum that appears for $A=6$). The strict FM critical point remains a minimum until the line-out crosses the transverse FM curve. Afterwards, the strict FM critical point becomes a saddle, and the only stable minimum is FM-like. The system can remain close to this FM-like state until the line-out crosses the lower off-diagonal bifurcation curve again, at which point the FM-like minimum ceases to exist.

In this scenario, the system does not alternate between a strict FM state and an AFM-like state, but rather between a strict FM state and an FM-like state. Even though the alignment parameter $\rho$ is therefore expected to remain above the threshold $\rho=1/2$ and the alignment state changes only slightly, the sequence of chambers traversed by the line-out still implies a rapid and periodic change in the relative orientation of the slider moments along this trajectory. We therefore consider the line-out at $A=8$ to belong to the CP regime as well.

In fact, by comparing the magnetic-friction peak obtained from numerically solving the equations of motion with the reduced decomposition of the $(A,\tilde{v}t)$-plane, we find that the region of enhanced dissipation is bounded by the smallest value of $A$ on the lower off-diagonal bifurcation curve and the $ A $ value for which the lower off-diagonal and the transverse FM bifurcation curves merge. Hence, this regime corresponds precisely to the layer separations for which an alternating switch between alignment states is possible by the mechanisms described in this section. Therefore, we now formally define the CP regime as the layer separation range bounded by the lower off-diagonal curve and the transverse FM curve, in which sudden changes in the alignment parameter $\rho$ can occur.

Note that a small interval, $8.8 \lesssim A \le 10$, exhibits a different behavior. In this regime, delineated by the maximum of the transverse FM curves and their intersection with the lower off-diagonal curves, the dynamics are determined by a smooth alternation between the strict FM minimum and the FM-like minimum. This smooth transition occurs because their energy values must coincide at the crossing of the bifurcation set. Otherwise, an alternating transition between alignment states would be impossible given the Morse signature of the chambers. Here, the change in alignment state proceeds smoothly rather than suddenly or spontaneously, though the alignment parameter may exhibit minor oscillations. Due to the absence of sudden energy jumps, we do not treat this interval as part of the conventional CP regime, although a distinguished might not be clear without further studies of this regime which we leave for future works.

\subsection{Numerical verification of the chamber decomposition}
\label{SEC:NUMERICS}

The previous subsections provided a complete prediction of the system’s energy landscape and dynamics. In the following, we verify our findings using two different strategies. First, we quantitatively reproduce the predicted behavior shown in Fig.~\ref{fig:Landscape} by numerically tracking the critical points with different alignment states and analyzing the eigenvalues of the corresponding Hessian for each time $t$. Secondly, we numerically solve the equations of motion of the simplified model to check whether their behavior corresponds to the predicted dynamical regime.

\begin{figure*}
    \centering
    \includegraphics[width=\linewidth]{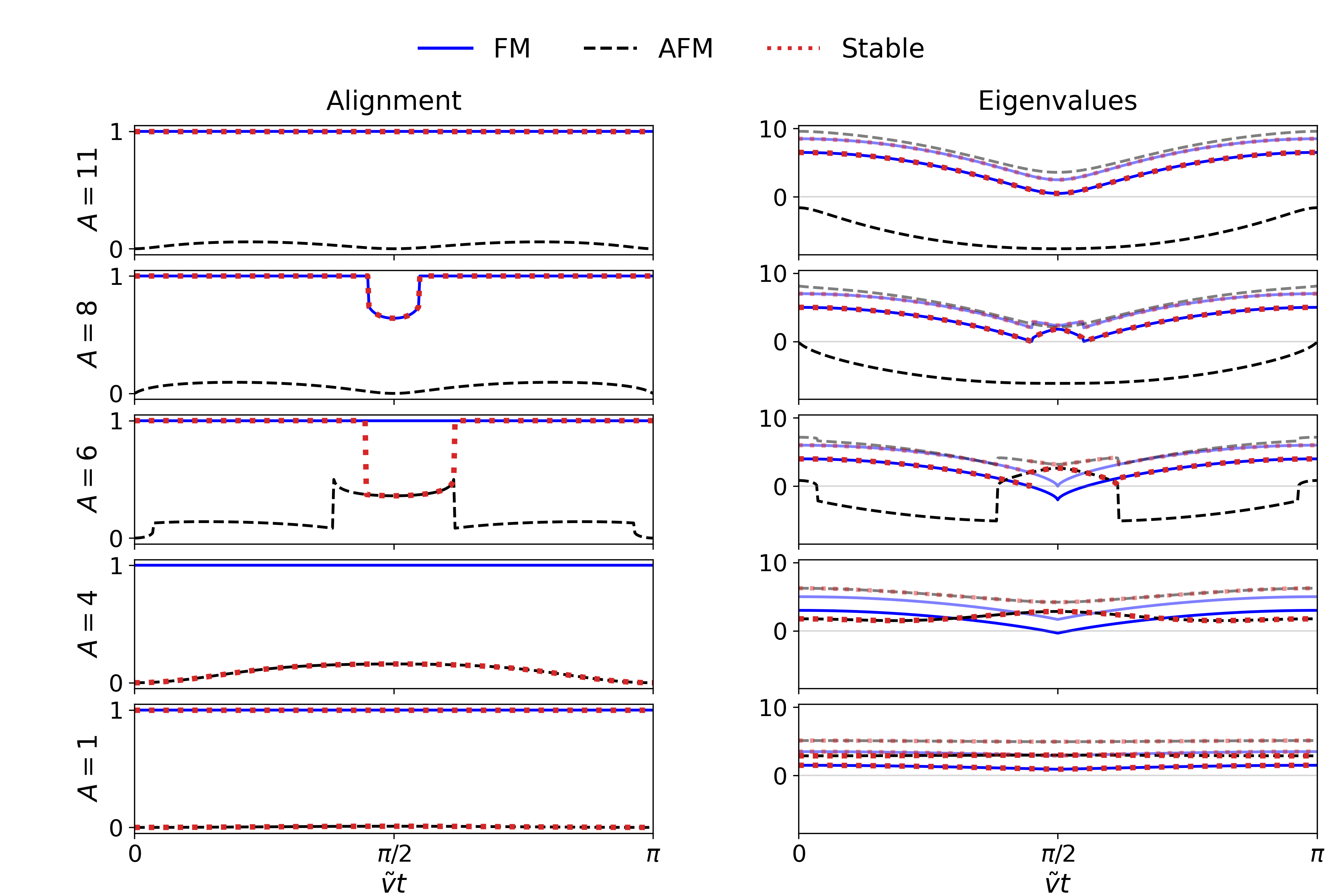}
    \caption{Numerical validations for the expected dynamical behavior at $A\in\{1,4,6,8,11\}$, shown from top to bottom as $A=11,8,6,4,1$. Shown is the alignment parameter $ \rho $ as a function of the driving phase $\tilde v t$ as well as the corresponding eigenvalues of the Hessian of the reduced energy. For each fixed-$A$ line-out, we follow an FM or FM-like critical point (blue solid lines) and an AFM-like critical point (black dashed lines). The tracked critical point is chosen to be a minimum whenever one exists and the corresponding saddle otherwise. The Hessian eigenvalues indicate the resulting changes in stability. For $A=1$, more than the two depicted branches exist; of the two FM branches, we show the one with lower energy. The red dotted lines indicate the quasi-static state predicted by the chamber decomposition, where we assume that the system remains on its current critical point while it is stable and switches to the other available critical point when the current one loses stability. }
    \label{fig:eval_lineouts}
\end{figure*}

The chamber decomposition maps out the stable states for any parameter pair $(A,\tilde{v}t)$. For fixed $A$, an approximate trajectory can be constructed by continuing minima in time and evaluating relative orientation of the slider moments. However, it is also possible to solve the inverse problem: for a given relative orientation of the rotor moments, we can numerically track selected critical points corresponding to a particular alignment and check their stability along the trajectory. This is done for different values of $A$ in Fig.~\ref{fig:eval_lineouts}. For a given alignment state, this figure shows the behavior of $\rho$ along selected critical points over the interval $\tilde{v}t\in[0,\pi]$, together with the corresponding Hessian eigenvalues. Note that here we group FM and FM-like states as FM states, and AFM-like states as AFM states.

Each tracked critical point is chosen to be the minimum the system is expected to occupy for a given relative orientation whenever possible, and the corresponding saddle otherwise. If a critical point ceases to exist, we switch to another critical point with the corresponding relative orientation. The red dotted curves show the trajectories predicted by our analysis: the system remains in its current minimum while it is stable, and transitions to another stable critical point only after the current one loses stability. The Hessian eigenvalues indicate precisely where these losses of stability occur. Fig.~\ref{fig:eval_lineouts} is therefore a quantitative realization, and thus a confirmation of the schematic picture in Fig.~\ref{fig:Landscape}. The analysis of the eigenvalues corroborates our prediction based on the line-outs through the chamber decomposition, as the times at which the stability changes align with the derived bifurcation curves. This agreement thereby confirms our findings from the previous subsection.

Next, we solve the equations of motion~\eqref{eq:EOMFULL} of the system numerically and compare the results with our predictions. To this end, we use the parameters $A\in{1,4,6,8,11}$, $N=20$, $\tilde{v}=1\,\mathrm{s}^{-1}$, and $\gamma=0.0011\,\mathrm{s}$. For the numerical integration, we again use the implicit fifth-order Runge--Kutta method of the Radau IIA family~\cite[Sec.~IV.8]{HairerWanner1996} and apply relative and
absolute error tolerances of $\texttt{rtol}=10^{-12}$ and $\texttt{atol}=10^{-14}$, respectively. Note that, here, we do not analyze systems that are in their time-periodic steady states; instead, the first periods illustrate how the system approaches its periodic dynamics. To check whether a particular layer separation $A$ can exhibit different dynamical behaviors depending on the initial state, we consider again both 
\begin{equation}
\theta_j(0)=\eta_j
\qquad
\text{(FM initial condition)},\label{eq:FM_init}
\end{equation}
and
\begin{equation}
\theta_j(0)=
\begin{cases}
\eta_j, & j\ \text{odd},\\
\pi+\eta_j, & j\ \text{even},
\end{cases}
\qquad
\text{(AFM initial condition)},\label{eq:AFM_init}
\end{equation}
where $\eta_j$ are independent, uniformly distributed random variables drawn from $[-0.01,0.01]$. As discussed before, the random perturbations break the exact symmetry of the strict FM state, allowing the numerical solution to leave symmetry-protected trajectories. Since any solution that starts with $\theta_j(0) \neq \theta_{j+1}(0)$ for any $j$ cannot exactly reach a state where all angles are equal in finite time, any statements about states being in a strict FM state are to be read as 'asymptotically close' to a strict FM state. The alignment of the $N$-degree-of-freedom version of the simplified model is defined using the nearest-neighbor average alignment
\begin{equation}
\bar\rho(t)
=
\frac{1}{N}
\sum_{j=1}^{N}
\cos^2\!\left(
\frac{\theta_j(t)-\theta_{j+1}(t)}{2}
\right),
\end{equation}
with periodic boundary condition $\theta_{N+1}=\theta_1$, as a proxy for the alignment parameter $\rho$, which we introduced for the two-angle reduction.

\begin{figure*}
    \centering
    \includegraphics[width=\linewidth]{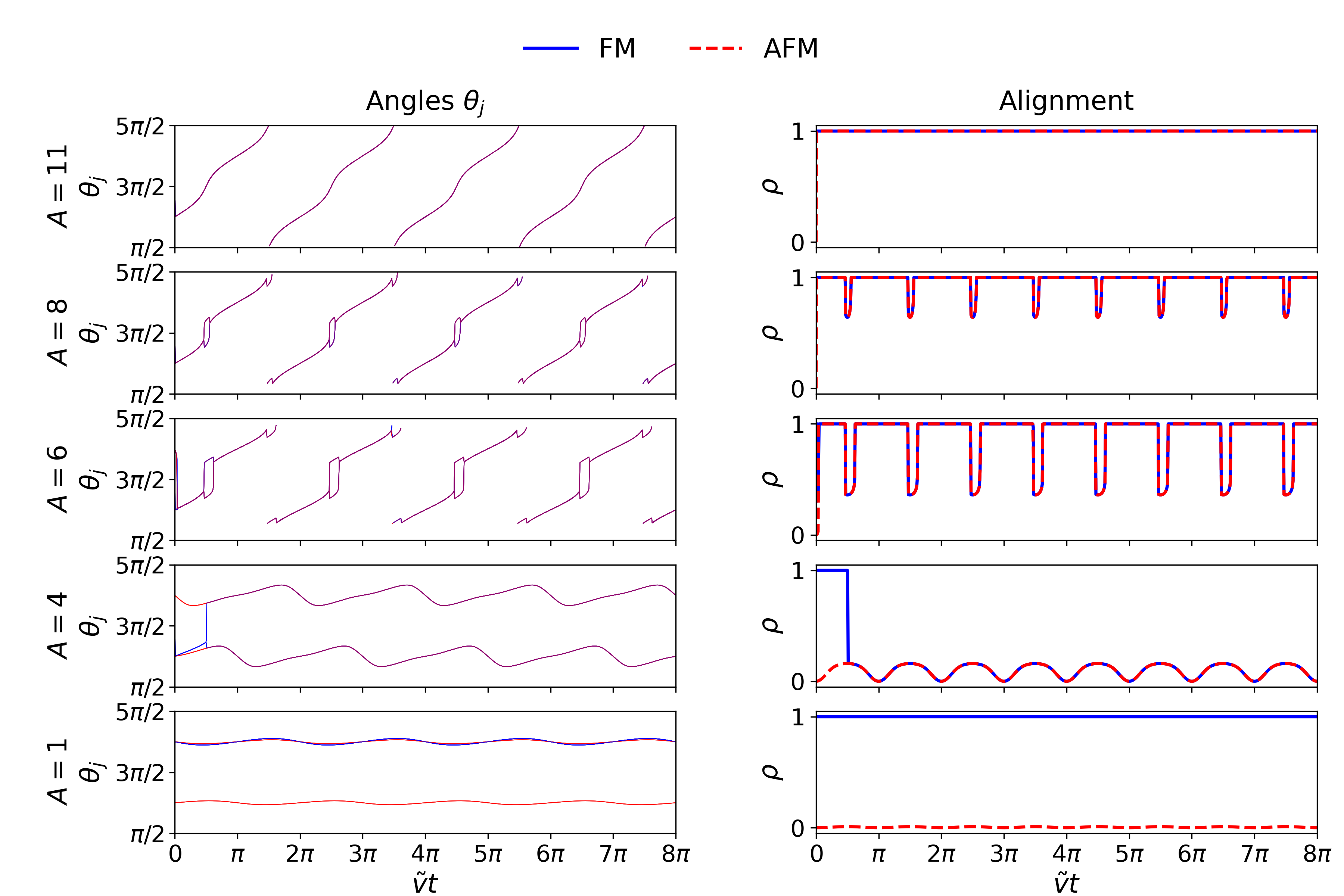}
    \caption{Numerical solutions of the full equations of motion~\eqref{eq:EOMFULL} for $N=20$ and $\gamma \tilde{v}=0.0011 $, integrated over four complete periods of the drive ($ \tilde{v} $ is set to unity for the numerical integration). Rows correspond, from top to bottom, to $A=11,8,6,4,1$. Depicted are the rotor angles $\theta_j$ as well as the corresponding mean alignment $\bar\rho$. Blue and red curves correspond to FM and AFM initial conditions, respectively, each perturbed by independent noise $\eta_j\in[-0.01,0.01]$. The first periods primarily shows relaxation toward the asymptotic time-periodic steady-state, whereas the later periods reveal the alignment states corresponding to the dynamical regime.}
    \label{fig:Numerics_full}
\end{figure*}

The results are shown in Figs.~\ref{fig:Numerics_full}. As predicted in the previous section, both stable FM and AFM-like states are observed at $A=1$, whereas only the strict FM response remains at $A=11$. At $A=6$ and $A=8$, the simulations reproduce the CP regime, with different degrees of variation in the alignment. Notably, the chamber decomposition also predicts the transient stability of the FM state at $A=4$, followed by a one-way transition to the AFM-like state in which the system subsequently remains. The long-time behavior of the alignment parameter in Fig.~\ref{fig:Numerics_full} corresponds directly to the stable trajectories predicted in Fig.~\ref{fig:eval_lineouts}.

\section{Magnetic friction and asymptotic behavior} \label{SEC:FRICTION}

The previous section focused on the different dynamical regimes that can emerge in these simplified systems. Another characteristic feature of the governing equations of motion, however, is the emergence of a distinct dissipation peak at intermediate values of the layer separation parameter $A$. In the following, we derive explicit estimates for the magnetic friction associated with the identified different dynamical regimes.

While we are unable to provide exact closed-form solutions for the full magnetic friction, we can rigorously derive useful approximations. Specifically, we apply a standard perturbative ansatz to obtain explicit expressions for the system's dynamics up to order $O(A)$ in the AFM regime and $O(1/A^3)$ in the FM regime. Using these expressions, we then compute the corresponding contributions to the average friction force $\langle F_x^{\textrm{mag}}\rangle$. Note that the expansion orders for each regime are chosen to yield the first non-trivial terms of the magnetic friction with respect to the layer separation parameter $A$. For the CP regime, we leverage our insights regarding the nature of the instability transition between different alignment states to estimate the energy dissipated due to the overdamped nature of the system.

\subsection{AFM regime or large layer separations}

For larger layer separations, corresponding to small $A$, the system remains close to the AFM configuration. In this limit, the external drive acts only as a weak perturbation to the stable AFM state $\theta_{j+1}^{(0)} \approx \theta_j^{(0)} + \pi$ [wrapped back into $[0,2\pi)$]. We therefore expand the full solution to first order in $A$ as
\begin{equation}
\theta_j(t) = \theta_j^{(0)} + A\psi_j(t) + O(A^2).
\end{equation}
Substituting this expansion into Eq.~\eqref{eq:EOMFULL}, applying trigonometric identities, and retaining only terms linear in $A$, we obtain the differential equation
\begin{equation}
\gamma\dot{\psi}_j = -8\psi_j + \psi_{j+1} + \psi_{j-1} + \sin(\theta_j^{(0)} - \tilde{v}t)
\end{equation}
for the first-order correction $\psi_j(t)$.

\begin{figure}[ht!]
    \centering
    \includegraphics[width=1\linewidth]{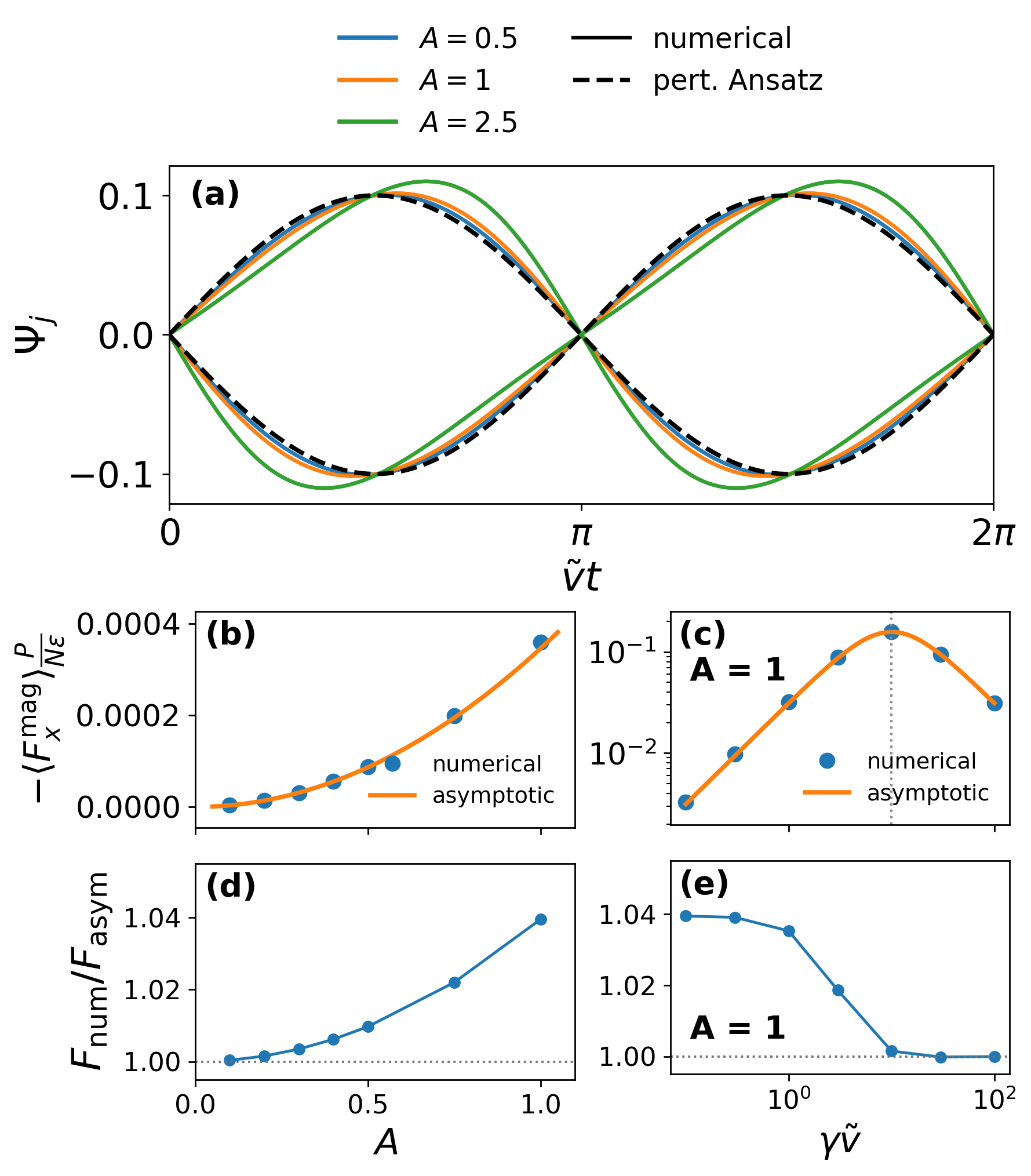}
    \caption{Comparison of the asymptotic estimates for small $A$ and results from numerical integration. The numerical integration are performed for $A=1$, $N=20$ and $\gamma \tilde{v}=0.0011$ if not stated otherwise.  (a) First order correction $\psi_j= (\theta_j-\theta_j^{(0)})/A$ for even and odd $j$. Also shown is the numerical and asymptotic magnetic friction as functions of $A$ (b) and $\gamma\tilde{v}$ (c), as well as the ratios of these frictions as functions of $A$ (d) and of $\gamma\tilde{v}$ (e). }
    \label{fig:AFM_assymptotics}
\end{figure}

To solve this equation, we introduce a complex auxiliary differential equation
\begin{equation}
\gamma\dot{\tilde{\psi}}_j = -8\tilde{\psi}_j + \tilde{\psi}_{j+1} + \tilde{\psi}_{j-1} + e^{i\theta_j^{(0)} - i\tilde{v}t}.
\end{equation}
We then adopt the ansatz
\begin{equation}
\tilde{\psi}_j(t) = (-1)^j c_j e^{-i\tilde{v}t},
\end{equation}
which reflects the alternating structure of the unperturbed AFM state. Substituting this relation into the auxiliary equation yields
\begin{align}
(8 - i\gamma \tilde{v} )\, c_j + c_{j+1} + c_{j-1} = (-1)^j e^{i\theta_j^{(0)}} .
\end{align}
Because the unperturbed AFM state possesses a staggered structure, the right-hand side is independent of $j$, allowing us to assume site-independent coefficients $c_{j+1} = c_{j-1} = c_j =: c$. Consequently, the coefficient satisfies
\begin{equation}
c = \frac{(-1)^N e^{i\theta_N^{(0)}}}{10 - i\gamma \tilde{v} } = -\frac{ e^{i\theta_1^{(0)}}}{10 - i\gamma \tilde{v} }.
\end{equation}
The physical first-order correction to the original equation of motion is recovered by taking the imaginary part of the auxiliary solution, $\psi_j(t) = \textrm{Im}[\tilde{\psi}_j(t)]$. This yields
\begin{equation}
\psi_j(t) = \frac{(-1)^{j+1}}{100 + \gamma^2 \tilde{v}^2}
\left[
10\sin(\theta_1^{(0)} - \tilde{v} t)
+ \gamma \tilde{v}\,\cos(\theta_1^{(0)} - \tilde{v} t)
\right],
\end{equation}
where we used the fact that $N$ is even. Figure~\ref{fig:AFM_assymptotics} shows $ \psi_j(t) $ for different $ A $. It is a good approximation for the resulting oscillations for $ A \lesssim 1 $.

We can estimate the dissipation in the AFM regime using our result for $ \psi_j(t) $. Using trigonometric identities, we can use our first-order solution to approximate
\begin{align}
\sin(\theta_j(t) - \tilde{v}t)
\approx
\sin(\theta_j^{(0)} - \tilde{v}t)
+ A\,\psi_j(t)\,\cos(\theta_j^{(0)} - \tilde{v}t).
\end{align}
Multiplying by $\varepsilon A^2 \dot{\psi}_j / P$, neglecting terms of order $O(A^3)$, and integrating over one full period in time, all oscillatory contributions vanish. Based on Eq.~\eqref{eq:MAGFRIC}, this yields the average magnetic friction 
\begin{align}
\langle F_x^{\textrm{mag}} \rangle \big\vert{}_{\textrm{AFM}}
\approx
-\frac{N \pi A^2 \gamma \tilde{v}}{(100 + \gamma^2 \tilde{v}^2)} \frac{\varepsilon}{P}.
\end{align}
of the AFM regime. The dissipation scales as $A^2 \propto 1/h^6$ and decreases as $1/\tilde{v}$ in the high-velocity limit. The magnetic friction is therefore highly sensitive to the layer separation $h$ and decreases with increasing sliding velocity after reaching a peak at $\gamma\tilde{v} = 10$ (which is outside the quasi-static regime; for the asymptotic expansion, however, this restriction is not mandatory). The $ A^2 $ dependence of the magnetic friction is expected for observables derived from the energy within a linear response framework as applied here. In particular, the $ 1/h^6 $ scaling is connected to the dipole nature of the interactions. The velocity dependence, on the other hand, contrasts with conventional models of sliding friction, where dissipation typically increases with relative velocity. In the context of magnetic friction, such behavior has been reported before~\cite{Fusco2008}. 

Figure~\ref{fig:AFM_assymptotics}~b) to e) show a comparison between magnetic friction obtained from numerical integration and from the derived asymptotic friction law with respect to the $ A $ and the $ \tilde{v} $ dependence.  The integration was done for $A=1$, $N=20$ and $\gamma \tilde{v}=0.0011$ if not stated otherwise. The perturbation results reproduce the magnetic friction remarkably well, especially the velocity dependence. 

\subsection{FM regime or small layer separations}

For small layer separations, the parameter $A$ becomes large. In this regime, the slider moments are basically in a strict FM state, which allows us to simplify Eq.~\eqref{eq:EOMFULL} by decoupling the rotational degrees of freedom. Since all moments align, we have $\theta_j(t) \approx \theta(t)$ for all $j$, and the dynamics reduce to the single differential equation
\begin{equation}
\gamma \dot{\theta}
= -3 \sin(2\theta) + A \sin(\theta - \tilde{v}t).
\end{equation}
To characterize the response to the external drive, we introduce the phase lag $\phi = \theta - \tilde{v}t$, which obeys the equation of motion
\begin{equation}
\sin(\phi)
= \frac{1}{A}
\left[
\gamma(\dot{\phi} + \tilde{v})
+ 3 \sin(2\phi + 2\tilde{v}t)
\right].
\end{equation}
In the limit $A \to \infty$, the right-hand side vanishes, giving the zeroth-order solution $\phi^{(0)} = \pi$ (reflecting the anti-alignment of the moments relative to the external drive; see also Appendix~\ref{APP:Lag}).

\begin{figure}[ht!]
    \centering
    \includegraphics[width=1\linewidth]{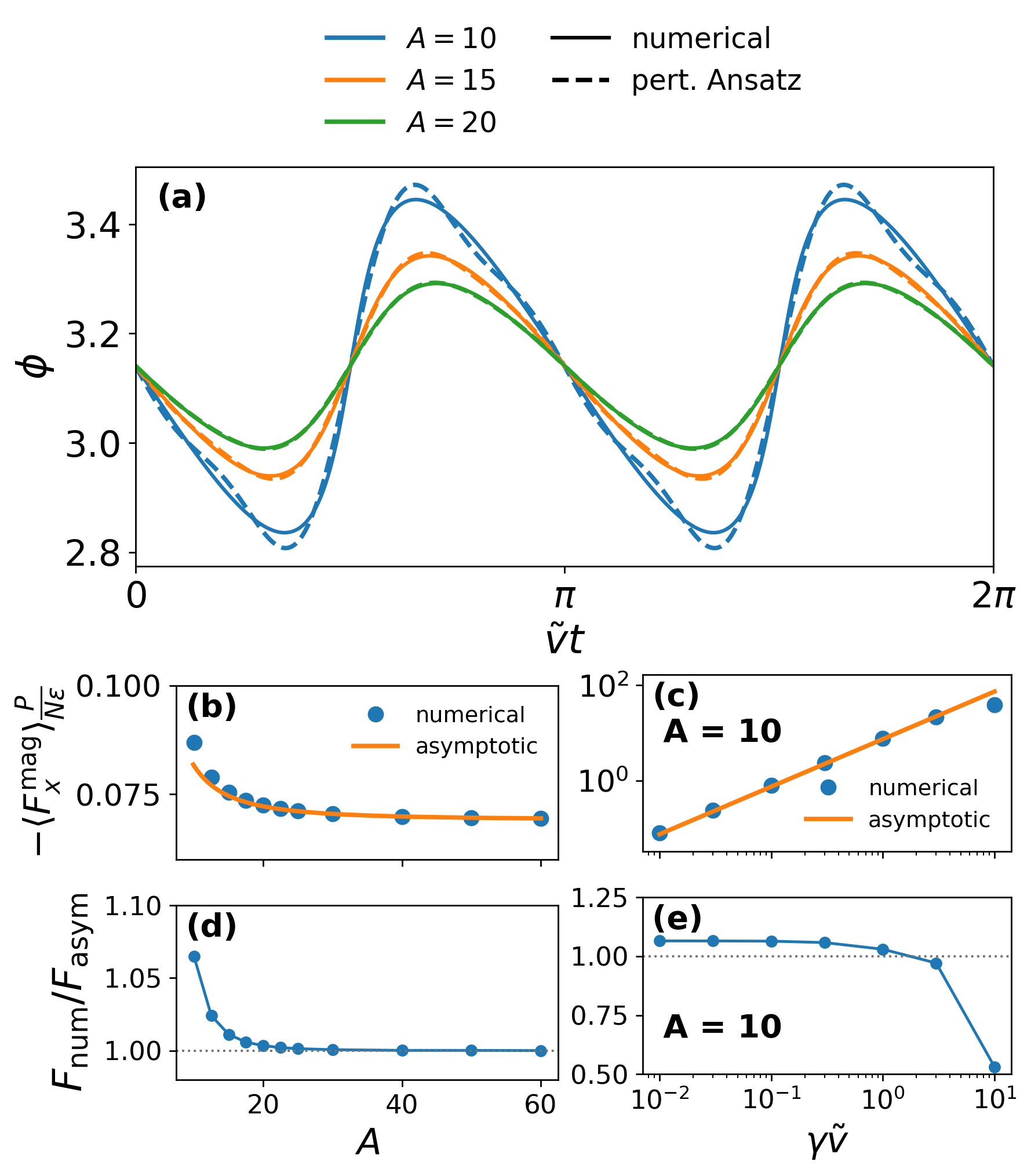}
    \caption{Comparison of the asymptotic estimates and the results of numerical integration for $A\in [10,60]$. All runs use $A=10$, $N=20$ and $\gamma \tilde{v}=0.0011$ if not stated otherwise. (a) Phase lag $\phi_j$. Also shown is a compares between numerical and asymptotic magnetic friction as functions of $A$ (b) and $\gamma\tilde{v}$ (c) as well as ratios of these frictions as functions of $A$ (d) and of $\gamma\tilde{v}$ (e). }
    \label{fig:FM_assymptotics}
\end{figure}

To make further progress, we choose the perturbative ansatz
\begin{equation}
    \phi(t)
    = \pi +\frac{\phi^{(1)}}{A}
    + \frac{\phi^{(2)}}{A^2}
    + \frac{\phi^{(3)}}{A^3}
    + O\!\left(\frac{1}{A^4}\right).
\end{equation}
Substituting this expansion into the differential equation for $\phi$, expanding consistently in powers of $1/A$, and matching terms order by order, we obtain
\begin{align}
    \phi^{(1)}(t)
    &= -\gamma \tilde{v} - 3 \sin(2\tilde{v} t), \\
    \phi^{(2)}(t)
    &= 12 \gamma \tilde{v} \cos(2\tilde{v} t)
       + 9 \sin(4\tilde{v} t),
\end{align}
and
\begin{align}
    \phi^{(3)}(t)
    &= -\frac{1}{6} \gamma^3 \tilde{v}^3 - \frac{81}{4} \gamma \tilde{v} +\left( \frac{57}{2} \gamma^2 \tilde{v}^2 + \frac{81}{8} \right) \sin(2\tilde{v}t) \nonumber \\ &-\frac{351}{4} \gamma \tilde{v} \cos(4\tilde{v}t) -\frac{315}{8} \sin(6\tilde{v}t).
\end{align}
Here, only algebraic substitutions and differentiation are needed to recover $ \phi^{(n)}(t) $. Our perturbation result for $ \phi(t) $ compared to numerical integration in dependence on $ A $ is depicted in Fig.~\ref{fig:FM_assymptotics}~a). The derived corrections for the phase lag recovers the dynamics sufficiently for $ A \gtrsim 15 $.

From the phase lag, we can again estimate the magnetic friction. We begin with the integral
\begin{align}
&\int_0^{2\pi/\tilde{v}}
\sin(\theta_j - \tilde{v} t)\,\dot{\theta}_j,\mathrm{d}t
\nonumber \\
&=
-\left.\cos(\phi(t))\right|_0^{2\pi/\tilde{v}}
+ \tilde{v}
\int_0^{2\pi/\tilde{v}}
\sin(\phi(t))\,\mathrm{d}t .
\end{align}
Since all derived orders of $\phi(t)$ are periodic with respect to $2\pi/\tilde{v}$, the boundary term vanishes, leaving only the second integral. Expanding the sine up to order $O(1/A^3)$, most terms resulting from $\phi^{(1)}$, $\phi^{(2)}$, and $\phi^{(3)}$ vanish upon integration due to periodicity. Specifically, the first-order term contributes $2\pi \gamma \tilde{v} / A$, while the second-order term yields no net contribution. The next nonvanishing correction arises from the third-order expansion of the sine, where the total integrand contribution is given by $-\phi^{(3)}(t)/A^3 + [\phi^{(1)}(t)]^3 / 6A^3$. Among these terms, only those independent of trigonometric factors or proportional to $\sin^2$ and $\cos^2$ survive temporal integration, yielding a contribution of $36\pi \gamma \tilde{v} / A^3$. Substituting this result into the expression for the magnetic friction, Eq.~\eqref{eq:MAGFRIC}, we obtain
\begin{equation}
\langle F_x^{\textrm{mag}} \rangle \big\vert{}_{\textrm{FM}}
\approx
-\left[2\pi N \gamma \tilde{v}
+ \frac{36\pi N \gamma \tilde{v}}{A^2}\right] \frac{\varepsilon}{P}.
\end{equation}
The first term corresponds to the dissipation due to Stokes drag, $N v \Gamma k^2$, for smoothly rotating degrees of freedom and agrees with the result reported in Refs.~\cite{Gu2026, Magiera2009}. The second term represents a small height-dependent correction $\propto h^6$, originating from the $-3\cos(2\theta_j)/2$ contribution in the simplified Hamiltonian. Remarkably, in the FM regime, the magnetic friction increases linearly with $\tilde{v}$, which is qualitatively distinct from the behavior observed in the AFM regime. A comparison to results from numerical integration ($A=10$, $N=20$ and $\gamma \tilde{v}=0.0011$ if not stated otherwise) with respect to the $ A $ and $ \tilde{v} $ dependence shown in Fig.~\ref{fig:FM_assymptotics}~(b) to (e) confirms the derived asymptotic friction law.

\subsection{CP regime and intermediate layer separations}
\label{subsec:quasistatic-jump-friction}

More involved is an estimation of the magnetic friction peak at intermediate layer separations. Here, our analysis of the energy landscape provides an intuitive picture of the dissipation mechanism, which we can use to derive the friction. As indicated in Fig.~\ref{fig:Landscape}, the defining feature of the CP regime is the occurrence of relaxation processes, wherein the system jumps from one critical point of the energy landscape to another. Such jumps inherently occur from energetically higher to lower states. Unlike in the underdamped case, where the system can oscillate around a stable critical point after falling from a higher energy state, the overdamped simplified model must dissipate all excess energy as it smoothly relaxes toward its new stable minimum. The energy dissipated through this mechanism accounts directly for the enhanced magnetic sliding friction.

We can use this result to obtain quantitative approximations for the dissipated magnetic friction in the CP regime. As discussed in Appendix~\ref{APP:MF}, the dissipated energy can be written as
\begin{align}
    E_{\textrm{diss}} &= - \int_0^{2\pi/\tilde{v}} \frac{\partial}{\partial t} \mathcal{H}((\theta^{(\gamma \tilde{v})}_j(t))_j;A, \tilde{v}t) \, \textrm{d}t \nonumber \\
    &= - \frac{\varepsilon N}{2} \int_0^{2\pi/\tilde{v}} \frac{\partial}{\partial t} \mathcal{H}_r(\vec{p}^{(\gamma\tilde{v})}(t);A, \tilde{v}t) \, \textrm{d}t,
\end{align}
where we used the two-angle reduction $\mathcal{H} = \varepsilon N \mathcal{H}_r / 2$ and assume that the system is already in a steady periodic state at time $ t = 0 $. The superscript ${(\gamma\tilde{v})}$ denotes the periodic solution at finite but small $\gamma\tilde{v}$. In the quasi-static limit, the lag between the system state and the minimum of the energy landscape is small, allowing us to write
\begin{equation}
    \vec{p}^{(\gamma\tilde{v})}(t) \approx \vec{p}^*(t) + \delta \vec{p}(t),
\end{equation}
with $\delta \vec{p}(t) = O(\gamma\tilde{v})$ (see, for instance, Appendix~\ref{APP:Lag}, where $\delta \vec{p}(t)$ is explicitly estimated for the FM regime). Here, $\vec{p}^*(t)$ denotes a piecewise continuous function constructed from the critical points of the energy landscape that are followed by $ \vec{p}^{(\gamma\tilde{v})}(t) $, satisfying
\begin{equation}                      \nabla_p\Hr\!\left(\vec{p}^\ast(t);A,\tilde{v}t\right)=0. \label{eq:EDissGradZero}
\end{equation}
Using this expansion, we estimate the dissipated energy via
\begin{align}
     E_{\textrm{diss}} &\approx - \frac{\varepsilon N}{2} \int_0^{2\pi/\tilde{v}} \frac{\partial}{\partial t} \mathcal{H}_r(\vec{p}^*(t);A, \tilde{v}t) \, \textrm{d}t \nonumber \\ &-  \frac{\varepsilon N}{2} \int_0^{2\pi/\tilde{v}} \nabla_p \left[\frac{\partial}{\partial t} \mathcal{H}_r(\vec{p}^*(t);A, \tilde{v}t)\right] \cdot \delta \vec{p}(t) \, \textrm{d}t. \label{eq:EExpand}
\end{align}
Since $\delta \vec{p}(t) = O(\gamma\tilde{v})$ and $\nabla_p \left[\partial \mathcal{H}_r/ \partial t\right] = O(\tilde{v})$ by construction of the drive, the second term on the right-hand side of Eq.~\eqref{eq:EExpand} is of order $O(\gamma \tilde{v})$ after integration. Hence, we obtain
\begin{equation}
     E_{\textrm{diss}} \approx - \frac{\varepsilon N}{2} \int_0^{2\pi/\tilde{v}} \frac{\partial}{\partial t} \mathcal{H}_r(\vec{p}^*(t); A, \tilde{v}t) \, \textrm{d}t + O(\gamma \tilde{v}), \label{eq:EDissJumps}
\end{equation}
where the corrections mostly account for the Stokes-drag contributions.

Because of property~\eqref{eq:EDissGradZero}, we can replace the partial time derivative in Eq.~\eqref{eq:EDissJumps} with the total time derivative. Hence, since $\vec{p}^*(t)$ contains jumps and is piecewise continuous between them, we can evaluate the dissipated energy directly from the energy differences at the jumps. We find
\begin{align}
     E_{\textrm{diss}} &\approx - \frac{\varepsilon N}{2} \int_0^{2\pi/\tilde{v}} \frac{\textrm{d}}{\textrm{d} t} \mathcal{H}_r(\vec{p}^*(t);A, \tilde{v}t) \, \textrm{d}t + O(\gamma \tilde{v}) \nonumber \\ &= - \frac{\varepsilon N}{2} \sum_k \left[ \mathcal{H}_r(\vec{p}^+_k) - \mathcal{H}_r(\vec{p}_k^-) \right] + O(\gamma \tilde{v}),
\end{align}
where the sum runs over all jumps between critical points within a period, $\vec{p}^+_k$ denotes the critical point immediately before jump $k$ (i.e., the \textit{pre-jump state}), and $\vec{p}_k^-$ denotes the critical point immediately after jump $k$ (i.e., the \textit{post-jump state}). Note that the integral does not vanish even though the system is periodic because $ \mathcal{H}_r $ is evaluated at the discontinuous function $ \vec{p}^*(t) $ instead of the continuous solution $ \vec{p}^{(\gamma \tilde{v})}(t) $ of the equations of motion (cf. Appendix~\ref{APP:MF}). Also, since the jumps in energy are static properties of $ \mathcal{H}_r $, the first term is of order $ O(1) $ with respect to the dependence on $ \gamma \tilde{v} $.

To make further progress, we must compute the energy jumps identified by our chamber decomposition. Figure~\ref{fig:BuL} provides the necessary information for this calculation. For a fixed value of $A$, the horizontal line-out through Fig.~\ref{fig:BuL} determines whether jumps occur: they take place only when the tracked minimum terminates or become unstable at a bifurcation curve. The type of curve crossed identifies the pre-jump critical point, while the minimum structure in the adjacent chamber identifies the post-jump critical point.

Only two independent mechanisms occur: first, the loss of transverse stability of the strict FM minimum at $B_{\rm FM}^{\perp}$. Let $(\tilde{v}t)_\perp(A)$ be the corresponding intersection phase, obtained from Eqs.~\eqref{eq:fm_transverse_curve_sin} and \eqref{eq:fm_transverse_curve_cos}, or equivalently from Eq.~\eqref{eq:FM_param_curves} with $\alpha=\perp$. The pre-jump angle $\chi_T$ is the associated value of $\chi_+$, while the post-jump off-diagonal minimum $(\chi_{+,\perp}^{-},\zeta^{-}_{\perp})$ is the stable solution to Eqs.~\eqref{eq:interior_cp_1} and \eqref{eq:interior_cp_2} evaluated at the same $(A,(\tilde{v}t)_\perp)$. The corresponding reduced energy drop is
\begin{equation}
\Delta\Hr^{\perp}(A)
=
\Hr^{\rm FM}\!\left(\chi_T;A,(\tilde{v}t)_\perp\right)
-
\Hr\!\left(\chi_{+,\perp}^{-},\zeta^{-}_{\perp};A,(\tilde{v}t)_\perp\right),
\label{eq:dHr_perp}
\end{equation}
where the two energies are evaluated using Eqs.~\eqref{eq:Hr_FM_branch} and \eqref{eq:Hr_chi_zeta}, respectively.

The second mechanism is the disappearance of the tracked off-diagonal minimum at the lower off-diagonal bifurcation curve. Let $(\tilde{v}t)_{\rm low}(A)$ be the corresponding intersection phase. The pre-jump state $(\chi_{+,\rm low}^{+},\zeta_b)$ is determined by the lower branch of Eq.~\eqref{eq:offdiag_xpm} together with Eqs.~\eqref{eq:offdiag_param_A} and \eqref{eq:offdiag_param_t}. The post-jump strict FM angle $\chi_{+,\rm low}^{-}$ is the stable solution to Eq.~\eqref{eq:FM_cp_condition} at the same $(A,(\tilde{v}t)_{\rm low})$, with stability governed by Eq.~\eqref{eq:FM_hessian_diag}. The energy drop is
\begin{align}
\Delta\Hr^{\rm low}(A)
&=
\Hr\!\left(\chi_{+,\rm low}^{+},\zeta_{\rm low}^{+};(\tilde{v}t)_{\rm low},A\right)
\nonumber \\
&-
\Hr^{\rm FM}\!\left(\chi_{+,\rm low}^{-};A,(\tilde{v}t)_{\rm low}\right).
\label{eq:dHr_low}
\end{align}

Equations~\eqref{eq:dHr_perp} and \eqref{eq:dHr_low} provide implicit analytic expressions for the two possible energy drops that must be dissipated. The chamber decomposition specifies which stable branch is followed and, consequently, which solution to the stationarity equations must be selected. By the reflection and half-period symmetries of $\mathcal{H}_r$, each independent energy drop occurs twice over a full period of the drive. Hence, the total dissipated energy per period is
\begin{equation}
E_{\textrm{diss}}
=
-N\varepsilon
\left[
\Delta\Hr^{\perp}(A)
+
\Delta\Hr^{\rm low}(A)
\right]
+
O(\gamma\tilde{v}),
\label{eq:Ejump_total}
\end{equation}
with a term omitted whenever the corresponding boundary crossing is absent from the particular line-out. 

\begin{figure}
    \centering
    \includegraphics[width=\linewidth]{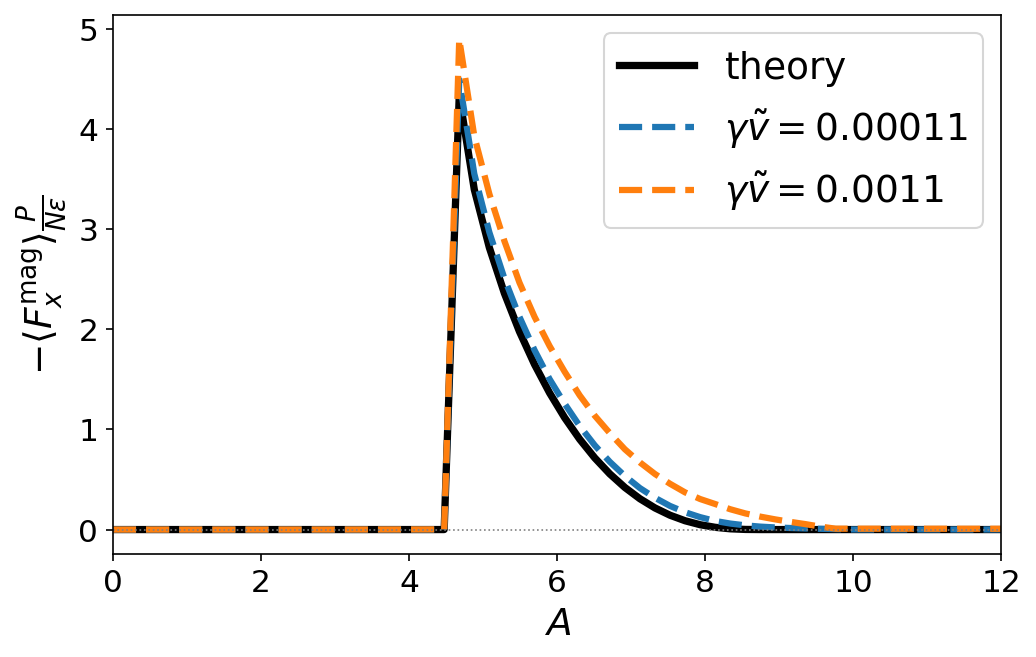}
    \caption{Comparison of the magnetic friction calculated from the energy drops at quasi-static jumps with the magnetic friction obtained by direct integration of Eq.~\eqref{eq:EOMFULL}. The latter approaches the quasi-static prediction as $\gamma\tilde v$ decreases. }
    \label{fig:jump_friction} 
\end{figure}

Using the derived dissipated energy, we can approximate the magnetic friction via~\footnote{This result also hints at why magnetic friction in the CP regime might be largely independent of the specific microscopic model for shaft friction as suggested earlier~\cite{Gu2026}: the dominant contribution to the total dissipation is governed directly by the energy drops. The precise mechanism by which the excess energy after the jump is transferred to the thermal bath is secondary. For instance, the energy may be dissipated via microscopic Stokes friction under the assumption of overdamped rotational degrees of freedom, or mediated by the excitation of overdamped collective modes that distribute the energy across a large ensemble of rotors, each dissipating a small fraction.}
\begin{equation}
    \langle F_x^{\textrm{mag}} \rangle \big\vert{}_{\textrm{CP}} \approx - N
\left[
\Delta\Hr^{\perp}(A)
+
\Delta\Hr^{\rm low}(A)
\right] \frac{\varepsilon}{P}. \label{eq:ExCPMF}
\end{equation}
Figure~\ref{fig:jump_friction} shows the friction obtained from an automated sweep over $A$ and compares it with the average friction obtained from numerical integrations of the full system~\eqref{eq:EOMFULL}. The numerical integrations were performed for $\gamma \tilde{v}=0.0011 $ and $\gamma \tilde{v}=0.0001 $ starting from AFM initial conditions (perturbed by uniform random numbers as before). Friction values were averaged over several driving periods following an initial two-period relaxation phase to allow the system to reach its time-periodic steady state. All other parameters were kept fixed. The magnetic friction calculated directly from the energy jumps successfully reproduces the friction peak, showing increasingly excellent quantitative agreement as $\gamma\tilde{v} \to 0$.

The variations in the upper boundary of the magnetic friction peak interval with $\gamma \tilde{v}$ are directly tied to the distinguished interval $8.8 \lesssim A \le 10$. In this regime, the enhanced dissipation arises not from abrupt energy jumps, but from smooth transitions between strict FM and FM-like alignment states. The resulting magnetic friction in this region is expected to be of order $O(\gamma \tilde{v})$ and is therefore not captured by our leading-order derivations. In the following, we neglect any onset of the magnetic friction peak driven by such $ O(\gamma \tilde{v})$ contributions and focus exclusively on the CP friction in the limit $\gamma \tilde{v} \to 0$.

Notably, our results in Eqs.~\eqref{eq:Ejump_total} and \eqref{eq:ExCPMF} establish a direct conceptual bridge between the non-equilibrium dynamics of the CP regime and the classical stick-slip friction mechanism~\cite{Baumberger2006, Persson2000, Ruina1983, Berman1996}, identifying the CP regime as its magnetic analogue: within each chamber, the rotor array follows a stable minimum (the stick phase), whereas at the bifurcation boundaries the tracked minimum loses stability, causing the system to relax dynamically to a new state (the slip phase). The dissipated energy per period is thus given by the sum of these quasi-static energy drops, in direct analogy to the mechanical work released during slip events in classical stick-slip friction. Stick-slip states have previously been linked to magnetic friction~\cite{Komatsu2023}. In that work, however, stick-slip dynamics were considered for the translation of the sliding layer as a whole within an externally driven Ising spin system.

Equation~\eqref{eq:ExCPMF} provides a rigorous representation of the $O(1)$ contribution (with respect to $\gamma \tilde{v}$) to the magnetic friction in the CP regime. Together with the chamber decomposition, this gives an algorithm to estimate the magnetic friction up to order $ O(1) $ for any $A$ at accuracies only limited by the accuracy of the implicit solver used. However, unlike the asymptotic results obtained for the other regimes, the overall magnitude of the friction cannot be immediately read off from this expression. In the following, we therefore present several explicit approximations that serve as closed-form expressions, providing direct quantitative access to the dissipation in the CP regime. For this, we define the total dimensionless energy drop entering both $E_{\textrm{diss}}$ and $\langle F_x^{\textrm{mag}} \rangle$ as
\begin{equation}
    \mathcal{J}(A) = \Delta\Hr^\perp(A)+\Delta\Hr^{\mathrm{low}}(A).
\label{eq:cp_jump_function}
\end{equation}
Note that all quantities comprising $\mathcal{J}(A)$ are fully determined by the chamber decomposition and the associated relations derived earlier. Exact derivations of the approximations presented below are provided in Appendix~\ref{app:proxies}.

First, note that $\mathcal{J}(A)$ is only non-zero, when any jumps actually occur. This only happens for $A\in [A_-,A_*]$, where 
\begin{align}
A_- \approx 4.635,
\label{eq:cp_Aminus}
\end{align}
marks the minimum $A$-value on the lower off-diagonal curve and 
\begin{equation}
A_*
=
\sqrt{\frac{1100+40\sqrt{601}}{27}}
\approx
8.778,
\label{eq:cp_Astar}
\end{equation}
marks the maximum $A$-value where it merges with the transverse FM curve~\footnote{Using heuristic arguments, Ref.~\cite{Gu2026} estimated the magnetic friction peak of the CP regime to occur at roughly $A \approx 5$. This value falls directly within our rigorously derived layer separation interval for the CP regime.}. For than $A<A_-$, stable trajectories never intersect zone 1 in Fig.~\ref{fig:BuL}~a) and thus AFM-like states are stable and no jump into an FM state happens. For $A\gg A_*$, critical points corresponding to strict FM states are the exclusive stable state. 

In $[A_-,A_*]$, we can expand $ \mathcal{J}(A) $ at the boundaries. At the lower boundary ~\footnote{In Ref.~\cite{Gu2026}, heuristic arguments were used to estimate a magnetic friction peak magnitude of roughly $ \langle	F_x^{\textrm{mag}}\rangle \approx - 6[N-1]\,\varepsilon/P$, where the factor $ N-1 $ comes from the open boundaries used in Ref.~\cite{Gu2026}. In contrast, our rigorous derivation yields a peak value of $ \langle F_x^{\textrm{mag}} \rangle \approx -5N\,\varepsilon/P$. While these two expressions agree exactly for $N = 6$, the heuristic estimate of Ref.~\cite{Gu2026} overestimates the magnetic friction by $20\%$ in the limit $N \to \infty$ (under the assumptions of periodic boundaries).}
\begin{align}
\mathcal J_-(A)
&\approx
4.976-
3.014\sqrt{A-A_-}, \qquad
A\downarrow A_-,
\label{eq:cp_lower_proxy}
\end{align}
whereas at the upper boundary, we have 
\begin{equation}
\mathcal J_*(A)
\approx
0.104\,(A_*-A)^3,
\qquad
A\uparrow A_*.
\label{eq:cp_endpoint_behavior}
\end{equation}
The square-root dependence at the lower boundary may be viewed as the leading term of a \textit{Puiseux expansion}~\cite{Kuznetsov2023,Wall2004}. The resulting scaling at the boundaries of the CP regime indicates that a description of magnetic friction is beyond the scope of linear response frameworks. 

Note that both Eq.~\eqref{eq:cp_lower_proxy} and Eq.~\eqref{eq:cp_endpoint_behavior} are local approximations that naturally deteriorate towards the other end of the interval. Expansions of $\mathcal{J}_-$ to higher orders of $\sqrt{A-A_-}$ are eventually competitive approximations on most of $[A_-,A_*]$, but struggle to capture the $(A_*-A)^3$ behavior with a reasonable number of terms. Higher-order Taylor series at $A_*$ or inside $(A_-,A_*)$ provide only local improvements and do not converge on the full interval due to the identified $\sqrt{A-A_-}$ behavior of $\mathcal{J}$. Because of this, the maximum error of any Taylor approximation does not necessarily decrease with higher-order approximations.

For a quick and accurate explicit approximation, we therefore provide a \textit{constrained multipoint Pad\'e-type rational interpolation}~\cite{BakerGravesMorris1996,Buslaev2013}. For this, we define the abbreviation
\begin{equation}
s=\sqrt{\frac{A-A_-}{A_*-A_-}},
\qquad
0\leq s\leq1.
\label{eq:cp_scaled_coordinate}
\end{equation}
This square-root coordinate correctly resolves the behavior at the lower boundary. To impose the upper cubic zero as well, we assume
\begin{equation}
\mathcal J_{[m/n]}(A)
=
(1-s^2)^3\frac{P_m(s)}{Q_n(s)},
\end{equation}
with 
\begin{align}
P_m(s)&=\sum_{k=0}^m \alpha_ks^k,\\
Q_n(s)&=1+\sum_{k=1}^n \beta_ks^k.
\label{eq:cp_rational_form}
\end{align}
This rational estimate is obtained directly from the implicit system, without first constructing a local power series. For this, we choose $m+n+1$ \textit{Chebyshev-distributed nodes}, which provide the usual endpoint clustering for global approximation on a finite interval~\cite{Trefethen2020},
\begin{equation}
s_i
=
\frac{1}{2}
\left[
1+
\cos\left(
\frac{(i+\tfrac12)\pi}{m+n+1}
\right)
\right],
\qquad
i=0,\ldots,m+n,
\label{eq:cp_nodes}
\end{equation}
set $A_i=A_-+(A_*-A_-)s_i^2$, and evaluate
$\mathcal J_i=\mathcal J(A_i)$ from
Eq.~\eqref{eq:cp_implicit_friction}. Requiring 
$\mathcal J_{[m/n]}(A_i)=\mathcal J_i$ gives
\begin{equation}
(1-s_i^2)^3\sum_{k=0}^m \alpha_ks_i^k
-
\mathcal J_i\sum_{k=1}^n \beta_ks_i^k
=
\mathcal J_i,
\label{eq:cp_linear_fit}
\end{equation}
which is a linear system for the $m+n+1$ unknown coefficients. The resulting \textit{multipoint Pad\'e-type approximation} is admissible only if
\begin{equation}
Q_n(s)\neq0
\qquad
\text{for all }s\in[0,1],
\label{eq:cp_pole_condition}
\end{equation}
since a zero of $Q_m$ would generate a spurious pole.

\begin{figure}[ht!]
    \centering
    \includegraphics[width=1\linewidth]{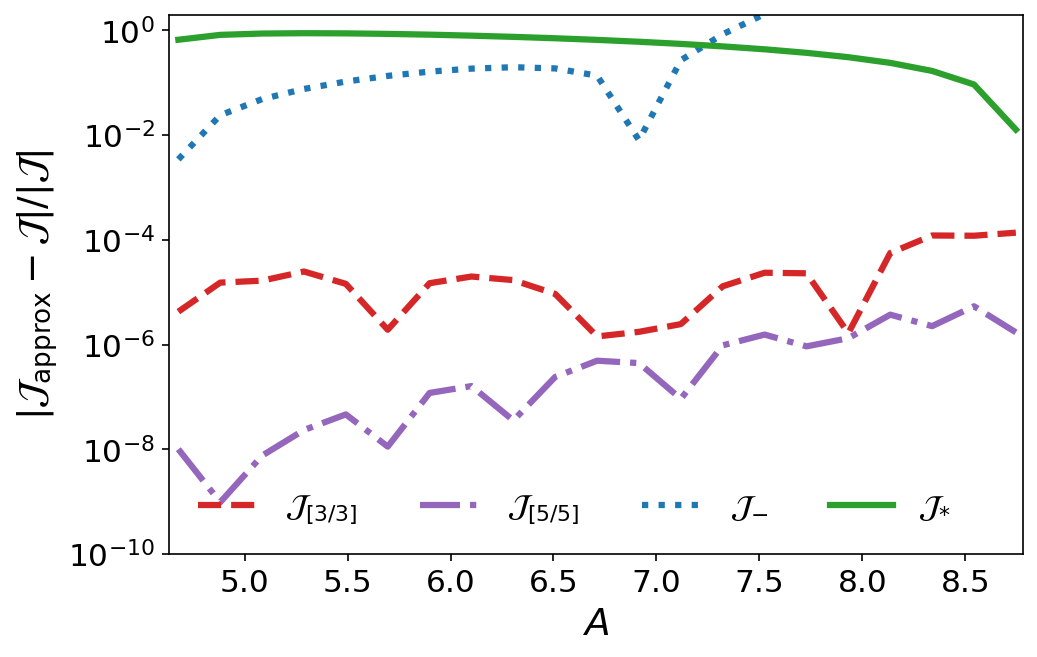}
    \caption{Comparison of relative differences between the provided explicit approximations of $\mathcal{J}$ and the results from the energy drops obtained via implicit solve. }
    \label{fig:relative_friction}
\end{figure}

For $m=n=3$, the pole-free approximation is specified by
\begin{align}
P_3(s)
={}&
4.975912864275528 \nonumber \\
&-4.641152742433373\,s
\nonumber\\
&+
2.553486904795277\,s^2 \nonumber \\
&-1.550335590883650\,s^3,
\\
Q_3(s)
={}&
1
+0.3032598453406820\,s
\nonumber\\
&-
1.888267337852114\,s^2 \nonumber \\
&+0.7651534006123918\,s^3.
\label{eq:cp_33}
\end{align}
For $m=n=5$, we obtain
\begin{align}
P_5(s)
={} & 
4.975766415704477 \nonumber \\
&-4.274194970406318\,s
\nonumber\\
&-
0.1730907043784124\,s^2 \nonumber \\
&-0.3543330035300636\,s^3
\nonumber\\
&+
1.096588408089551\,s^4 \nonumber \\
&-0.2490546748077928\,s^5,
\\
Q_5(s)
={}&
1
+0.3740733914140734\,s
\nonumber\\
&-2.297807656107382\,s^2
\nonumber\\
&-
0.05560186936526916\,s^3  \nonumber \\
&+2.004251070603650\,s^4
\nonumber\\
&-
0.8873164067638182\,s^5.
\label{eq:cp_55}
\end{align}
Both denominators are nonzero in $[0,1]$. To check the accuracy, we show a comparison between the results obtained from the energy drops and from the presented estimated in Fig.~\ref{fig:relative_friction}. In detail, we generated an independent reference sweep of $20\,001$ points uniformly spaced in $\zeta_{\rm low}^{+}\in(\zeta_0,1)$, evaluated $A=A_L(\zeta_{\rm low}^{+})$, continued the post-jump state selected by the chamber decomposition, and compared the rational expressions with Eq.~\eqref{eq:cp_implicit_friction}. The largest observed relative residuals $\epsilon_\infty^{[m/n]} = \sup_{A\in [A_-,A_*)} \lvert\mathcal{J}_{[m/n]}-\mathcal{J}\rvert/\lvert\mathcal{J}\rvert$  are approximately
\begin{equation}
\epsilon_\infty^{[3/3]}
=1.37\times10^{-4},
\qquad
\epsilon_\infty^{[5/5]}
=5.27\times10^{-6}.
\label{eq:cp_fit_errors}
\end{equation}
Note that these estimates of the residuals are dense-grid validation errors rather than rigorous uniform bounds.

\section{Discussion and Conclusion} \label{SEC:DISCUSSION}

When a rigid magnetic rotor array slides over a commensurate magnetic substrate, the system exhibits distinct dynamical behaviors and non-monotonic energy dissipation depending on the layer separation. The core phenomena can already be captured by a simplified model that considers only nearest-neighbor interactions and coarse-grains the orientations of the magnetic moments along the lines of the slider parallel to the sliding direction. In this work, we analytically derived both the emergence of these different dynamical regimes and the resulting magnetic friction for this simplified model.

We find that the parameter plane spanned by layer separation and time can be decomposed into distinct chambers within which the number, Morse type, and alignment state of the energy landscape's critical points remain invariant. These chambers are separated by bifurcation and alignment curves that can be rigorously calculated and used to predict the system's dynamical behavior. In particular, we identify line-outs within the parameter plane that correspond to the system's time evolution, along which the stability of the critical points alternates. This rigorously demonstrates the existence of the CP regime, in which the global alignment of the moments alternates during sliding.

Additionally, we derive analytical approximations for the magnetic friction for the different dynamical regimes. For asymptotically large and small layer separations, we employ a perturbation ansatz to calculate the first nontrivial terms of the magnetic friction. For the CP regime, we determine the expected dissipation by tracking the system’s perturbation-triggered jumps from unstable critical points to stable minima within the energy landscape.

Our results provide an alternative, energy-based explanation for the magnetic friction peak observed in the CP regime. When the current alignment state of the slider magnets becomes unstable, the system is forced to relax into a lower local minimum of the energy landscape. Because the system and its dynamical modes are highly overdamped, no kinetic excitations or oscillations occur; instead, the system relaxes directly to the new stable critical point (apart from the small lag produced by the sliding motion). Since this new critical point lies at a lower energy level, the energy difference between the states before and after the jump must be dissipated. This sudden energy release accounts precisely for the observed peak in magnetic friction. Notably, this interpretation establishes the CP regime as a magnetic analogue of classical stick-slip friction, where the chambers correspond to stick phases and the bifurcation curves trigger slip events.

The chamber decomposition also provides insight into the nature of the moment alignment transitions within the CP regime. Since these transitions primarily occur between saddle points and minima of the energy landscape, the system requires a slight perturbation to trigger the jump along the unstable direction. For the $N$-degree-of-freedom version of the model, this perturbation can stem from small deviations in the initial conditions of individual lines. In numerical integrations of the two-degree-of-freedom reduction, uncertainties in the numerical precision are enough to induce these jumps. For implementations of the full $N \times N$ slider, the necessary perturbations can also arise from boundary effects, transients, or experimental uncertainties in the system parameters, which represent the dominant mechanisms directly influencing the energy landscape or the system’s motion within it.

Although the studied Hamiltonian and equations of motion are motivated by the slider-substrate setup, the strong assumptions and approximations preclude a direct, one-to-one correspondence with the real physical system (see Appendix~\ref{APP:Limitations}). Nevertheless, this toy model provides a highly accessible framework that captures the same dynamical phases as those observed in experiments and simulations. In particular --- as demonstrated in this work --- the model is simple enough to be treated analytically, yielding deep insights into the underlying mechanisms driving this behavior. Furthermore, similar behavior is expected in entirely different physical contexts that can be mapped onto a related Hamiltonian. Consequently, our results may be applicable to a broader range of systems beyond magnetic friction experiments.

Note that the same chamber decomposition of the parameter plane can be applied to extensions of the simplified model that account for long-range interactions beyond nearest neighbors (see Appendix~\ref{app:robustness_two_angle}). In these cases, however, the bifurcation and alignment curves can no longer be derived entirely analytically. Nonetheless, numerical tests show that all three dynamical regimes persist even when considering the full, infinite set of pair interactions, confirming the robustness of our findings.

Interestingly, with full pair interactions, the layer separation interval for the CP regime shrinks significantly. This might indicate that in the physical system, additional perturbations play a more prominent role than in the simplified model. Fluctuations in the system parameters, open boundary effects, and transient excitations in a physical realization of the slider-substrate setup might allow the system to overcome finite energy barriers in the landscape before a critical point becomes fully unstable. Furthermore, the assumption of synchronous moment alignment along each line of the slider may artificially restrict the size of the layer separation interval that exhibits the CP regime.

Beyond explaining the macroscopic mechanics of rotor arrays, the scale-free nature of our simplified model ensures that these findings have far-reaching implications across multiple length scales. The analytical boundaries and friction scaling laws derived here can serve as a theoretical blueprint for the design of synthetic magnetic metamaterials~\cite{Aghamiri2025, Concha2018}, microfluidic actuators~\cite{Chiriac2023}, or nanoscale spintronic devices~\cite{Shukla2025, Lone2024} operating under relative motion. By showing that the complex interplay between layer separation and sliding dynamics can be precisely engineered via chamber-bifurcation pathways, our work opens new avenues for controlling dissipation and structural phases in artificial tribological systems from the macroscopic down to the atomic scale.

\begin{acknowledgments}

The authors thank M. Kienzler for generating the sketch of the setup in Fig.~\ref{fig:sket}, as well as G. Jung, T. Kiechl, J. Lopez Garcia and Y. Tang for helpful feedback and fruitful discussions.  This work was supported by the Center for Exascale Monte-Carlo Neutron Transport (CEMeNT), a PSAAP-III project funded by the Department of Energy under Grant DE-NA003967 and by the Center for Advancement of the Radiation Resilience of Electronics(CARRE) under the Department of Energy (DoE) National Nuclear Safetey Administration (NNSA) award number DE--NA0004268. Anthropic Claude Sonnet 4.5 was used to assist with writing and organizing code for numerical simulations, repeated computational runs, and figure generation. OpenAI ChatGPT 5.4 and 5.5 were used as exploratory tools during parts of the analytical derivations and to assist with polishing and condensing the manuscript text. All AI-generated code, calculations, derivation steps, and written material were independently reviewed, tested, and revised by the authors. The authors take full responsibility for the accuracy, validity, and presentation of the results.

\end{acknowledgments}

\appendix

\section{Dynamical behavior for non-generic initial conditions} \label{app:InitialCheck}

In the main text, we focus primarily on perturbed FM and AFM initial conditions, such as those defined in Eqs.~\eqref{eq:FM_init} and \eqref{eq:AFM_init}. These choices represent the two characteristic relative alignments of the rotor moments and are directly motivated by configurations observed in experiments. Here, we briefly discuss numerical tests with more general initial conditions to verify that the resulting long-time dynamics do not depend sensitively on these particular choices. We consider $N = 20$, $\gamma\tilde{v} = 0.0011$, and $A \in \{1, 4, 6, 8, 11\}$, matching the parameter sets used in the numerical investigations of Sec.~\ref{SEC:NUMERICS}.

First, we consider FM and AFM initial conditions characterized by an arbitrary overall orientation. For the FM case, we initialize the angles according to
\begin{align}
\theta_j(0)=\theta_0+\eta_j,
\end{align}
while for the AFM case we use
\begin{equation}
\theta_j(0)=
\begin{cases}
\theta_0+\eta_j, & j\ \text{odd},\\
\theta_0+\pi+\eta_j, & j\ \text{even},
\end{cases}
\end{equation}
where $\theta_0$ is drawn uniformly from $[0, 2\pi)$ and $\eta_j$ represents the small random perturbation employed throughout the main text. Following a brief initial transient relaxation, these systems exhibit numerical behavior that is indistinguishable from the corresponding reference configurations centered around $\theta_1(0) = 0$ or $\theta_1(0) = \pi$. Consequently, the specific initial conditions adopted in the main text serve as fully representative, convenient choices for more generally oriented FM and AFM states.

At small amplitudes $A$, the energy landscape permits an additional strict FM initial state. In this regime, two stable strict FM minima coexist, meaning that in addition to initializing all angles near $0$, the system can alternatively be prepared with all angles near $\pi$. This feature is consistent with the energy landscape analyzed in Sec.~\ref{SEC:LANDSCAPE}; for example, the $A = 1$ cross-section exhibits two distinct strict FM minima alongside an AFM-like minimum. However, this secondary strict FM state disappears as $A$ increases and thus plays no role in the CP or large-$A$ FM regimes.

As a less structured test, we also initialize every angle independently and uniformly on $[0, 2\pi)$. Following initial fluctuations, the rotor angles dynamically segregate into two distinct groups that are numerically indistinguishable from the two-angle reduction employed in our analytical treatment. However, strictly speaking, a generic solution cannot collapse onto the exact two-angle configuration in finite time due to the uniqueness of solutions to Eq.~\eqref{eq:EOMFULL}; instead, the angles approach this limiting structure to within numerical precision. Occasionally, one or two rotors remain detached from the two dominant sublattices. Such minor deviations have a negligible effect on the bulk dynamics and are comparable to departures from ideal two-angle behavior induced by open boundary conditions in $ y $-directions, as discussed in Appendix~\ref{app:robustness_two_angle}.

Finally, periodic boundary conditions may permit specially structured states for specific values of $N$. For instance, the structure of Eq.~\eqref{eq:EOMFULL} suggests that repeating configurations commensurate with divisors of $N$ could, in principle, exhibit dynamics not captured by the two-angle reduction. While we have not systematically constructed or tested such states, leaving open the possibility of specialized behavior for carefully engineered initial conditions, our random initialization tests provide no indication that these configurations are reached generically. Consequently, they are unlikely to play a significant role in physical realizations of the rotor array, where perfectly structured initial states and exact periodicity are absent. A comprehensive classification of all such states for arbitrary $N$ lies beyond the scope of this work.

\section{Limitations of the simplified model} \label{APP:Limitations}

To emphasize the limits of applicability of our simplified model with respect to a physical realization of the slider-substrate system, we detail the primary approximations used in our derivations and compare them with the physical parameters of macroscopic rotor-array experiments. Although the model is intended as a minimal representation to describe collective dynamics, factors such as macroscopic friction laws, higher-order multipolar interactions, and finite-size effects introduce quantitative shifts in the energy landscape. While these factors influence the specific boundaries of the dynamical regimes, they are not expected to alter the qualitative structure of the energy landscape or the underlying mechanisms governing the system's dynamic behavior.

A key difference between the macroscopic rotor-array experiments and the simplified model lies in the treatment of shaft-friction dissipation. The model uses a Stokes-like drag term to represent shaft friction, whereas a macroscopic implementation such as Ref.~\cite{Gu2026} might be better described by a Coulomb-type friction law with a normal load that varies with layer separation (due to changes in the magnetic forces pressing the rotors against their shafts). Consequently, the simplified model’s dependence on sliding velocity is expected to deviate from that of a macroscopic implementation. However, employing a microscopic dissipation law proportional to the local velocity of the moments is common in minimal tribological models, including the well-established \textit{Prandtl-Tomlinson model}~\cite{Popov2012} or the Frenkel-Kontorova model~\cite{Braun1999,Floria1996}. Moreover, while the Stokes-type term is not an ideal representation of dissipation in the macroscopic rotor array, it is consistent with rotational friction experienced by microscopic magnetic moments in a viscous fluid at the colloidal scale. Also, a similar microscopic friction law is frequently used in other theoretical model systems studying magnetic friction~\cite{Fusco2008, Magiera2009, Magiera2011, Magiera2014, Magiera2013, Magiera2011-2}. Consequently, our system can also be interpreted as a model for such a realization. 

Compared to the true magnetic field generated by the substrate, which is a square lattice of fixed magnets, the rotating effective field used in the simplified model is a strong idealization. While the substrate interaction should indeed be periodic in time and orientations of the slider moments, additional contributions are expected when modeling the coupling between the two layers in greater detail. In this sense, the substrate term employed in the simplified model resembles more closely an externally applied rotating magnetic field or an approximation based on neglecting higher order harmonics. Nevertheless, because the dominant features of the dynamics arise from the periodic nature of the drive, this simplified representation of the substrate as a periodic external field is sufficient for the scope of our work.

A further minor limitation arises from the use of pure dipole-dipole interactions to represent the millimeter-sized magnets in rotor-array experiments. Because real magnets have macroscopic dimensions and a distinct geometry, higher-order multipolar contributions to the magnetic interaction are expected. However, as already discussed in Ref.~\cite{Gu2026}, these higher-order terms can be neglected within a good approximation --- at least for large $ P $ and small $ A $.

Note that some aspects presented in the main text rely on periodic boundary conditions or bulk assumptions. For instance, the coefficients of the reduced energy functions as presented here and the used definitions of the FM and AFM states are only globally well-defined under even $ N $ and periodic boundaries. Compared with experimental realizations, an application of these simplifications corresponds to assuming that boundary effects do not strongly influence the bulk behavior. However, such effects (in particular, the open boundaries of the experiments and the simulations of Ref.~\cite{Gu2026}) may in fact play an essential role in the observed dynamics of the macroscopic rotor array because they clearly perturb the system. On the other hand, we find that our results discussed in this work are robust against opening the boundary in the $y$-direction, and that applying periodic boundary conditions in the $y$-direction serves primarily as a modeling convenience (see Appendix~\ref{app:robustness_two_angle}).

We anticipate that the most significant simplification in our model is the assumption that each rotor interacts only with its nearest neighbors. Because dipolar interactions decay as $1/r^3$, magnetic moments beyond the nearest neighbors are expected to still influence the dynamics of the macroscopic rotor array. To roughly estimate the error introduced by neglecting these longer-range interactions, we can compare the nearest-neighbor interaction result with the total interaction energy of a single magnetic moment in a two-dimensional square lattice for the two stable equilibrium configurations (in-plane FM and AFM) without the substrate.

\begin{figure}[ht!]
\includegraphics[width=0.45\textwidth]{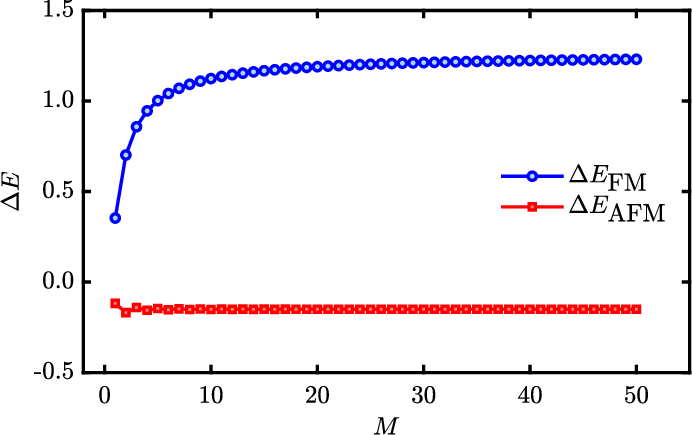}
\caption{Relative deviation between the total interaction energy of one moment with all other moments inside $M$ concentric square layers surrounding it and the nearest neighbor approximation in dependence on $ M $. The blue curve corresponds to a ferromagnetic arrangement of the slider moments and the red curve corresponds to an antiferromagnetic arrangement of the slider moments. The energy terms are calculated in the absence of a substrate.}
\label{fig:NN}
\end{figure}

In a square lattice with spacing $P$, each moment has a nearest-neighbor coupling energy
\begin{align}
    E_{\textrm{FM}} &= -2\tilde{\varepsilon}, \\
    E_{\textrm{AFM}} &= -6\tilde{\varepsilon}
\end{align}
for the in-plane FM state ($E_{\textrm{FM}}$) and the AFM state ($E_{\textrm{AFM}}$), where $\tilde{\varepsilon} = \mu_0 m_0^2 / (4\pi P^3) = \varepsilon / N$. To estimate the influence of the longer-range interactions, we compute the total interaction energy of one moment with all other moments inside $M$ concentric square layers surrounding it. The corresponding energies for the FM and AFM states are
\begin{align}
    E^{(M)}_{\textrm{FM}} &= \tilde{\varepsilon} 
    \sum\limits_{\substack{k,n=-M \\ (k,n)\neq (0,0)}}^{M}
    \frac{k^2 - 2n^2}{(n^2 + k^2)^{5/2}}, \\
    E^{(M)}_{\textrm{AFM}} &= \tilde{\varepsilon} 
    \sum\limits_{\substack{k,n=-M \\ (k,n)\neq (0,0)}}^{M}
    (-1)^k \frac{k^2 - 2n^2}{(n^2 + k^2)^{5/2}}.
\end{align}
We then evaluate the relative deviations
\begin{align}
    \Delta E_{\textrm{FM}} &= \frac{E^{(M)}_{\textrm{FM}} - E_{\textrm{FM}}}{E_{\textrm{FM}}}, \\
    \Delta E_{\textrm{AFM}} &= \frac{E^{(M)}_{\textrm{AFM}} - E_{\textrm{AFM}}}{E_{\textrm{AFM}}}
\end{align}
and take the limit $M \to \infty$. This yields $\Delta E_{\textrm{FM}} = 1.25$ and $\Delta E_{\textrm{AFM}} = -0.15$. The dependences of $\Delta E_{\textrm{FM}}$ and $\Delta E_{\textrm{AFM}}$ for finite $M$ are shown in Fig.~\ref{fig:NN}, highlighting that the limiting values are reached rather quickly for small $M$. These results show that the simplified model overestimates the interaction energy of the AFM state and significantly underestimates the interaction energy of the FM state (by more than a factor of two).

Although the difference between the nearest-neighbor approximation and the full pair-interaction energy appears large, the overall structure of the energy function for arbitrary $M$ remains similar to Eq.~\eqref{eq:FULLH} when restricting the system to two degrees of freedom  $\theta_e$ and $\theta_o$. The only changes are additional terms proportional to $\cos(\theta_e + \theta_o)$ and modified coefficients that arise when taking more interaction partners into account (further details are provided in Appendix~\ref{app:robustness_two_angle}).

Because the energy function retains a similar form for different values of $M$, the analysis presented in this paper can be directly extended to systems that include more interaction pairs. Numerical tests show that varying $M$ mainly modifies the layer separations intervals for which the three dynamical regimes occur. Even in the limits $N \to \infty$ and $M \to \infty$, suitable values of $A$ and $\gamma \tilde{v}$ can still be found that reproduce the characteristic dynamics of each regime (although the parameter range for the CP regime becomes notably narrow). However, the dissipative behavior becomes distinctly more complex. Longer-range interactions preserve the energy-landscape mechanism, but change the branch energies and bifurcation locations independently; consequently, the jump-induced energy losses and the detailed magnetic-friction curve must be recomputed for each interaction range. See Appendix~\ref{app:robustness_two_angle}; a detailed analysis of the magnetic friction for $ N \to \infty $ and $ M \to \infty $ is outside the scope of this work~\footnote{For $ N \to \infty $ and $ M \to \infty $, the magnetic friction diverges by definition. Here, only the magnetic friction per slider moment is a reasonable observable.}. 

Finally, our assumption that all magnetic moments within a sublattice remain always parallel during the time evolution is also a strong simplification. In general, this approximation is reasonable in the quasi-static regime, where the system is effectively constrained to either the AFM or FM state (or transitions only between two critical points on negligible time scales). However, for sufficiently large rotor arrays, this assumption may break down. In such systems, small local flips of individual magnetic moments could generate excitations that propagate through the array and contribute to magnetic friction. We expect these excitations to be transient, arising primarily during transitions between the two different ordered states, starting, for example, from open boundaries. Such effects should not influence the structure of the general energy landscape, which is the main focus of our analysis.

It is important to note that all the limitations discussed above arise from a direct comparison with macroscopic experimental setups such as that of Ref.~\cite{Gu2026}. They should therefore be understood only as shortcomings in modeling that particular system with the proposed equations of motion. This does not preclude the existence of other physical systems with coupled degrees of freedom that are accurately described by our dynamical framework. The results presented in this work apply to any system that can be energetically mapped onto the simplified model. Note also that the equations of motion of the simplified model~\eqref{eq:gradient_flow} are sufficient to reproduce the three dynamical regimes observed in experiments and many-body simulations and are therefore highly likely to be sufficient to understand the mechanisms of the observed behavior.

\section{Dissipation and sliding friction in periodic systems} \label{APP:MF}

In Ref.~\cite{Gu2026}, it was derived that magnetic friction can be defined by Eq.~\eqref{eq:MAGFRIC} whenever the system is periodic with the external drive. In particular, this implies that the dissipation is determined solely by the torque exerted by the substrate on the slider moments and by the angular velocities of those moments, independent of the specific model used to describe shaft friction (which together describe the work done on the slider by the substrate). In the following, we demonstrate that this result holds explicitly for our simplified model and derive equivalent definitions of the dissipation.

The equations of motion of the simplified model assume a Stokes-type shaft friction of the form $ -\Gamma \dot{\theta}_j $, implying that the dissipative power of each rotor is given by $ -\Gamma \dot{\theta}_j^2 $. The total energy dissipated over one period of the external drive is therefore
\begin{equation}
    E_{\textrm{diss}} = -\Gamma \sum_j \int_0^{2\pi/\tilde{v}} 
    \dot{\theta}_j^{\,2}(t)\, \mathrm{d}t .
\end{equation}
The corresponding magnetic friction is then given by 
$\langle F_x^{\textrm{mag}} \rangle = E_{\textrm{diss}} / P$. Here, we assume that the system has already reached its time-periodic steady state at $ t = 0 $.

We can show that this natural definition of friction is equivalent to Eq.~\eqref{eq:MAGFRIC} for our systems by multiplying the equations of motion~\eqref{eq:EOMFULL} by $\dot{\theta}_j$, integrating over time, and summing over all moments. This yields
\begin{align}
    &-\frac{E_{\textrm{diss}}}{\varepsilon} = -3 \sum_j
    \int_0^{2\pi/\tilde{v}} \sin(2\theta_j)\,\dot{\theta}_j\,\mathrm{d}t \nonumber\\
    &\quad + \sum_j \int_0^{2\pi/\tilde{v}} 
    \left[ \sin(\theta_j-\theta_{j+1})\,\dot{\theta}_j 
    + \sin(\theta_j-\theta_{j-1})\,\dot{\theta}_j \right] \mathrm{d}t \nonumber\\
    &\quad + A\sum_j \int_0^{2\pi/\tilde{v}} 
    \sin(\theta_j-\tilde{v} t) \, \dot{\theta}_j\,\mathrm{d}t .
    \label{eq:EDISS}
\end{align}
The first term simplifies because
\begin{equation}
    \sin(2\theta_j)\,\dot{\theta}_j 
    = -\frac{\mathrm{d}}{\mathrm{d}t} \frac{1}{2}\cos(2\theta_j).
\end{equation}
For the coupling terms, we shift the summation index and apply periodic boundary conditions to obtain
\begin{align}
    &\sum_j \left[ \sin(\theta_j-\theta_{j+1})\,\dot{\theta}_j 
    + \sin(\theta_j-\theta_{j-1})\,\dot{\theta}_j \right] \nonumber\\
    =& \sum_j \left[ \sin(\theta_j-\theta_{j+1})\,\dot{\theta}_j 
    - \sin(\theta_j-\theta_{j+1})\,\dot{\theta}_{j+1} \right] \nonumber\\
    =& -\frac{\mathrm{d}}{\mathrm{d}t} \sum_j \cos(\theta_j-\theta_{j+1}).
\end{align}
Thus, the first two terms on the right-hand side of Eq.~\eqref{eq:EDISS} reduce to evaluations at the boundary of the integrals. Because the system is periodic with period $2\pi/\tilde{v}$, these boundary contributions vanish. We therefore obtain
\begin{equation}
    E_{\textrm{diss}} = -\varepsilon A \sum_j 
    \int_0^{2\pi/\tilde{v}} \sin(\theta_j-\tilde{v} t)\, \dot{\theta}_j \, \mathrm{d}t,
\end{equation}
which directly yields Eq.~\eqref{eq:MAGFRIC}.

\begin{table*}[ht]
\centering
\caption{Numeric comparison of the different integrals contained in the formulas for dissipation. The integrals are evaluated for various values of $A\in\{1,4,6,8,11\}$ and different initial conditions (IC) (FM: $0=\theta_j=\theta_{j+1}$; AFM: $0 = \theta_j=\theta_{j+1}+\pi$, each perturbed by a uniform random number $\eta_j\in[-0.01,0.01]$. The comparison is done for systems with $N=20 $ and $ \gamma \tilde{v}= 0.0011 $. Each integral is evaluated over a period $[\tau,\tau+2\pi/\tilde{v}]$ and averaged over multiple starting points $\tau \in [4\pi/\tilde{v},6\pi/\tilde{v}]$, i.e. we report $\frac{\tilde{v}}{2\pi} \int_{4\pi/\tilde{v}}^{6\pi/\tilde{v}} \int_\tau^{\tau+2\pi/\tilde{v}}(\cdot)\,\textrm{d}t\,\textrm{d}\tau$.}

\label{TAB:EDISS}
\begin{tabular}{c | c | c | c | c}
\toprule
$\quad A\quad$ &$\quad$ IC $\quad$ &
$ $\quad$\displaystyle A\int \sum_j \sin(\theta_j-\tilde{v}t)\,\dot{\theta}_j\,dt$ $\quad$ &
$ $\quad$ \displaystyle \gamma \int \sum_j \dot{\theta}_j^{\,2}\,dt$ $\quad$ &
$ $ \quad $ \displaystyle A\int \sum_j \sin(\theta_j-\tilde{v}t)\,dt$ $ \quad $ \\
\midrule
1  & FM  & $2.018947\times 10^{-3}$ & $2.018948\times 10^{-3}$ & $2.018948\times 10^{-3}$ \\
1  & AFM & $7.184587\times 10^{-4}$ & $7.184589\times 10^{-4}$ & $7.184589\times 10^{-4}$ \\
\addlinespace
4  & FM  & $2.296053\times 10^{-2}$ & $2.296052\times 10^{-2}$ & $2.296052\times 10^{-2}$ \\
4  & AFM & $2.296052\times 10^{-2}$ & $2.296052\times 10^{-2}$ & $2.296052\times 10^{-2}$ \\
\addlinespace
6  & FM  & $3.398356\times 10^{1}$ & $3.315224\times 10^{1}$ & $3.407353\times 10^{1}$ \\
6  & AFM & $3.411337\times 10^{1}$ & $3.414253\times 10^{1}$ & $3.410911\times 10^{1}$ \\
\addlinespace
8  & FM  & $5.743078$ & $5.779477$ & $5.728856$ \\
8  & AFM & $5.711672$ & $5.759762$ & $5.697104$ \\
\addlinespace
11 & FM  & $1.654722\times 10^{-1}$ & $1.654868\times 10^{-1}$ & $1.654868\times 10^{-1}$ \\
11 & AFM & $1.654784\times 10^{-1}$ & $1.654868\times 10^{-1}$ & $1.654868\times 10^{-1}$ \\
\bottomrule
\end{tabular}
\end{table*}

To numerically confirm this result, we evaluated both definitions of the dissipated energy (the one based on the Stokes-law and the one based on the work acted on the slider) and compared them directly. The corresponding values for different $A$ are listed in Table~\ref{TAB:EDISS}. We find that the two definitions of dissipation agree within expected numerical deviations, demonstrating that the identity holds for our system.

Note that the derived identity also extends beyond periodic boundary conditions and is already a result of the fact that $ \mathcal{H}_I $ is a potential for the corresponding torques that enter the equations of motion. For instance, using open boundary conditions in $ y $-direction, the nearest-neighbor interaction term still satisfies
\begin{align}
&\sum_{j=1}^{N}
(1-\delta_{jN}) \sin(\theta_j-\theta_{j+1})
\nonumber \\+& \sum_{j=1}^{N}
(1-\delta_{j1}) \sin(\theta_j-\theta_{j-1})
\dot\theta_j \nonumber \\
=&-\frac{\textrm{d}}{\textrm{d}t}\sum_{j=1}^{N-1}\cos(\theta_j-\theta_{j+1}),
\label{eq:open_boundary_final}
\end{align}
More generally, for any finite-range gradient system the internal interaction contribution is the total time derivative of the corresponding internal energy. Therefore, the identity relating the dissipated energy to the work done by the substrate continues to hold for time-periodic steady states. What changes under open boundaries in $ y $-direction is the exact dynamical closure of the two-angle subspace, not the underlying energy-balance relation.

Finally, there is a third identity for the dissipated energy, which is especially useful when evaluating the magnetic friction numerically. This identity is also used to estimate the magnetic friction in the CP regime by tracking the jumps between instable and stable critical points. Calculating the time derivative of the full energy function, we obtain
\begin{equation}
    \frac{\textrm{d}}{\textrm{d}t} \mathcal{H} = \sum_j \frac{\partial \mathcal{H}}{\partial \theta_j} \dot{\theta} + \frac{\partial \mathcal{H}}{\partial t}.
\end{equation}
Inserting the equations of motion~\eqref{eq:gradient_flow} yields
\begin{equation}
    \frac{\textrm{d}}{\textrm{d}t} \mathcal{H} = - \sum_j \Gamma \dot{\theta}_j^2 + \frac{\partial \mathcal{H}}{\partial t}.
\end{equation}
Because $ \theta_j $ is periodic with the same period $ \mathcal{H} $ is explicitly periodic in time, we know that $ \langle \textrm{d} \mathcal{H} / \textrm{d}t\rangle = 0 $ when averaging with respect to one substrate period during the steady state (even for a setup that is not periodic in $ \theta_j $, if the system is in a stable steady state, $ \langle \textrm{d} \mathcal{H} / \textrm{d}t\rangle $ must vanish). Hence, we find
\begin{equation}
     -\Gamma \sum_j \int_0^{2\pi/\tilde{v}} 
    \dot{\theta}_j^{\,2}(t)\, \mathrm{d}t  = -\int_0^{2\pi/\tilde{v}} \frac{\partial \mathcal{H}}{\partial t} \, \textrm{d}t.
\end{equation}
Using that 
\begin{equation}
    \frac{\partial \mathcal{H}}{\partial t} = \varepsilon A \tilde{v} \sum_j \sin(\theta_j-\tilde{v} t), 
\end{equation}
the dissipated energy is therefore
\begin{align}
    E_{\textrm{diss}} &= -\int_0^{2\pi/\tilde{v}}  \frac{\partial \mathcal{H}}{\partial t}\, \textrm{d}t. \nonumber \\
    &-\varepsilon A \tilde{v} \sum_j \int_0^{2\pi/\tilde{v}}  \sin(\theta_j-\tilde{v} t) \, \textrm{d}t.
    \label{eq:MAGNETICFRICTIONROBUST}
\end{align}
Strikingly, this identity for the friction only depends on $ \theta_j $ and not on the corresponding angular velocities. This is the reason why this relation is especially robust for numerical evaluation in situations where the angular velocities spike due to a jump from one critical point to another (for instance, in the CP regime). Note that Eq.~\eqref{eq:MAGNETICFRICTIONROBUST} is equivalent to the definition for the friction utilized in other established magnetic friction literature, where similar derivations have been performed~\cite{Fusco2008, Magiera2009, Magiera2011, Magiera2014, Magiera2013, Magiera2011-2}. To check this third identity for the dissipation, we compare its results for numerically solved systems to the other two definitions. The results are again shown in Tab.~\ref{TAB:EDISS}, confirming that all three formulas reproduce the same results. 

\section{Lag between the system and the energy landscape in the FM regime} \label{APP:Lag}

To describe the system deep within the FM regime, we can assume that $ A \rightarrow \infty $. Neglecting the first three terms on the right-hand side of Eq.~\eqref{eq:EOMFULL} fully decouples the equations of motion, yielding the simplified differential equation
\begin{equation}
\gamma \dot{\theta}_j \approx A \sin(\theta_j - \tilde{v}t).
\end{equation}
For the steady-state solution, we use the ansatz
\begin{equation}
\theta_j(t) = \pi + \tilde{v}t + c, \label{eq:DEEPFM}
\end{equation}
where $c$ (with $\vert{}c\vert{} \ll 1$) represents a small constant lag between the magnetic moment and the minimum of the energy landscape at $\pi + \tilde{v}t$. Substituting this ansatz into the simplified equation of motion yields
\begin{equation}
c = \textrm{arcsin}\left[-\frac{\gamma \tilde{v}}{A}\right] \approx -\frac{\gamma \tilde{v}}{A}.
\end{equation}
This lag is essential for the result to satisfy the equation of motion in the quasi-static regime. It is consistent with the estimates of Ref.~\cite{Gu2026}. For a discussion of such a lag in a system of Heisenberg spins, see Ref.~\cite{Magiera2009}.

We can also easily derive the magnetic friction corresponding to the dynamics of the form of Eq.~\eqref{eq:DEEPFM}. The energy dissipated per period of sliding is given by (see Appendix~\ref{APP:MF})
\begin{align}
    E_{\textrm{diss}} &= - \Gamma N \int_0^{2\pi/\tilde{v}} \dot{\theta}_j^2(t) \, \textrm{d}t = -2\pi N \Gamma \tilde{v}.
\end{align}
This results into the magnetic friction $ {\langle F_x^{\textrm{mag}} \rangle = -2\pi N \gamma \tilde{v} \, \varepsilon / P} $, which is exactly the zeroth order contribution (with respect to $ A $) of the magnetic friction in the FM regime (see Sec.~\ref{SEC:FRICTION}).

\section{Strict FM states and symmetry breaking}
\label{APP:Protected}

All discussions in the main text assume that the observed trajectories follow stable critical points of the energy landscape. An exception occurs when the system is initialized in an exactly strict FM state given by $\theta_{j+1}(0)=\theta_j(0)$ modulo $2\pi$. For such initial conditions, the simplified model admits trajectories that never leave a strict FM alignment.

\begin{figure*}
    \centering
    \includegraphics[width=\linewidth]{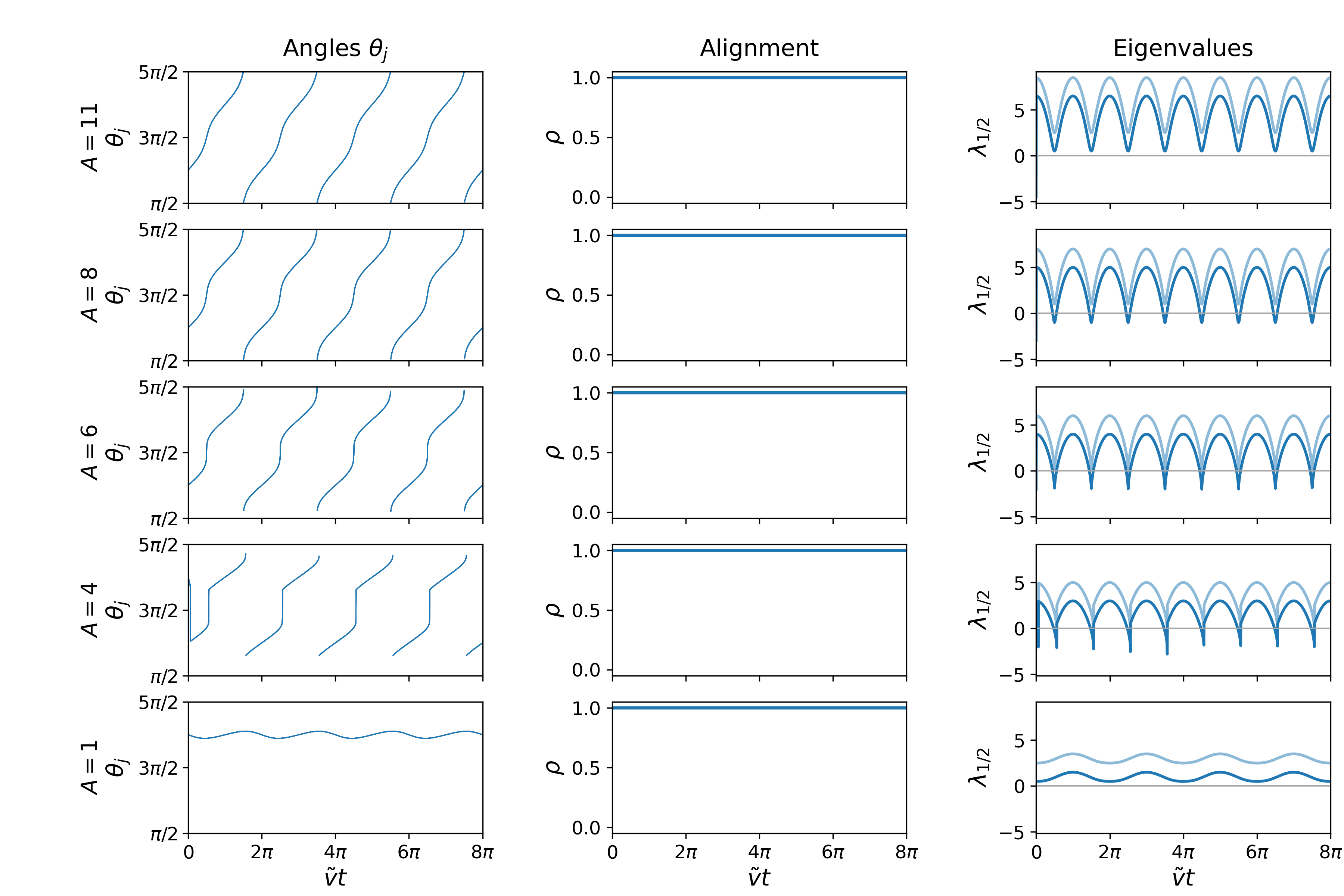}
    \caption{Full-system numerical integration initialized in a perfect strict FM state. The mean alignment remains $\bar\rho=1$, although the Hessian eigenvalues show that the state is not always stable. Parameters were set to $N=20$ and $\gamma \tilde{v}=0.0011 $ and initial condition were $\theta_j=0$ for all $j$ ($ \tilde{v} $ is set to unity for the numerical integration). No symmetry-breaking noise was added to the initial state.}
    \label{fig:strict_FM_simulation}
\end{figure*}

Initializing all moments with the same angle, all rotors experience the same torque at all times $ t $. Their relative angles remain zero, so the system stays perfectly aligned and
\begin{equation}
\bar\rho(t)=1
\end{equation}
for the entire time-frame of observation. The dynamics reduces to a single equation of motion,
\begin{equation}
\gamma\dot{\theta}
=
-3\sin(2\theta)
+
A\sin(\theta-\tilde vt).
\end{equation}
Numerical integrations of this relation as well as the corresponding alignment parameter $ \rho $ are depicted in Fig.~\ref{fig:strict_FM_simulation} for $N=20$, $\gamma\tilde{v}=0.0011 $ and initial condition $\theta_j=0$ for all $j$. In contrast to the main text, no symmetry-breaking noise was added to the initial state.

Note that the results of the numerical integration do not imply that the strict FM state is always stable. As also shown in Fig.~\ref{fig:strict_FM_simulation}, one of the Hessian eigenvalues can become negative while $\bar\rho$ remains equal to one. The negative eigenvalue indicates instability to perturbations that break the exact FM symmetry. However, if no such perturbation is present in an idealized system such as our simplified model, the unstable mode is never excited. These trajectories are therefore protected by the exact symmetry of the initial condition rather than by energetic stability.

\begin{figure}
    \centering
    \includegraphics[width=\linewidth]{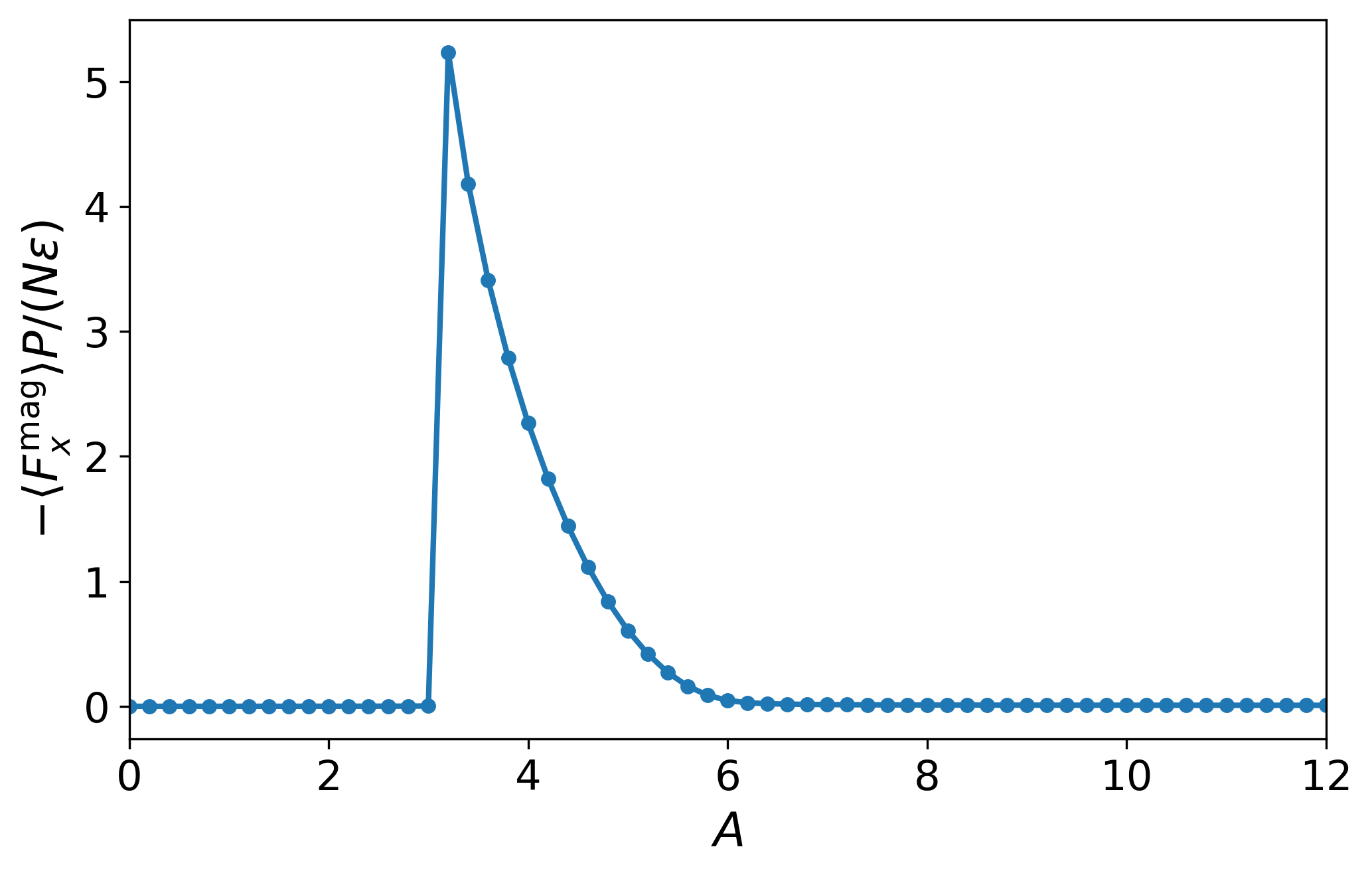}
    \caption{Magnetic friction obtained from perfectly symmetric strict FM initial conditions. The data is obtained through evaluating the dissipation of numerical integration of the equations of motion using $N=20$, and $\gamma \tilde{v}=0.0011 $, and initial condition were $\theta_j=0$ for all $j$ ($\tilde{v}$ is set to unity for the numerical integration). No symmetry-breaking random numbers was added to the initial state.}
    \label{fig:strict_FM_friction}
\end{figure}

Strikingly, even this symmetry-protected special case retains a non-monotonic friction response as a function of $A$. Figure~\ref{fig:strict_FM_friction} shows the sliding friction obtained when the dynamics remain confined to the strict FM set. The resulting peak is shifted to smaller $A$ compared with the generic behavior discussed in the main text.

The observed dynamical behavior and dissipation mechanism follows directly from the chamber decomposition in Fig.~\ref{fig:BuL}~a). Perfect strict FM symmetry constrains the system to $\theta_e=\theta_o$, so only the tangential Hessian eigenvalue $\lambda_\parallel$ determines whether the followed state is stable under the allowed motion. The strict FM saddles shown in Fig.~\ref{fig:BuL}~a) are unstable only in the transverse direction and therefore remain stable under the enforced symmetry.

Crossing the transverse FM curve changes the stability against symmetry-breaking perturbations, but it does not affect the exactly symmetric trajectory. Only the tangential FM curve changes the relevant eigenvalue $\lambda_\parallel$. Crossing this curve can therefore force the trajectory to leave its current strict FM branch and rapidly move to another strict FM state. The common rotor angle changes during such a jump, while the alignment remains $\bar\rho=1$.

The range in which these jumps are possible follows from the minimum and maximum values of $A$ along the tangential FM curve. From Eqs.~\eqref{eq:fm_tangential_curve_sin} and \eqref{eq:fm_tangential_curve_cos},
\begin{equation}
A^2
=
9\sin^2(2\chi_+)
+
36\cos^2(2\chi_+),
\end{equation}
and hence
\begin{equation}
3\leq A\leq6.
\end{equation}
This agrees with the interval in which the observed friction peak in Fig.~\ref{fig:strict_FM_friction} emerges. For $A<3$, the tangential stability of the followed critical point is not lost. For $A>6$, the system follows a single strict FM critical point continuously through the driving cycle. In either case, no symmetry-constrained jump occurs.

The strict FM set is an invariant set of the deterministic equations of motion. By uniqueness, a trajectory that is exactly strict FM at one time must remain strict FM for its entire evolution. It therefore cannot leave this set spontaneously, even when the corresponding state is unstable to symmetry-breaking perturbations. The converse also follows from uniqueness. If the symmetry is broken initially, the trajectory cannot reach an exactly strict FM state at any finite time. Otherwise, evolving the solution backward from that time would imply that it had always been strict FM. A strict FM state may nevertheless be attractive: nearby trajectories can approach it arbitrarily closely while it is transversely stable, without ever reaching exact equality of all rotor angles.

The symmetry-protected solutions are therefore mathematically valid but nongeneric. In the numerical integrations discussed in the main text, small perturbations are added to the initial angles so that the trajectory is not confined to the strict FM set. Even perturbations at the level of numerical precision are sufficient. They remain small while the strict FM state is transversely stable and grow when that state becomes unstable, producing the typical CP trajectories described in the main text.

\section{Discussion on why degenerate critical point mark the chamber boundaries} \label{APP:Implicite}

To understand why degenerate critical points form the boundaries of the chambers, one can analyze how nondegenerate critical points behaves as the parameters $(A,\tilde{v}t)$ are varied: Suppose $\vec{p}_0$ is a nondegenerate critical point at $(A_0,\tilde{v}t_0)$. Then $F(\vec{p}_0;A_0,\tilde{v}t_0)=0$, and the Hessian $ D_pF(\vec{p}_0;A_0,\tilde{v}t_0)=D_p^2\Hr(\vec{p}_0;A_0,\tilde{v}t_0)$ is invertible. By the \textit{implicit function theorem}, for $(A,\tilde{v}t)$ sufficiently close to $(A_0,\tilde{v}t_0)$, there exists a unique function $(A,\tilde{v}t) \mapsto \vec{p}(A,\tilde{v}t)$ satisfying
\begin{align}
    F(\vec{p}(A,\tilde{v}t);A,\tilde{v}t)&=0,\\
    \vec{p}(A_0,\tilde{v}t_0)&=\vec{p}_0.
\end{align}
Physically, $\vec{p}(A,\tilde{v}t)$ corresponds to the same critical point of the energy landscape, smoothly continued as the parameters vary. This becomes more clear while keeping in mind that each nondegenerate critical point of the energy function is isolated (see also Appendix~\ref{APP:Number}). Consequently, a neighborhood of $(A_0,\tilde{v}t_0)$ lies within the branch of the minimum described by $\vec{p}(A,\tilde{v}t)$, and the number of nondegenerate critical points cannot change at $(A_0,\tilde{v}t_0)$~\footnote{The implicit-function argument holds for each nondegenerate critical point. One can use the intersect of the resulting neighborhoods as an environment around $ (A_0, \tilde{v}t_0) $ where the number of degenerate critical points cannot change.}. Moreover, the eigenvalues of the Hessian vary smoothly with $\vec{p}$, $A$, and $\tilde{v}t$ as well. Since the signs of the Hessian eigenvalues fix the Morse type, sufficiently small changes in the parameters cannot alter it and there is a neighborhood of $ (A_0,\tilde{v}t_0) $ where the Morse type of $\vec{p}(A,\tilde{v}t)$ does not change. Thus, $(A_0,\tilde{v}t_0)$ cannot lie on a chamber boundary if all critical points that exists for this parameter set are nondegenerate.

The requirement that a degenerate critical point must exist at $(A_0,\tilde{v}t_0)$ for the total number of critical points to change can also be understood more intuitively. Imagine two critical points, $\vec{p}_1(A,\tilde{v}t)$ and $\vec{p}_2(A,\tilde{v}t)$, as distinct states on the $ (\theta_e, \theta_o) $-plane (i.e., \textit{the angle space}). As $(A,\tilde{v}t)$ varies and these points merge at $(A_0,\tilde{v}t_0)$ in the angle space, the energy landscape becomes flat in the direction connecting them; at the moment of collision, moving infinitesimally along this direction produces no restoring force to linear order. More formally, since $F(\vec{p}_1(A,\tilde{v}t);A,\tilde{v}t)=F(\vec{p}_2(A,\tilde{v}t);A,\tilde{v}t)=0$ for two nearby critical points, the directional derivative of $F$ along the vector of the angle space connecting them must vanish in the limit $ (A,\tilde{v}t) \to (A_0,\tilde{v}t_0) $ where they merge. Because $D_pF = D_p^2\Hr$, this implies that the Hessian develops a zero eigenvalue in the collision direction.

Although implicit in the arguments above, we emphasize that both the number and the Morse type of the critical points within a chamber are fixed, since all critical points away from the set defined by the bifurcation curves are nondegenerate. Indeed, each nondegenerate critical point can be continued uniquely under a small parameter change by the implicit-function argument above, and its Morse type remains unchanged because no Hessian eigenvalue crosses zero. This local statement extends to an entire chamber as follows: fix a connected component of the complement of the bifurcation set in the $(A,\tilde{v}t)$-plane. Any two points in this component can be joined by a path that does not intersect the bifurcation curves. Cover this path by small neighborhoods on which the implicit-function theorem applies. By concatenating the resulting local continuations, each critical point at one end of the path is matched with a unique critical point at the other end. Since the path never crosses a degenerate critical point, no critical point can change Morse type, merge with another critical point, or disappear along the way.

\section{Number of critical points in each chamber} \label{APP:Number}

That each chamber contains a finite number of critical points can be seen as follows: for fixed $(A,\tilde{v}t)$, the critical points form the zero set
\begin{equation}
\mathcal{C}_{A,\tilde{v}t}
=
\{p\in\mathbb{T}^2 : \nabla_p\Hr(p;A,\tilde{v}t)=0\}.
\label{eq:critical_set}
\end{equation}
This set is closed because $\nabla_p\Hr$ is continuous. Since the angle space $\mathbb{T}^2$ is a compact set, $\mathcal{C}_{A,\tilde{v}t}$ is compact as well. If all critical points are nondegenerate, each of them is isolated, meaning that each represents a distinct local state of the frozen energy landscape rather than belonging to a continuum of critical points. Here we use the standard property of compactness that every infinite subset of a compact metric space has an accumulation point, according to the Bolzano-Weierstrass theorem.  A compact set consisting only of isolated points is finite. Consequently, on each connected component of the complement of the bifurcation set, the numbers of minima, saddles, and maxima remain finite and constant.

\section{Tracking the change of alignment state when passing the alignment curve} \label{APP:Alignment}

To determine which type of critical point changes its alignment state when crossing $L_{1/2}$, we have to examine the Morse type of the critical point on the curve itself. For this purpose, we use the description in the variables $(\chi_+,\chi_-)$. At $\chi_-=\pi/4$ and along the corresponding curve defined by the stationarity equations \eqref{eq:align_1} and \eqref{eq:align_2}, the Hessian takes the form
\begin{align}
H_{1/2}(t)
=
\begin{pmatrix}
4-6\cos(2\tilde{v}t) & -12\sin(2\tilde{v}t)\\
-12\sin(2\tilde{v}t) & 4-6\cos(2\tilde{v}t)
\end{pmatrix}.
\end{align}
Its eigenvalues are therefore
\begin{align}
\mu_\pm(t)
=
4-6\cos(2\tilde{v}t)\pm 12\sin(2\tilde{v}t).
\label{eq:alignment_eigs}
\end{align}
To find where the curve $L_{1/2}$ intersects the bifurcation set (i.e., where the critical point on $L_{1/2}$ changes its Morse type), we determine where the Hessian becomes degenerate. As discussed in the main text, degeneracy occurs precisely when one of the eigenvalues vanishes, which happens at the two values
\begin{align}
\tilde{v}t_1 &= \arctan\!\left(\frac{\sqrt{41}-6}{5}\right),\\
\tilde{v}t_2 &= \arctan\!\left(\frac{\sqrt{41}+6}{5}\right).
\label{eq:t12_alignment}
\end{align}
Consequently, $L_{1/2}$ splits into three segments: for $0<\tilde{v}t<\tilde{v}t_1$, the critical point on $L_{1/2}$ is a maximum; for $\tilde{v}t_1<\tilde{v}t<\tilde{v}t_2$, it is a saddle; and for $\tilde{v}t_2<\tilde{v}t<\pi/2$, it is a minimum. Away from the intersections with the bifurcation set, crossing $L_{1/2}$ changes the alignment state of a single continued critical-point branch, without changing its Morse type. According to the segment of $L_{1/2}$ being crossed, this branch is a maximum, a saddle, or a minimum.

However, it is not only important to know what kind of critical point changes its alignment state, but also in which direction the relative orientation of the slider moments changes when one crosses $L_{1/2}$. We restrict ourselves to moving over $ L_{1/2} $ in the $ A $-direction. For this purpose, we analyze the stationarity equations for fixed $\tilde{v}t$ and differentiate $\rho=\zeta^2$ with respect to $A$. We find
\begin{align}
\frac{\mathrm{d}\rho}{\mathrm{d}A}
&=
\frac{2\sqrt{2}\,\bigl|3\cos(2\tilde{v}t)-2\bigr|}
{\left[4-6\cos(2\tilde{v}t)\right]^2-\left[12\sin(2\tilde{v}t)\right]^2}
\nonumber\\
&=
\frac{2\sqrt{2}\,\bigl|3\cos(2\tilde{v}t)-2\bigr|}
{\mu_+(t)\,\mu_-(t)}.
\label{eq:drhodA}
\end{align}
Since the numerator is strictly positive on $L_{1/2}$, the sign of $\mathrm{d}\rho/\mathrm{d}A$ is determined entirely by the determinant $\mu_+(t)\mu_-(t)$ of the Hessian along $L_{1/2}$. Therefore, on the maximum and minimum segments of $L_{1/2}$, i.e.\ for $\tilde{v}t\in[0,\tilde{v}t_1)$ and $\tilde{v}t\in[\tilde{v}t_2,\pi/2]$, one has $\mathrm{d}\rho/\mathrm{d}A>0$, so crossing toward larger $A$ moves the branch from AFM-like to FM-like. On the saddle segment, one has $\mathrm{d}\rho/\mathrm{d}A<0$, and the direction is reversed: crossing toward larger $A$ moves the branch from FM-like to AFM-like. In this sense, the change in relative orientation across the alignment curve is local and depends on which segment of $L_{1/2}$ is being crossed with respect to the $ A $-direction of the $ (A,\tilde{v}t) $-plane.

\section{Derivation of the quasi-static magnetic friction and its explicit approximations}
\label{app:proxies}

This appendix gives a self-contained discussion of the energy-jump based magnetic friction in the CP regime and derives the equations given in Sec.~\ref{SEC:FRICTION}. Within the two-angle reduction, the full energy is $\mathcal{H}=\varepsilon N\Hr/2$, and the reduced energy can be written as (cf. Eq.~\eqref{eq:Hr_chi_zeta})
\begin{align}
\Hr(\chi_+,\zeta;A,\tilde{v}t)
&=
\bigl(4-6\cos (2\chi_+)\bigr)\zeta^2
+2A\zeta\cos(\chi-\tilde{v}t)
\nonumber\\
&+3\cos (2\chi)-2,
\label{eq:app_Hr_chi_zeta}
\end{align}
where $\zeta=\cos(\chi_-)\in[0,1]$. The strict FM states lie on the boundary $\zeta=1$, corresponding to $\chi_-=0$. Restricting the reduced energy to this boundary gives
\begin{align}
\Hr^{\rm FM}(\chi_+;A,\tilde{v}t) &\equiv \Hr^{\rm FM}(\chi_+,1;A,\tilde{v}t) \nonumber\\ 
&=
2-3\cos (2\chi_+)+2A\cos(\chi_+-\tilde{v}t)
\label{eq:app_Hr_FM}
\end{align}
(cf. Eq.~\eqref{eq:Hr_FM_branch}).

\subsection{Reduction of the dissipation to energy drops}

In the quasi-static regime, the state remains close to a stable critical point of the instantaneous energy landscape until the followed minimum disappears or loses stability, as is standard for slowly driven dissipative systems~\cite{Berglund2000}. Let $\vec{p}^{\ast}(t)$ denote the resulting piecewise smooth function through the selected minima during sliding. Between jumps, it satisfies
\begin{equation}
    \nabla_p\Hr\bigl(\vec p^{\ast}(t);A,\tau\bigr)=0.
\label{eq:app_stationary_path}
\end{equation}
Consequently, between the jumps,
\begin{equation}
\frac{\mathrm d}{\mathrm dt}
\Hr\bigl(\vec p^{\ast}(t);A,\tau\bigr)
=
\frac{\partial}{\partial t}
\Hr\bigl(\vec p^{\ast}(t);A,\tau\bigr),
\end{equation}
because the contribution $\nabla_p\Hr\cdot\dot{\vec{p}}^{\ast}$ vanishes. The continuous contributions over one period cancel by periodicity. The leading order dissipation therefore consists only of energy drops. If $\vec{p}_k^{+}$ and $\vec{p}_k^{-}$ denote the states immediately before and after jump $k$, respectively, then
\begin{equation}
E_{\textrm{diss}}
=
-\frac{\varepsilon N}{2}
\sum_k
\left[
\Hr(\vec p_k^{+})-
\Hr(\vec p_k^{-})
\right]
+O(\gamma\tilde v).
\label{eq:app_Ediss_jumps}
\end{equation}
Here, we chose the sign convention where $E_{\textrm{diss}}<0$, while each reduced energy drop appearing in square brackets is positive.

A line-out of the CP regime contains two independent jumps in the time frame of half a period of the drive. The first occurs when the strict-FM minimum loses transverse stability at $B_{\rm FM}^{\perp}$; the second occurs when the subsequently followed off-diagonal minimum terminates at the lower off-diagonal bifurcation curve $B_{\rm off}^{-}$. Reflection symmetry produces one additional copy of each jump over a full period of the drive. We therefore define
\begin{equation}
\mathcal J(A)
:=
\Delta\Hr^{\perp}(A)
+
\Delta\Hr^{\rm low}(A),
\label{eq:app_cp_jump_function}
\end{equation}
where
\begin{align}
\Delta\Hr^{\perp}(A)
&=
\Hr^{\rm FM}
\bigl(\chi_T;A,(\tilde{v}t)_{\perp}\bigr) \nonumber \\
&-
\Hr
\bigl(\chi_{+,\perp}^{-},
      \zeta_{\perp}^{-};A,(\tilde{v}t)_{\perp}\bigr),
\label{eq:app_drop_perp}
\\
\Delta\Hr^{\rm low}(A)
&=
\Hr
\bigl(\chi_{+,\rm low}^{+},
      \zeta_{\rm low}^{+};A,(\tilde{v}t)_{\rm low}\bigr) \nonumber\\
&-
\Hr^{\rm FM}
\bigl(\chi_{+,\rm low}^{-};A,(\tilde{v}t)_{\rm low}\bigr).
\label{eq:app_drop_low}
\end{align}
The factor of two from the symmetry-related copies cancels the factor $1/2$ in $\mathcal{H}=\varepsilon N\Hr/2$. Hence,
\begin{align}
E_{\rm diss}
&=
-N\varepsilon\mathcal J(A)
+\mathcal O(\gamma\tilde v),
\\
\left\langle F_x^{\rm mag}\right\rangle
&=
-N\mathcal J(A)\frac{\varepsilon}{P}
+\mathcal O(\gamma\tilde v).
\label{eq:app_friction_from_J}
\end{align}
Thus, calculating the magnetic friction based on the energy drops reduces to locating four stationary states and evaluating the two energy differences in Eqs.~\eqref{eq:app_drop_perp} and \eqref{eq:app_drop_low}.

\subsection{Explicit construction for a fixed value of $A$}

We next give the complete calculation for a line-out of the $ (A,\tilde{v}t) $-plane for fixed $ A $. As before, the chamber decomposition identifies which solution is the minimum followed by the dynamics and which minimum is reached after a jump. First, we determine the states entering $\Delta\Hr^\perp(A)$ at the transverse loss of stability of the strict FM branch. We then determine the states entering $\Delta\Hr^{\rm low}(A)$ at the lower off-diagonal bifurcation.

\subsubsection{Transverse loss of stability of the strict-FM minimum}

A strict-FM critical point satisfies (cf. Eq.~\eqref{eq:FM_cp_condition})
\begin{equation}
A\sin(\chi-\tilde{v}t)=3\sin (2\chi_+).
\label{eq:app_FM_stationarity}
\end{equation}
At the transverse bifurcation, the Hessian eigenvalue $\lambda_{\perp}$ also vanishes. Denoting the strict-FM angle and the phase of the drive at the transverse bifurcation by $\chi_T$ and $(\tilde{v}t)_{\perp}$, respectively, combining Eq.~\eqref{eq:app_FM_stationarity} with $\lambda_{\perp}=0$ gives [cf. Eqs.~\eqref{eq:fm_transverse_curve_sin} and \eqref{eq:fm_transverse_curve_cos}]
\begin{align}
A\sin\bigl(\chi_T-(\tilde{v}t)_{\perp}\bigr)
&=3\sin(2\chi_T),
\label{eq:app_transverse_sin}
\\
A\cos\bigl(\chi_T-(\tilde{v}t)_{\perp}\bigr)
&=2\bigl(3\cos(2\chi_T)-2\bigr).
\label{eq:app_transverse_cos}
\end{align}
Set $c_T=\cos(2\chi_T)$. Squaring and adding Eqs.~\eqref{eq:app_transverse_sin} and \eqref{eq:app_transverse_cos} eliminates the phase of the drive:
\begin{align}
A^2
&=9\sin^2 (2\chi_T)
 +4\bigl(3\cos (2\chi_T)-2\bigr)^2
\nonumber\\
&=9(1-c_T^2)+4(3c_T-2)^2
\nonumber\\
&=27c_T^2-48c_T+25.
\end{align}
Therefore,
\begin{equation}
    c_T
    =
    \frac{8\pm\sqrt{3A^2-11}}{9}.
\end{equation}
Throughout the CP regime, the plus sign gives $c_T>1$ and is inadmissible. The physical transverse point is consequently
\begin{equation}
    c_T(A)
    =
    \cos (2\chi_T)
    =
    \frac{8-\sqrt{3A^2-11}}{9}.
\label{eq:app_cp_cT}
\end{equation}
Choosing the representative $\chi_T=\tfrac12\arccos (c_T)$, the phase at the jump is obtained directly from Eqs.~\eqref{eq:app_transverse_sin}--\eqref{eq:app_transverse_cos}:
\begin{equation}
(\tilde{v}t)_{\perp}
=
\chi_T-
\operatorname{atan2}
\left(
3\sin(2\chi_T),
2(3c_T-2)
\right),
\label{eq:app_tau_perp}
\end{equation}
modulo the time and angle symmetries used to map the result to the fundamental interval. Here, $\operatorname{atan2}(x,y)\in(-\pi,\pi]$ denotes the quadrant-resolved polar angle $\Theta$ of the vector $(x,y)$. 

At the same parameter set $(A,(\tilde{v}t)_{\perp})$, the post-jump off-diagonal state is a solution of the two interior stationarity equations (cf. Eqs.~\eqref{eq:interior_cp_1} and \eqref{eq:interior_cp_2}),
\begin{align}
A\cos\bigl(\chi_{+,\perp}^{-}-(\tilde{v}t)_{\perp}\bigr)
&=
2\zeta_{\perp}^{-}
\left(3\cos(2\chi_{+,\perp}^{-})-2\right),
\label{eq:app_perp_landing_1}
\\
A\zeta_{\perp}^{-}
\sin\bigl(\chi_{+,\perp}^{-}-(\tilde{v}t)_{\perp}\bigr)
&=
3\left(2(\zeta_{\perp}^{-})^2-1\right)
\sin(2\chi_{+,\perp}^{-}).
\label{eq:app_perp_landing_2}
\end{align}
These equations may have several roots, but the chamber decomposition identifies the relevant one. The state reached by the jump is the continuation of the unique minimum in at least one of the chambers adjacent to the bifurcation curve. By continuity of the critical points and their energies, its limiting state on the curve remains an energy minimum among the critical points. Thus, $(\chi_{+,\perp}^{-},\zeta_{\perp}^{-})$ is selected as the roots that minimize the reduced energy.
 
\subsubsection{Termination of the off-diagonal minimum at $B_{\rm off}^{-}$}

The lower off-diagonal bifurcation curve is most conveniently parameterized by the pre-jump value $\zeta\in(0,1]$. The Hessian-degeneracy condition gives (cf. Eq.~\eqref{eq:offdiag_xpm})
\begin{equation}
 x_{\pm}(\zeta)
 =
 \cos (2\chi_+)
 =
\frac{
4\zeta^2
\pm
\sqrt{160\zeta^6+324\zeta^4+108\zeta^2+9}
}
{3(8\zeta^2+1)}.
\label{eq:app_xpm}
\end{equation}
The relevant curve is the minus branch, which corresponds to $B_{\rm off}^{-}$. We define
\begin{equation}
 x_L(\zeta):=x_-(\zeta),
 \qquad
 \chi_L(\zeta):=\frac12\arccos (x_L(\zeta)).
\end{equation}
The two stationarity equations determine the corresponding parameter values through
\begin{align}
C_L(\zeta)
&:=
2\zeta\bigl(3x_L(\zeta)-2\bigr),
\\
S_L(\zeta)
&:=
\frac{3(2\zeta^2-1)\sqrt{1-x_L(\zeta)^2}}{\zeta},
\end{align}
so that
\begin{align}
&A_L(\zeta)
:=
\sqrt{C_L(\zeta)^2+S_L(\zeta)^2}
\nonumber\\
&=
\left[
4\zeta^2\bigl(3x_L(\zeta)-2\bigr)^2
+
\frac{9(2\zeta^2-1)^2\bigl(1-x_L(\zeta)^2\bigr)}{\zeta^2}
\right]^{1/2},
\label{eq:app_AL}
\\
&(\tilde{v}t)_L(\zeta)
:=
\chi_L(\zeta)-
\operatorname{atan2}\bigl(S_L(\zeta),C_L(\zeta)\bigr).
\label{eq:app_tauL}
\end{align}
Thus, $\zeta$ parameterizes the lower off-diagonal bifurcation curve: $A_L(\zeta)$ gives the corresponding value of $A$, while $(\tilde{v}t)_L(\zeta)$ gives the phase of the drive. Equations~\eqref{eq:app_AL} and \eqref{eq:app_tauL} are obtained by specializing the general parametric representation in Eqs.~\eqref{eq:offdiag_param_A} and \eqref{eq:offdiag_param_t} to the minus branch and eliminating $\chi_+$ in favor of $\zeta$. 

For arbitrary but fixed $A$, the pre-jump state at the lower off-diagonal boundary is found from
\begin{equation}
    A_L(\zeta_{\rm low}^{+})=A.
\label{eq:app_lower_inverse}
\end{equation}
These are the intersections between $B_{\rm off}^{-}$ and vertical line-outs. Such intersections, i.e. solutions, exist for $A$ between the lowest point of $B_{\rm off}^{-}$ and its highest point $B_{\rm off}^{-}$ defined as  $A_-$ and $A_*$, respectively. Equation~\eqref{eq:app_lower_inverse} generally has two roots. The chamber decomposition is essential here: the trajectory is determined by the larger-$\zeta$ root. It connects to the minimum reached after crossing $B_{\rm FM}^{\perp}$ and is the minimum that terminates when the line-out reaches $B_{\rm off}^{-}$. The smaller-$\zeta$ solution belongs to a different segment of the bifurcation set and is not the pre-jump state for the energy drop in the CP regime. Once the larger root has been selected,
\begin{equation}
\chi_{+,\rm low}^{+}
=
\chi_L(\zeta_{\rm low}^{+}),
\qquad
(\tilde{v}t)_{\rm low}
=
(\tilde{v}t)_L(\zeta_{\rm low}^{+}).
\label{eq:app_lower_departure}
\end{equation}

The post-jump strict-FM angle solves
\begin{equation}
A\sin(\chi_{+,\rm low}^{-}-(\tilde{v}t)_{\rm low})
=
3\sin (2\chi_{+,\rm low}^{-}).
\label{eq:app_lower_FM_landing}
\end{equation}
Again, this equation can have several roots. We know from the chamber decomposition that there are only two strict-FM states, up to symmetries, one of which is a maximum. Thus, the selected strict-FM state has to be the energetically lower one.

\subsection{Implicit magnetic friction law of the CP regime}

The energy evaluation simplifies at an off-diagonal critical point. From the first stationarity equation,
\begin{equation}
A\cos(\chi_+-\tilde{v}t)
=
2\zeta(3\cos (2\chi_+)-2).
\end{equation}
Substitution into Eq.~\eqref{eq:app_Hr_chi_zeta} gives
\begin{align}
\Hr(\chi_+,\zeta;A,\tilde{v}t)
&=
(4-6\cos (2\chi_+))\zeta^2
\nonumber \\&
+4\zeta^2(3\cos (2\chi_+)-2)
+3\cos (2\chi_+)-2
\nonumber\\
&=
(3c-2)(1+2\zeta^2).
\end{align}
It is therefore useful to define
\begin{equation}
    G(x_1,x_2):=(3x_1-2)(1+2x_2^2),
\label{eq:app_G}
\end{equation}
for which
\begin{equation}
    \Hr(\chi_+,\zeta;A,\tau)=G(\cos (2\chi_+),\zeta)
\label{eq:app_stationary_energy}
\end{equation}
at every off-diagonal critical point.

The strict-FM pre-jump energy at the transverse event is also explicit. Using Eq.~\eqref{eq:app_transverse_cos} in Eq.~\eqref{eq:app_Hr_FM},
\begin{align}
\Hr^{\rm FM}(\chi_T;A,(\tilde{v}t)_{\perp})
&=
2-3c_T+4(3c_T-2)
\nonumber\\
&=
9c_T-6
=
2-\sqrt{3A^2-11}.
\label{eq:app_transverse_energy}
\end{align}
Combining Eqs.~\eqref{eq:app_drop_perp}, \eqref{eq:app_drop_low}, \eqref{eq:app_stationary_energy}, and \eqref{eq:app_transverse_energy} gives the complete implicit jump function,
\begin{align}
\mathcal J(A)
={}&
2-\sqrt{3A^2-11}
-
G\left(
\cos 2\chi_{+,\perp}^{-}(A),
\zeta_{\perp}^{-}(A)
\right)
\nonumber\\
&+
G\left(
 x_L(\zeta_{\rm low}^{+}(A)),
 \zeta_{\rm low}^{+}(A)
\right)
\nonumber\\
&-
\Hr^{\rm FM}
\left(
\chi_{+,\rm low}^{-}(A);
A,(\tilde{v}t)_{\rm low}(A)
\right).
\label{eq:cp_implicit_friction}
\end{align}
For a fixed $A$, this equation requires only the finite-dimensional stationary solutions described above. No numerical integration of the equations of motion is needed.

In practical terms, the magnetic friction of the CP regime is calculated as follows: first solve Eq.~\eqref{eq:app_cp_cT} and Eq.~\eqref{eq:app_tau_perp}; then solve Eqs.~\eqref{eq:app_perp_landing_1}--\eqref{eq:app_perp_landing_2} for the chamber-selected off-diagonal post-jump minimum; solve Eq.~\eqref{eq:app_lower_inverse} for the larger-$\zeta$ root; solve Eq.~\eqref{eq:app_lower_FM_landing} for the chamber-selected strict-FM post-jump minimum; finally evaluate Eq.~\eqref{eq:cp_implicit_friction} and insert the result into Eq.~\eqref{eq:app_friction_from_J}. 

\subsection{Boundaries of layer separation interval of the CP regime}

The main text made the claim that the jumps contributing to the magnetic friction only occur for $A\in [A_-,A_*]$, where $A_-$  and $A_*$ were declared to be the minimal and maximal $A$-values on $B_{\rm off}^{-}$. Equivalently, $A_- = \min_{\zeta}[A_L(\zeta)]$ and $A_+= \max_\zeta [A_L(\zeta)]$.
This can be read of directly from the chamber decomposition in Fig.~\ref{fig:BuL}, which allows us to deduce that for $A\le A_-$ and $A\ge 10$ some states exist for all $t$, leading to jump-free trajectories. Similarly one can read off that for $A\in[A_*,10]$ the transitions dictated by the quasi-static dynamics have to be jump-free as well, given that negative energy drops are non-physical. The next two paragraphs establish how $A_-$ and $A_*$ can be analytically calculated.

\subsubsection{Lower boundary}

Equation~\eqref{eq:app_AL} gives the $A$-coordinate of the lower off-diagonal bifurcation curve $B_{\rm off}^{-}$ as a function of its parameter $\zeta\in(0,1]$. Hence, the lower boundary is
\begin{equation}
    A_- = \min_{0<\zeta\leq 1} [A_L(\zeta)].
\end{equation}
Since $A_L(\zeta)\geq0$ by Eq.~\eqref{eq:app_AL}, minimizing $A_L^2(\zeta)$ is equivalent to minimizing $A_L(\zeta)$.

Using the minus branch $x_L(\zeta)=x_-(\zeta)$ from Eq.~\eqref{eq:app_xpm}, define
\begin{equation*}
    \Delta(\zeta)
    :=160\zeta^6+324\zeta^4+108\zeta^2+9.
\end{equation*}
Direct substitution into Eq.~\eqref{eq:app_AL} gives the explicit one-variable expression
\begin{align}
A_L^2(\zeta)
={}&
\frac{4}{(8\zeta^2+1)^2}
\Bigl[
864\zeta^6-84\zeta^4+36\zeta^2+9
\nonumber\\
&\hspace{0mm}
+\bigl(32\zeta^4-4\zeta^2+2\bigr)\sqrt{\Delta(\zeta)}
\Bigr].
\label{eq:app_AL_squared_y}
\end{align}
We introduce the two even polynomials
\begin{align*}
P(\zeta)
&:=576\zeta^6+216\zeta^4-38\zeta^2-9,
\\
Q(\zeta)
&:=2560\zeta^{10}+4096\zeta^8+1752\zeta^6
  +128\zeta^4-72\zeta^2-9.
\end{align*}
Differentiating Eq.~\eqref{eq:app_AL_squared_y} directly yields
\begin{equation}
\frac{\mathrm d}{\mathrm d\zeta}A_L^2(\zeta)
=
\frac{96\zeta}{(8\zeta^2+1)^3\sqrt{\Delta(\zeta)}}
\left[
P(\zeta)\sqrt{\Delta(\zeta)}+2Q(\zeta)
\right].
\end{equation}
At an interior minimum, $\zeta>0$, and every factor outside the square brackets is nonzero. The stationary-point condition is therefore
\begin{equation*}
P(\zeta)\sqrt{\Delta(\zeta)}+2Q(\zeta)=0.
\end{equation*}
Squaring this equation and factoring gives
\begin{align}
0
={}&P(\zeta)^2\Delta(\zeta)-4Q(\zeta)^2
\nonumber\\
={}&-5(2\zeta^2+1)(2\zeta^2+3)(8\zeta^2+1)^3(10\zeta^2+3)
\nonumber\\
&\quad
\left(256\zeta^8-384\zeta^6+16\zeta^4+54\zeta^2-9\right).
\label{eq:app_degeneracy_xy}
\end{align}
All factors preceding the final parenthesis in Eq.~\eqref{eq:app_degeneracy_xy} are nonzero for real $\zeta$. Thus, every interior stationary point satisfies the even degree-eight equation
\begin{equation*}
256\zeta^8-384\zeta^6+16\zeta^4+54\zeta^2-9=0.
\end{equation*}
For the purpose of solving this equation, one may now set $y=\zeta^2$, which reduces it to
\begin{equation}
256y^4-384y^3+16y^2+54y-9=0.
\label{eq:app_endpoint_quartic}
\end{equation}
Equation~\eqref{eq:app_endpoint_quartic} has two roots in $0<y<1$,
\begin{align}
    y_0&=0.2106609145241876\ldots, \\
    y_1&=0.3209745470481302\ldots .
\end{align}
Because Eq.~\eqref{eq:app_degeneracy_xy} was obtained by squaring, these roots must be checked in the unsquared stationary-point condition. Only $y_0$ satisfies that condition; $y_1$ is an artificial root introduced by squaring. The opposite-sign roots of the degree-eight equation are excluded because the physical parameter satisfies $\zeta>0$. Hence,
\begin{equation*}
    \zeta_0=\sqrt{y_0}
    =0.4589781198752156\ldots .
\end{equation*}
There are no other stationary points in $0<\zeta<1$, and direct evaluation of Eq.~\eqref{eq:app_AL} at $\zeta=1$ and in the limit $\zeta\downarrow0$ gives values larger than $A_L(\zeta_0)$. Therefore, $\zeta_0$ gives the global minimum,
\begin{align}
A_-
&:=A_L(\zeta_0)
\nonumber\\
&=4.634583466980286\ldots .
\label{eq:app_Aminus}
\end{align}

\subsubsection{Upper boundary}

At the upper boundary, the relevant lower off-diagonal branch reaches the boundary $\zeta=1$ and merges with the transverse strict-FM branch. Equation~\eqref{eq:app_xpm} gives
\begin{equation}
    x_*:=x_-(1)
    =
    \frac{4-\sqrt{601}}{27}.
\end{equation}
Substituting $\zeta=1$ and $x=x_*$ into Eq.~\eqref{eq:app_AL} yields
\begin{align}
A_*
&=
A_L(1)
\nonumber\\
&=
\left[
4(3x_*-2)^2+9(1-x_*^2)
\right]^{1/2}
\nonumber\\
&=
\sqrt{\frac{1100+40\sqrt{601}}{27}}
=
8.778365775883558\ldots .
\label{eq:app_Astar}
\end{align}
At this point, $\chi_*:=\tfrac12\arccos (x_*)$ and the phase obtained from Eq.~\eqref{eq:app_tauL} agrees with the transverse phase in Eq.~\eqref{eq:app_tau_perp}. Thus, the two bifurcation events coalesce.

The contribution of the jump to the magnetic friction is therefore only defined on
\begin{equation}
    A_-<A<A_*.
\label{eq:app_CP_interval}
\end{equation}
For $A<A_-$, the line-out does not complete the alternating jump cycle. For $A>A_*$, the corresponding branches connect smoothly and there is no $\mathcal O(1)$ jump contribution (with respect to $ \gamma \tilde{v} $).

\subsection{Why the magnetic friction a the lower bound shows square root behavior}

The square-root behavior arises because $\mathcal{J}$ depends smoothly on $\zeta^+{\rm low}$ for $A\in[A-,A_*]$, whereas $\zeta^+{\rm low}$ is obtained by locally inverting $A_L(\zeta)$. Since $A_L(\zeta)$ has a quadratic minimum at $\zeta=\zeta_0$, with $A_L(\zeta_0)=A_-$, this inversion produces a leading square-root dependence on $A-A_-$.

Write
\begin{equation}
    \delta\zeta:=\zeta_{\rm low}^{+}-\zeta_0>0.
\end{equation}
The positive sign is fixed by the chamber decomposition because the pre-jump state is the larger $\zeta$ root of Eq.~\eqref{eq:app_lower_inverse}. Since $\frac{\textrm{d}A_L}{\textrm{d}\zeta} (\zeta_0)=0$ and $\frac{\textrm{d}^2A_L}{\textrm{d}\zeta^2} (\zeta_0)>0$, 
\begin{equation}
A_L(\zeta_0+ \delta\zeta)
=
A_-
+\frac12 a_2\delta\zeta^2
+\frac16 a_3 \delta\zeta^3
+O( \delta\zeta^4),
\label{eq:app_AL_lower_Taylor}
\end{equation}
where
\begin{equation}
    a_2=\frac{\textrm{d}^2A_L}{\textrm{d}\zeta^2}(\zeta_0)
    =53.65064324950696\ldots .
\end{equation}
Series reversion of Eq.~\eqref{eq:app_AL_lower_Taylor} gives
\begin{equation}
 \delta\zeta
=
\sqrt{\frac{2}{a_2}}\sqrt{A-A_-}
-
\frac{a_3}{3a_2^2}(A-A_-)
+
O\bigl((A-A_-)^{3/2}\bigr).
\label{eq:app_lower_series_reversion}
\end{equation}
Thus, any quantity that varies linearly with $ \delta\zeta$ necessarily develops a square-root dependence on $A-A_-$. This is the leading Puiseux behavior of the implicitly defined cruve~\cite{Wall2004}.

Now, since $\mathcal{J}$ is smooth, one may write
\begin{equation}
\mathcal J(\zeta_0+ \delta\zeta)
=
J_0+J_1 \delta\zeta+O( \delta\zeta^2),
\end{equation}
with
\begin{align}
J_0
&=
4.975766521411717\ldots,
\\
J_1
&=
\left.\frac{\mathrm d\mathcal J}{\mathrm d\zeta}\right|_{\zeta_0}
=
-15.61081481170818\ldots .
\end{align}
Combining this expansion with Eq.~\eqref{eq:app_lower_series_reversion} yields
\begin{align}
\mathcal J(A)
={}&
J_0
+J_1\sqrt{\frac{2}{a_2}}\sqrt{A-A_-}
+O(A-A_-)
\nonumber\\
={}&
4.975766521411717 \nonumber \\
&-3.014068394247566\sqrt{A-A_-}
\nonumber\\
&+O(A-A_-),
\qquad A\downarrow A_-.
\label{eq:app_lower_proxy}
\end{align}
This derivation explains both the square root and the choice of its positive branch. The nonanalyticity is not caused by a singular energy function; it appears because the smooth parameterization by $\zeta$ is inverted at the minimum of $A_L(\zeta)$.

\subsection{Why the magnetic friction at the upper boundary begins at third order}

At the upper boundary, $A=A^\ast$, the transverse strict-FM bifurcation curve and the lower off-diagonal bifurcation curve occur at the same time. The chamber decomposition furthermore shows that the two jumps connect the same two followed critical points in opposite directions, depending on the direction in which the trajectory passes through the corresponding part of the cycle. Since each physical jump must lower the energy, the two limiting energy differences can only both be admissible if they vanish. Thus,
\begin{align}
    J(A^\ast)=0,
\end{align}
which immediately explains the zeroth-order cancellation.

There is also a useful local interpretation of why the next two orders vanish. The strict-FM state corresponds to $\chi_-=0$ with $\zeta=\cos(\chi_-)$. At the upper endpoint, the Hessian eigenvalue associated with the $\chi_+$ direction remains nonzero. The implicit-function theorem therefore allows $\chi_+$ to be eliminated locally in favor of $\chi_-$. Since the energy is invariant under $\chi_-\rightarrow-\chi_-$, the resulting energy difference between the nearby off-diagonal and strict-FM critical points is even,
\begin{align}
    \Delta H_{\rm eff}
      =a_2\chi_-^2+a_4\chi_-^4+a_6\chi_-^6+O(\chi_-^8).
    \label{eq:upper-effective-energy}
\end{align}
The coefficient $a_2$ is simply the quadratic curvature of the energy in the transverse direction. More precisely, expansion of the reduced energy about the strict-FM state gives
\begin{align}
    a_2=\frac{1}{2}\lambda_\perp,
\end{align}
where $\lambda_\perp$ is the transverse Hessian eigenvalue introduced above. At the transverse strict-FM bifurcation curve,
$\lambda_\perp=0$, and hence $a_2=0$. Physically, the energy landscape has therefore lost its quadratic restoring term for deviations away from the strict-FM state.

The cancellation of the quartic term uses the second bifurcation curve that reaches the same endpoint. After eliminating $\chi_+$,
$\partial_{\chi_-}^2\Delta H_{\rm eff}$ is the local curvature of the energy along the remaining direction. The off-diagonal critical point lies on a bifurcation curve and is therefore both stationary and degenerate, so along this curve
\begin{align}
    \partial_{\chi_-}\Delta H_{\rm eff}=0,
    \qquad
    \partial_{\chi_-}^2\Delta H_{\rm eff}=0 .
\end{align}
Using Eq.~\eqref{eq:upper-effective-energy}, these conditions give, for $\chi_-\neq0$,
\begin{align}
    0&=\frac{1}{\chi_-}\partial_{\chi_-}\Delta H_{\rm eff}
      =2a_2+4a_4\chi_-^2+6a_6\chi_-^4+O(\chi_-^6),\\
    0&=\partial_{\chi_-}^2\Delta H_{\rm eff}
      =2a_2+12a_4\chi_-^2+30a_6\chi_-^4+O(\chi_-^6).
\end{align}
Subtracting the two equations and taking the bifurcation curve to the upper endpoint, $\chi_-\rightarrow0$, yields $a_4=0$. Thus the first possible energy drop is
\begin{align}
    \Delta H_{\rm eff}=O(\chi_-^6).
\end{align}
Because
\begin{align}
    1-\zeta=1-\cos(\chi_-)=\frac{\chi_-^2}{2}+O(\chi_-^4),
\end{align}
the energy drop, and therefore the jump function, can first appear at order $(1-\zeta)^3$.

The above argument is intentionally presented in condensed form. Each step can be made fully rigorous by carrying out the corresponding local implicit-function and derivative calculations, but doing so produces a considerably longer and tedious derivation without changing the physical content of the result.

For completeness, the same conclusion can also be obtained directly from the implicit equations defining the critical states. Set
\begin{align}
    \delta\zeta:=1-\zeta^+_{\rm low}\geq0 .
\end{align}
Expanding all critical states entering Eq.~\eqref{eq:cp_implicit_friction} in powers of
$\delta\zeta$, inserting these expansions into the corresponding
stationarity and Hessian degeneracy conditions, and repeatedly differentiating the implicit equations determines the coefficients recursively. Matching equal powers in Eq.~\eqref{eq:cp_implicit_friction} then gives
\begin{align}
    J(1-\delta\zeta)
      =J_0^{(\ast)}
       +J_1^{(\ast)}\delta\zeta
       +J_2^{(\ast)}\delta\zeta^2
       +J_3^{(\ast)}\delta\zeta^3
       +O(\delta\zeta^4),
\end{align}
with
\begin{align}
    J_0^{(\ast)}=J_1^{(\ast)}=J_2^{(\ast)}=0,
    \qquad
    J_3^{(\ast)}=130.0663396613375\ldots .
\end{align}
Thus, the zeroth-order cancellation follows immediately from the
chamber decomposition, while the local energy argument explains the additional cancellations; alternatively, all three can be verified directly by recursive implicit differentiation and coefficient matching.

Finally, because $ \partial A_L(1) / \partial \zeta \neq0$,
\begin{align}
    \delta\zeta
      =\frac{A^\ast-A}{a_1}
       +O\!\left((A^\ast-A)^2\right),
    \qquad a_1=\partial A_L(1) / \partial\zeta,
\end{align}
and therefore
\begin{align}
\mathcal J(A)
={}&
\frac{J^{(*)}_3}{a_1^3}(A_*-A)^3
+ O\bigl((A_*-A)^4\bigr)
\nonumber\\
={}&
0.1043554086690065\,(A_*-A)^3
\nonumber\\
&+O\bigl((A_*-A)^4\bigr),
\qquad A\uparrow A_*.
\label{eq:app_upper_proxy}
\end{align}
as stated in the main text.

\subsection{Global rational approximations}

The expansions at the boundaries of the CP regime are local. A Taylor series in $A$ cannot efficiently represent the full interval because $\mathcal J(A)$ has a square-root branch behavior at $A_-$. A convenient global coordinate is therefore
\begin{equation}
    s
    =
    \sqrt{\frac{A-A_-}{A_*-A_-}},
    \qquad 0\leq s\leq1.
\label{eq:app_scaled_coordinate}
\end{equation}
Near $A_-$, the coordinate $s$ is proportional to $\zeta_{\rm low}^{+}-\zeta_0$, so the expansion at the lower boundary becomes an ordinary power series in $s$. Moreover,
\begin{equation}
    A_*-A=(A_*-A_-)(1-s^2),
\end{equation}
so the cubic upper-boundary behavior can be enforced exactly by a factor $(1-s^2)^3$. We consequently use a constrained multipoint Pad\'e-type form~\cite{BakerGravesMorris1996,Buslaev2013,AptekarevYattselev2015}
\begin{equation}
\mathcal J_{[m/n]}(A)
=
(1-s^2)^3\frac{P_m(s)}{Q_n(s)},
\label{eq:app_rational_form}
\end{equation}
with
\begin{equation}
P_m(s)=\sum_{k=0}^{m}\alpha_k s^k,
\qquad
Q_n(s)=1+\sum_{k=1}^{n}\beta_k s^k.
\end{equation}
The normalization $Q_n(0)=1$ removes the irrelevant common scaling of numerator and denominator.

Choose $m+n+1$ Chebyshev-distributed~\footnote{Any distribution of $m+n+1$ distinct points on $[0,1]$ leads to the same kind of linear system for the coefficients $\alpha_i$ and $\beta_i$. Chebyshev are not provably best for this specific application, but outperformed any other distributions that we tested.} points on $[0,1]$~\cite{Trefethen2020}, 
\begin{equation}
 s_i
 =
 \frac12\left[
 1+\cos\left(
 \frac{(i+\tfrac12)\pi}{m+n+1}
 \right)
 \right],
 \qquad
 i=0,\ldots,m+n,
\label{eq:app_nodes}
\end{equation}
and map them to
\begin{equation}
    A_i=A_-+(A_*-A_-)s_i^2.
\end{equation}
At each $A_i$, calculate the chamber-selected implicit value $\mathcal J_i=\mathcal J(A_i)$ from Eq.~\eqref{eq:cp_implicit_friction}. The interpolation condition $\mathcal J_{[m/n]}(A_i)=\mathcal J_i$ is equivalent to
\begin{equation}
(1-s_i^2)^3\sum_{k=0}^{m}\alpha_k s_i^k
-
\mathcal J_i\sum_{k=1}^{n}\beta_k s_i^k
=
\mathcal J_i,
\label{eq:app_linear_fit}
\end{equation}
which is linear in the $m+n+1$ unknown coefficients $\alpha_k$ and $\beta_k$. The resulting expression is admissible only if
\begin{equation}
    Q_n(s)\neq0
    \qquad\text{for all }s\in[0,1],
\label{eq:app_pole_condition}
\end{equation}
since a zero of the denominator would introduce a spurious pole in the physical interval.

For $m=n=3$, the pole-free approximation is
\begin{align}
P_3(s)
={}&
4.975912864275528 \nonumber\\
&-4.641152742433373\,s
\nonumber\\
&+2.553486904795277\,s^2
\nonumber\\
&-1.550335590883650\,s^3,
\\
Q_3(s)
={}&
1+0.3032598453406820\,s
\nonumber\\
&-1.888267337852114\,s^2
\nonumber\\
&+0.7651534006123918\,s^3.
\label{eq:app_33}
\end{align}
For $m=n=5$,
\begin{align}
P_5(s)
={}&
4.975766415704477 \nonumber\\
&-4.274194970406318\,s
\nonumber\\
&-0.1730907043784124\,s^2
\nonumber\\
&-0.3543330035300636\,s^3
\nonumber\\
&+1.096588408089551\,s^4
\nonumber\\
&-0.2490546748077928\,s^5,
\\
Q_5(s)
={}&
1+0.3740733914140734\,s
\nonumber\\
&-2.297807656107382\,s^2
\nonumber\\
&-0.05560186936526916\,s^3
\nonumber\\
&+2.004251070603650\,s^4
\nonumber\\
&-0.8873164067638182\,s^5.
\label{eq:app_55}
\end{align}
Both denominators remain nonzero on $[0,1]$. An independent reference sweep of $20\,001$ points, uniform in $\zeta_{\rm low}^{+}\in(\zeta_0,1)$, gives the largest observed relative residuals $\epsilon_\infty^{[m/n]} = \sup_{A\in [A_-,A_*)}\lvert\mathcal{J}_{[m/n]}-\mathcal{J}\rvert/\lvert\mathcal{J}\rvert$  as
\begin{equation}
\epsilon_\infty^{[3/3]}
=1.37\times10^{-4},
\qquad
\epsilon_\infty^{[5/5]}
=5.27\times10^{-6}.
\label{eq:app_fit_errors}
\end{equation}
These values are dense-grid validation errors, not rigorous uniform error bounds.

\section{Boundary conditions and longer-range interactions}
\label{app:robustness_two_angle}

In this appendix, we briefly discuss how longer-range interactions and open boundary conditions along the $y$-direction affect the properties of the simplified model. Our purpose is not to exhaustively classify the dynamic behavior and dissipation under these modifications, but rather to identify which aspects of our analysis remain applicable and which conclusions are specific to the model presented here. 

The following discussion supports viewing the periodic nearest-neighbor model as the simplest analytically tractable member of a broader class of rotor-array models. Longer-range interactions and open boundaries in $ y $-direction introduce quantitative changes and variations in the dynamics close to the edges of the slider, but the energy-landscape framework remains applicable. The precise chamber boundaries and friction response are model dependent and must be recalculated for each extended system.

\subsection{Longer-range interactions within the slider}

We first discuss the influence of including additional dipole-dipole interactions instead of using only nearest neighbors. For this, we continue to assume that the system can be described by the two-angle reduction $\theta_e$ and $\theta_o$. For any symmetric finite set of interaction pairs, direct summation gives a reduced energy function of the form
\begin{align}
H_r^{(M)}
&=
-a_M\left[\cos(2\theta_e)+\cos(2\theta_o)\right]
+b_M\cos(\theta_e-\theta_o)
\nonumber\\
&-c_M\cos(\theta_e+\theta_o)
+A\left[
\cos(\theta_e-\tilde vt)
+\cos(\theta_o-\tilde vt)
\right] \nonumber \\
&+ \textrm{const}.
\label{eq:appC-general-two-angle-energy}
\end{align}
Here, $M$ denotes the retained interaction range, and $a_M$, $b_M$, and $c_M$ are lattice sums over the included interaction pairs between slider moments. The nearest-neighbor model studied in the main text corresponds to
\begin{equation}
(a_M,b_M,c_M)
=
\left(\frac{3}{2},2,0\right).
\end{equation}
Longer-range interactions modify all three coefficients and generally produce a nonzero $c_M$. They therefore add a coupling proportional to $\cos(\theta_e+\theta_o)$ that is absent from the nearest-neighbor model.

Using
\begin{equation}
\chi_+
=
\frac{\theta_e+\theta_o}{2},
\qquad
\chi_-
=
\frac{\theta_e-\theta_o}{2},
\qquad
\zeta=\cos(\chi_-),
\end{equation}
Eq.~\eqref{eq:appC-general-two-angle-energy} becomes
\begin{align}
H_r^{(M)}(\chi_+,\zeta)
&=
\left(2b_M-4a_M\cos 2\chi_+\right)\zeta^2 \nonumber \\
&+2A\cos(\chi_+-\tilde vt)\zeta
+(2a_M-c_M)\cos 2\chi_+
\nonumber \\ &-b_M
+ \textrm{const}.
\label{eq:appC-general-zeta-energy}
\end{align}
The reduced energy therefore remains quadratic in $\zeta$. The same type of frozen-energy analysis used in the main text can still be applied: critical points can be continued, their Hessian eigenvalues can be evaluated, and bifurcation and alignment curves can be calculated. The resulting chamber boundaries must, however, be recomputed and might look very different than those of the nearest-neighbor model.

The infinite-range lattice sums are well-defined. In two dimensions, the number of interaction partners in a shell of radius $r$ grows as $r$, while the dipolar interaction decreases as $r^{-3}$. The absolute contribution from the omitted shells beyond a range $M$ is therefore bounded by
\begin{equation}
\sum_{r>M}O(r^{-2})=O(M^{-1}).
\end{equation}
More detailed calculations and tests show convergence of the reduced coefficients and indicate that the FM, CP, and AFM regimes remain present when the full pair interactions are included. Their parameter ranges change quantitatively. Also, the layer separation interval of the CP regime becomes narrower and the corresponding alignment states within the regime become more FM-like. A systematic classification of the resulting chamber structure and magnetic friction is left for future work.

\subsection{Open boundaries and the fixed layer}

Assigning a single angle $\theta_j$ to every rotor in row $j$ treats the slider as translationally uniform, and therefore effectively infinite or periodic, along the (sliding) $x$-direction. In the following, we discuss the influence of open boundaries while retaining this assumption. Hence, in this work, only the influence of open boundary conditions in the $y$-direction is considered.

For a slider containing $N$ rows and interactions extending over at most $M$ chain separations, opening the transverse boundary reduces the number of pairs separated by $u$ chains by the factor
\begin{equation}
1-\frac{u}{N}.
\end{equation}
For fixed $M \ll N$, the resulting correction to spatially averaged interaction coefficients is therefore of order $M/N$. Note, however, that local effects at the outermost chains can nevertheless remain of order one.

Open boundaries in the $y$-direction also make the rows close to the edges of the slider inequivalent to those in the bulk. Consequently, representing every even chain by $\theta_e$ and every odd chain by $\theta_o$ is no longer an exact reduction of the open-system dynamics. It remains useful for describing the bulk energy structure, but it does not resolve local motion at the open boundaries. Our numerical tests with open boundaries in the $y$-direction show that the qualitative dynamical regimes persist, although their transition values and detailed trajectories can change. As discussed in Appendix~\ref{APP:MF}, opening the boundary does not affect the definitions and identities used to calculate magnetic friction in our numerical integrations, since the torques between slider moments are still derived from a potential.

The substrate is treated differently. It is not represented as a finite lattice of individual magnets, but is coarse-grained into the spatially uniform rotating field represented by $ A\cos(\theta_j-\tilde vt) $. The substrate is therefore effectively assumed to be infinite or periodic and has no boundaries. The longer-range calculations above modify only the internal interactions within the slider; the effective substrate term is retained unchanged (if one neglects higher order harmonics). Finite substrate boundaries, spatial variations of the substrate field, and a fully finite slider that also loses translational symmetry along the sliding direction are not included. Their treatment requires returning to the complete two-layer geometry with $ N \times N$ degrees of freedom and is left for future work.

\bibliography{Quellen}

\end{document}